\documentclass[12pt]{article}  
\usepackage[a4paper,textwidth=490.0pt,textheight=703.1pt]{geometry}
\usepackage{slashed}
\usepackage{graphicx}
\usepackage{epsfig}
\usepackage{amsmath}
\usepackage{amssymb}
\usepackage{ulem}
\usepackage{mdwlist}
\usepackage{color}                                     
\usepackage{float}
\usepackage[rflt]{floatflt}
\usepackage{slashed}
\usepackage{cite}
\usepackage{enumitem}   
\usepackage{lineno}

\usepackage{amsthm}
\usepackage{enumitem}

\newtheorem{theorem}{Theorem}[section]
\newtheorem{proposition}[theorem]{{Proposition}}

\newtheorem{lemma}[theorem]{{Lemma}}
\newtheorem{remark}[theorem]{{Remark}}
\newtheorem{definition}[theorem]{{Definition}}

\newcommand{\ELI}{{\rm ELi}}

\usepackage[mathscr]{euscript}
 \let\mathscr\relax

\usepackage{array}

\usepackage[english]{babel}
\usepackage[latin1]{inputenc}
\usepackage[T1]{fontenc}
\usepackage{ae}

\usepackage{url}

\usepackage{amsmath, amsthm, amssymb}

\catcode`,\active
\newcommand*\pFq{\begingroup
        \catcode`\,\active
        \def ,{\mskip\pFqskip\relax}%
        \dopFq
}
\catcode`\,12

\usepackage{array}

\usepackage[english]{babel}
\usepackage[latin1]{inputenc}
\usepackage[T1]{fontenc}
\usepackage{ae}

\newcommand{\legendre}[2]{\genfrac{(}{)}{}{}{#1}{#2}}

\newcommand{\modd}{{\rm mod}}

\newcommand{\Li}{{\rm Li}}
\newcommand{\Mvec}{{\rm \bf M}}

\usepackage{rotating}

\usepackage{graphicx}

\newcounter{mmacnt}
\def\restartmma{\setcounter{mmacnt}{0}}
\restartmma \catcode`|=\active
\def|#1|{\mathrm{#1}}
\catcode`|=12
\newenvironment{mma}{
 \par\smallskip
 \catcode`|=\active
 \parskip=0pt\parindent=0pt 
 \small
 \def\In##1\\{%
   \def\linebreak{\hfill\break\null\qquad}%
   \refstepcounter{mmacnt}
   \hangindent=2.5em\hangafter=0
   \leavevmode
   \llap{\tiny\sffamily In[\arabic{mmacnt}]:=\kern.5em}%
   \mathversion{bold}\footnotesize$\displaystyle##1$\normalsize
   \mathversion{normal}\par
 }%
 \def\Print##1\\{%
   \def\linebreak{\hfill\break}%
   \hangindent=2.5em\hangafter=0
   \leavevmode ##1\par}%
 \def\Out##1\\{%
   \def\linebreak{$\hfill\break\null\hfill$}%
   \kern\abovedisplayskip\par
   \hangindent=2.5em\hangafter=0
   \leavevmode
   \llap{\tiny\sffamily Out[\arabic{mmacnt}]=\kern.5em}
   \footnotesize$\displaystyle##1$\normalsize\hfill\null\par
   \kern\belowdisplayskip
 }%
 \def\Warning##1##2\\{%
   \def\linebreak{\hfill\break}%
   \hangindent=2.5em\hangafter=0
   \leavevmode
   {\scriptsize##1 : ##2}\par}%
}{%
 \par\smallskip
}

\usepackage{color}

\newenvironment{fshaded}{%
\MakeFramed {\FrameRestore}
}%
{\endMakeFramed}

\usepackage{tikz}
\usetikzlibrary{matrix}

\allowdisplaybreaks[4]

\begin{document}
\setlength{\baselineskip}{0.515cm}
\sloppy
\thispagestyle{empty}

\begin{flushleft}
DESY 16--147  
\hfill  
\\
DO--TH 16/14  \\
May 2017\\
\end{flushleft}

\setcounter{table}{0}

\mbox{}
\vspace*{\fill}
\begin{center}

{\Large\bf Iterated Elliptic and Hypergeometric Integrals}

\vspace*{4mm}
{\Large \bf for Feynman Diagrams}

\vspace{2.5cm}
\large
J.~Ablinger$^a$,
J.~Bl\"umlein$^b$,
A.~De Freitas$^b$,
M.~van Hoeij$^c$,\\
E.~Imamoglu$^c$, 
C.G.~Raab$^d$,
C.-S.~Radu$^a$,
and C.~Schneider$^a$

\vspace{1cm}
\normalsize   
{\it  $^a$Research Institute for Symbolic Computation (RISC),}\\
{\it Johannes Kepler University,
Altenbergerstra\ss{}e 69, A-4040 Linz, Austria}

\vspace*{2mm}
{\it  $^b$Deutsches Elektronen--Synchrotron, DESY,}\\
{\it  Platanenallee 6, D-15738 Zeuthen, Germany}

\vspace*{2mm}
{\it  $^c$Department of Mathematics, Florida State University,} \\ 
{\it  208 Love Building, 1017 Academic Way,  Tallahassee, FL 32306-4510, USA}\\

\vspace*{2mm}
{\it  $^d$Institute for Algebra, Johannes Kepler University,
Altenbergerstra\ss{}e 69, A-4040 Linz, Austria}
\\

\end{center}

\normalsize
\vspace{\fill}
\begin{abstract}
\noindent 
We calculate 3-loop master integrals for heavy quark correlators and the 3-loop QCD corrections
to the $\rho$-parameter. They obey non-factorizing differential equations of second order with 
more than three singularities, which cannot be factorized in Mellin-$N$ space either. The solution 
of the homogeneous equations is possible in terms of convergent close integer power series as 
$_2F_1$ Gau\ss{} hypergeometric functions at rational argument. In some cases, integrals of this
type can be mapped to complete elliptic integrals at rational argument. This class of functions 
appears to be the next one arising in the calculation of more complicated Feynman integrals
following the harmonic polylogarithms, generalized polylogarithms, cyclotomic harmonic polylogarithms, 
square-root valued iterated integrals, and combinations thereof, which appear in simpler cases. 
The inhomogeneous solution of the corresponding differential equations can be given in terms of iterative
integrals, where the new innermost letter itself is not an iterative integral. A new class of iterative 
integrals is introduced containing letters in which (multiple) definite integrals appear as factors.
For the elliptic case, we also derive the solution in terms of integrals over modular functions and 
also modular forms, using $q$-product and series representations implied by Jacobi's $\vartheta_i$ 
functions and Dedekind's $\eta$-function. The corresponding representations can be traced back to 
polynomials out of Lambert--Eisenstein series, having representations also as elliptic polylogarithms, 
a $q$-factorial $1/\eta^k(\tau)$, logarithms and polylogarithms of $q$ and their $q$-integrals.  
Due to the specific form of the physical variable $x(q)$ for different processes, different representations 
do usually appear. Numerical results are also presented. 
\end{abstract}

\vspace*{\fill}
\noindent
\numberwithin{equation}{section}

\newpage
\section{Introduction}
\label{sec:1}

\vspace*{1mm}
\noindent
Many single scale Feynman integrals arising in massless and massive multi-loop calculations in Quantum
Chromodynamics (QCD) \cite{QCD} have been found to be expressible in terms of harmonic polylogarithms (HPLs) 
\cite{Remiddi:1999ew}, generalized harmonic polylogarithms \cite{Moch:2001zr,Ablinger:2013cf}, cyclotomic 
harmonic polylogarithms \cite{Ablinger:2011te}, square-root valued iterated integrals \cite{Ablinger:2014bra}, 
as well as more general functions, entering the corresponding alphabet in integral iteration. After taking 
a Mellin transform
\begin{eqnarray}
\label{eq:MELL}
\Mvec[f(x)](N) = \int_0^1 dx x^{N} f(x),
\end{eqnarray}
they can be equivalently expressed in terms of harmonic sums \cite{Blumlein:1998if,Vermaseren:1998uu}
in the simpler examples and finite sums of different kinds in the other cases 
\cite{Moch:2001zr,Ablinger:2013cf,Ablinger:2011te,Ablinger:2014bra}, supplemented by special numbers like 
the multiple zeta values \cite{Blumlein:2009cf} and others appearing in the limit $N \rightarrow \infty$ of the
nested sums, or the value at $x = 1$ of the iterated integrals in Refs.~\cite{Blumlein:1998if,Vermaseren:1998uu,
Remiddi:1999ew,Moch:2001zr,Ablinger:2013cf,Ablinger:2011te,Ablinger:2014bra}.

In many higher order calculations a considerable reduction of the number of integrals to be calculated is obtained using
integration by parts identities (IBPs) \cite{IBP}, which allow to express all required integrals in 
terms of a much smaller set of so called master integrals. Differential equations satisfied by these master integrals
\cite{DEQ,Caffo:1998du} can then be obtained by taking their derivatives with respect to the parameters of the problem and 
inserting the IBPs in the result. What remains is to solve these differential equations, given initial or boundary 
conditions, if possible analytically. One way of doing this is to derive an associated system of difference 
equations \cite{DIFFEQ,Ablinger:2010ty,Ablinger:2014nga,Ablinger:2015tua} after applying a mapping through a formal 
Taylor series or a Mellin transform. If these equations factorize to first order equations, we can use the algorithm
presented in Ref.~\cite{Ablinger:2015tua} for general bases to solve these systems analytically and to 
find the corresponding alphabets over which the iterated integrals or nested sums are built. The final solution in $N$ and
$x$ space is found by using the packages {\tt Sigma} \cite{SIG1,SIG2}, {\tt EvaluateMultiSums} and 
{\tt SumProduction}~\cite{EMSSP}. 

However, there are physical cases where full first order factorizations cannot
be obtained for either the differential equations in $x$ or the difference equations in $N$.\footnote{There 
are cases in which factorization fails in either $x$- or $N$-space, but not in 
both, cf.~\cite{Ablinger:2013eba}. This opportunity has to be always checked.} The next level of complexity 
is given by non-factorizable differential or difference equations of second order. Examples of this are the massive 
sunrise and kite integrals \cite{SABRY,TLSR1,Caffo:1998du,Caffo:2002ch,Laporta:2004rb,TLSR2,TLSR2a,Bailey:2008ib,
Broadhurst:2008mx,TLSR3,BLOCH2,TLSR4,Adams:2014vja,Adams:2015gva,Adams:2015ydq,TLSR5,Remiddi:2016gno,Adams:2016xah,
TSLR6}.

In the present paper, we will address the analytic solution of typical cases of this kind, related to a series of 
master integrals appearing in the 3-loop corrections of the $\rho$-parameter in \cite{Grigo:2012ji}. It turns 
out that these integrals are more general than those appearing in the sunrise and kite diagrams, due to the 
appearance of also the elliptic integral of the second kind, ${\bf E}(z)$, which cannot be transformed away.
The corresponding second order differential equations have more than three singularities, as in the case 
of the Heun equation \cite{HEUN}. For the sake of generality, we will seek solutions of the second order 
homogeneous differential equations which are given in terms of  Gau\ss{}' $_2F_1$ functions \cite{GAUSS} within 
the class of globally bounded solutions \cite{VANH1}, cf. also \cite{BOSTAN}. Here the parameters of the 
$_2F_1$ function are rational numbers and the argument is a rational function of $x$. The complete elliptic integrals 
${\bf K}(z)$ and ${\bf E}(z)$ \cite{TRICOMI,ELLINT,ABEL,JAC1}\footnote{As a convention, the modulus $k^2 = z$ is 
chosen in this paper, also used within the framework of {\tt Mathematica}.} are special cases of this class. 

The hypergeometric 
function obeys different relations like the Euler- and Pfaff-transformations \cite{HYP,SLATER}, the 24 Kummer 
solutions \cite{KUMMER,RIEMANN} and the 15 Gau\ss{}' contiguous relations \cite{HYP,SLATER}. There are more 
special transformations for higher than first order in the argument \cite{KUMMER,GOURSAT,VANHVID,VID2}. In the 
present case, equivalent $_2F_1$ representations are obtained by applying arithmetic triangle groups 
\cite{TAKEUCHI}. The corresponding algorithm has been described in Ref.~\cite{IVH} in its present most far reaching 
form. The relations of this type may be useful to transform a found solution into another one, which might be 
particularly convenient. In the case a function space of more solutions is considered, these relations have to be 
exploited to check the independence of the basis elements.

The main idea of the approach presented here is to obtain the factorization of a high order scalar difference or 
differential equation, after uncoupling \cite{Zuercher:94,ORESYS,NewUncouplingMethod} the corresponding linear 
systems, to all first order parts and its second order contributions. While the first order parts have been 
algorithmically 
solved in Ref.~\cite{Ablinger:2015tua}, the treatment of second order differential equations shall be 
automated.\footnote{For more involved physical problems also irreducible higher order differential equations may 
occur.} The class of $_2F_1$ solutions has an algorithmic automation to a wide extent \cite{IVH} and it seems that 
this class constitutes the next one following the iterated square-root valued letters in massive single-scale
3-loop integrals. Applying this method, we obtain the corresponding ${_2F_1}$ functions with (partly) fractional 
parameters and rational argument, and ir(rational) pre-factors, forming the {\it new letters} of the otherwise 
iterated integrals. These letters contain a {\it definite} integral by virtue of the integral representation of 
the ${_2F_1}$ function, which cannot be fully transformed into an integral depending on the follow-up integration 
variable only through its integration boundaries. In general, we have therefore to iterate new letters of this kind. 
Through this we obtain a complete algorithmic automation of the solution also when second order 
differential operators contribute, having $_2F_1$ solutions. 

As it will be shown, in a series of cases the reduction of the $_2F_1$ functions to complete elliptic integrals 
${\bf E}(r(z))$ and ${\bf K}(r(z))$ is possible. Therefore we also study special representations in terms of 
$q$-series, which have been obtained in the case of the sunrise graph, 
cf.~\cite{BLOCH2,Adams:2014vja,Adams:2015gva,Adams:2015ydq,Adams:2016xah}, before. More general representations
are needed for the integrals considered in the present paper and we describe the necessary extension. 

In performing a higher loop calculation, in intermediary steps usually more complicated nested 
integrals and sums occur than in the final result\footnote{For a simple earlier case, see e.g.
\cite{Vermaseren:2005qc,Ablinger:2010ty}.}. The various necessary decompositions of the problem that have 
to be performed, such as the integration by parts reduction and others, account for this in part. It appears 
therefore necessary to have full control on these occurring structures first, which finally may simplify 
in the result. Moreover, experience tells that in more general situations, more and more of these structures 
survive, cf. \cite{Ablinger:2014nga} in comparison to \cite{Vermaseren:2005qc}. If the mathematical properties 
of the quantities occurring are known in detail, various future calculations in the field will be more easily 
performed.

The paper is organized as follows. In Section~2 we present the linear systems of first order differential equations  
for master integrals in Ref.~\cite{Grigo:2012ji} which cannot be solved in terms of iterated 
integrals. We first perform a decoupling into a scalar second order equation and an associated equation for each 
system. Using the algorithm of Ref.~\cite{Ablinger:2015tua}, the non-iterative solution both in $x$- and 
Mellin-$N$-space is uniquely established. In Section~3 we first determine the homogeneous solutions of the 
second order equations, which turn out to be $_2F_1$ solutions~\cite{VANH1} and obey representations in terms of weighted 
complete elliptic integrals of first and second kind at rational argument. In Section~4 we derive the solutions in 
the inhomogeneous case, which are given by iterated integrals in which some letters are given by a higher 
transcendental function defined by a {\it non-iterative}, i.e. definite, integral in part. We present 
numerical representations for $x \in [0,1]$ deriving overlapping expansions around $x=0$ and $x=1$. The methods 
presented apply to a much wider class of functions than the ones being discussed here specifically. These need neither 
to have a representation in terms of elliptic integrals, nor of just a $_2F_1$ function. The respective letter 
can be given by {\it any} multiple definite integral. 

Owing to the fact that we have elliptic solutions in the present cases we may also try to cast the solution in 
terms of series in the nome 
\begin{eqnarray}
\label{eq:nome}
q = \exp(i\pi\tau),
\end{eqnarray}
where 
\begin{eqnarray}
\label{qeqa}
\tau = i \frac{{\bf K}(1-z(x))}{{\bf K}(z(x))}~~~~
\text{with}~~~~\tau \in \mathbb{H} =
\left\{z \in \mathbb{C}, {\sf Im}(z) > 0\right\}
\end{eqnarray}
denotes the ratio of two complete elliptic integrals of first kind and $z(x)$ is a rational function associated
to the elliptic curve of the problem. It is now interesting to see which closed form solutions the corresponding 
series in $q$ obey. All contributing quantities can be expressed in terms ratios of the Dedekind $\eta(\tau)$ 
function \cite{DEDEKINDeta}, cf.~Eq.~(\ref{eq:DEDmain}).  However, various building blocks are only  modular forms 
\cite{KF,KOECHER,MILNE,RADEMACHER,DIAMOND,SCHOENENBERG,APOSTOL,KOEHLER,ONO,MIYAKE,SERRE,MART1} up to  an additional 
factor of 
\begin{eqnarray}
\label{eq:etaF}
\frac{1}{\eta^k(\tau)},~~~~k > 0,~~~k \in \mathbb{N}. 
\end{eqnarray}
We seek in particular modular forms which have a representation in terms of Lambert--Eisenstein 
series \cite{LAMBERT,EISENSTEIN} and can thus be represented by elliptic polylogarithms \cite{ELLPOL}. However, the 
$\eta$-factor (\ref{eq:etaF}) in general remains. Thus the occurring $q$-integrands are modulated by a $q$-factorial 
\cite{SLATER,ERDELYI1} denominator. 

Structures of the kind  for $k > 0$ are frequent even in the early literature. A prominent case is given by the 
invariant $J$, see e.g. \cite{HURWITZ},
\begin{eqnarray}
J = \frac{G_2^3(q)}{216000 \Delta(q)}
\end{eqnarray} 
with $G_2(q)$ an Eisenstein series, cf.~Eq.(\ref{eq:EISEN1}), and the discriminant $\Delta$
\begin{eqnarray}
\Delta(q) = (2 \pi)^{12} q^2 \prod_{k=1}^\infty (1-q^{2k})^{24} = (2 \pi)^{12} \eta^{24}(\tau).
\end{eqnarray} 
In the more special case considered in
\cite{BLOCH2,Adams:2014vja,Adams:2015gva,Adams:2015ydq,Adams:2016xah} terms of this kind are not present.

For the present solutions, we develop the formalism in Section~5. We discuss possible extensions of integral classes to 
the present case in Section~6 and of elliptic polylogarithms \cite{ELLPOL}, as has been done previously in the  
calculation of the two-loop sunrise and kite-diagrams 
\cite{BLOCH2,Adams:2014vja,Adams:2015gva,Adams:2015ydq,Adams:2016xah}.
Here the usual variable $x$ is mapped to the nome $q$, expressing all contributing functions in the new variable.
This can be done for all the individual building blocks, the product of which forms the desired solution.
Section~7 contains the conclusions. 

In Appendix~\ref{sec:8} we briefly describe the algorithm finding for second 
order
ordinary differential equations $_2F_1$ solutions with a rational function argument. 
In Appendix~\ref{sec:10} we present for convenience details for the necessary steps
to arrive at the elliptic polylogarithmic representation in the examples of the sunrise and kite integrals 
\cite{BLOCH2,Adams:2014vja,Adams:2015gva,Adams:2015ydq,Adams:2016xah}. Here we compare some results given 
in Refs.~\cite{BLOCH2} and 
\cite{Adams:2014vja}. In Appendix~\ref{sec:11} we list a series of new sums, which simplify the recent results 
on the sunrise diagram of Ref.~\cite{Ananthanarayan:2016pos}. 

In the present paper we present the results together 
with all necessary technical details and we try to refer to the related mathematical literature as widely as 
possible, to allow a wide community of readers to apply the methods presented here to other problems. 
\section{The Differential Equations} 
\label{sec:2} 

\vspace*{1mm}
\noindent
The master integrals considered in this paper satisfy linear differential equations of second order
\begin{eqnarray}
\left[\frac{d^2}{dx^2} + p(x) \frac{d}{dx} + q(x)\right] \psi(x) = N(x)~, 
\label{eq:D2}
\end{eqnarray}
with rational functions $r(x) = p(x), q(x)$, which may be decomposed into
\begin{eqnarray}
r(x) = \sum_{k=1}^{n_r} \frac{b_k^{(r)}}{x - a_k^{(r)}},~~~~~a_k^{(r)},b_k^{(r)} \in \mathbb{Z}~.
\label{eq:rx}
\end{eqnarray}
The homogeneous equation is solved by the functions $\psi_{1,2}^{(0)}(x)$, which are linearly 
independent, i.e. their Wronskian $W$ obeys
\begin{eqnarray}
W(x) = \psi_1^{(0)}(x) \frac{d}{dx} \psi_2^{(0)}(x)
     - \psi_2^{(0)}(x) \frac{d}{dx} \psi_1^{(0)}(x) \neq 0~.
\end{eqnarray}
The homogeneous Eq.~(\ref{eq:D2}) determines the well-known differential equation for $W(x)$
\begin{eqnarray}
\frac{d}{dx} W(x) = -p(x) W(x)~,
\end{eqnarray}
which, by virtue of (\ref{eq:rx}), has the solution
\begin{eqnarray}
\label{eq:WR}
W(x) = \prod_{k=1}^{n_1} \left(\frac{1}{x-a_k^{(1)}}\right)^{b_k^{(1)}}~,
\end{eqnarray}
normalizing the functions $\psi_{1,2}^{(0)}$ accordingly.
A particular solution of the inhomogeneous equation (\ref{eq:D2}) is then obtained by Euler-Lagrange 
variation of constants \cite{EULLAG}  
\begin{eqnarray}
\label{eq:INHOM}
\psi(x) &=&~~\psi_1^{(0)}(x) \left[C_1 - \int dx~\psi_2^{(0)}(x) n(x)\right] 
 + \psi_2^{(0)}(x) \left[C_2 + \int dx~\psi_1^{(0)}(x) n(x) \right]~,
\label{eq:DI}
\end{eqnarray}
with 
\begin{eqnarray}
n(x) = \frac{N(x)}{W(x)} 
\end{eqnarray}
and two constants $C_{1,2}$ to be determined by special physical requirements. We will 
consider indefinite  integrals for the solution (\ref{eq:INHOM}), which allows for more singular 
integrands. For the class of differential equations under 
consideration, $N(x)$ can be expressed by harmonic polylogarithms and rational functions, $W(x)$ is
a polynomial, and the functions $\psi_{1,2}^{(0)}(x)$ turn out to be higher transcendental functions, which are
even expressible by complete elliptic integrals in the cases considered here. Therefore Eq.~(\ref{eq:DI}) 
constitutes a {\it nested} integral of known functions 
\cite{Remiddi:1999ew,Moch:2001zr,Ablinger:2013cf,Ablinger:2011te,Ablinger:2014bra} and elliptic 
integrals at rational argument.

We consider the systems of differential equations \cite{Grigo:2012ji} for the $O(\varepsilon^0)$ terms 
of the master integrals
\begin{eqnarray}
\renewcommand{\arraystretch}{1.6}
\frac{d}{dx}
\left(
\begin{array}{c}
f_{8a}(x) \\
f_{9a}(x) 
\end{array}
\right)
= \left(
\begin{array}{cc}
\frac{4}{x} & \frac{6}{x} \\
\frac{4(x^2-3)}{x(x^2-9)(x^2-1)} &
\frac{2(x^4-9)}{x(x^2-9)(x^2-1)} 
\end{array}
\right) \otimes
\left(
\begin{array}{c}
f_{8a}(x) \\
f_{9a}(x) 
\end{array}
\right)
+
\left(
\begin{array}{c}
N_{8a}(x) \\
N_{9a}(x) 
\end{array}
\right)
\end{eqnarray}
\renewcommand{\arraystretch}{1.0}
and
\begin{eqnarray}
\renewcommand{\arraystretch}{1.6}
\frac{d}{dx}
\left(
\begin{array}{c}
f_{8b}(x) \\
f_{9b}(x) 
\end{array}
\right)
= \left(
\begin{array}{cc}
\frac{4}{x} & \frac{2}{x} \\
\frac{4(3x^2-1)}{x(9x^2-1)(x^2-1)} &
\frac{2(9x^4-1)}{x(9x^2-1)(x^2-1)} 
\end{array}
\right) \otimes
\left(
\begin{array}{c}
f_{8b}(x) \\
f_{9b}(x) 
\end{array}
\right)
+
\left(
\begin{array}{c}
N_{8b}(x) \\
N_{9b}(x)
\end{array}
\right),
\end{eqnarray}
\renewcommand{\arraystretch}{1.0}
with
\begin{eqnarray}
N_{8a}(x) &=& \frac{15 (-13 - 16 x^2 + x^4)}{4 x} - 3 x (-24 + x^2) \ln(x) - 18 x \ln^2(x)
\\
N_{9a}(x) &=& \frac{1755 + 1863 x^2 - 1255 x^4 + 157 x^6}{12 x(x^2-9)(x^2-1)}
- \frac{x (324 - 145 x^2 + 15 x^4)}{(x^2-9)(x^2-1)} \ln(x)
\nonumber\\ &&
+ \frac{2 x (45 - 17 x^2 + 2 x^4)}{(x^2-9)(x^2-1)} \ln^2(x)
- \frac{16 x^3}{3(x^2-9)(x^2-1)} \ln^3(x)
\\
N_{8b}(x) &=& -\frac{15 (-1+16 x^2+13 x^4)}{4 x}+9 x (8+15 x^2) 
\ln(x) - 18 (x+6 x^3) \ln^2(x)
\\
N_{9b}(x) &=& 
-\frac{15-397 x^2+925 x^4+297 x^6}{4 x (9x^2-1) (x^2-1)}
+\frac{3 x \left(-36+35 x^2+195 x^4\right)}{(9x^2-1)(x^2-1)} \ln(x)
\nonumber\\ &&
+\frac{6 x \left(5+37 x^2-144 x^4\right)}{(9x^2-1)(x^2-1)} \ln^2(x)
+\frac{16 x^3 \left(-8+27 x^2\right)}{(9x^2-1)(x^2-1)} \ln^3(x).
\end{eqnarray}
By applying decoupling algorithms \cite{Zuercher:94,ORESYS,NewUncouplingMethod} one obtains the following 
scalar differential equation 
\begin{eqnarray}
\label{eq:one}
0 &=& 
\frac{d^2}{dx^2} f_{8a}(x) 
+\frac{9-30 x^2+5 x^4}{x(x^2-1)(9-x^2)} \frac{d}{dx} f_{8a}(x)
-\frac{8 (-3+x^2)}{(9-x^2)(x^2-1)} f_{8a}(x)
\nonumber\\ &&
-\frac{32 x^2}{(9-x^2)(x^2-1)} \ln^3(x)
+\frac{12 (-9+13 x^2+2 x^4)}{(9-x^2)(x^2-1)} \ln^2(x)
\nonumber\\ &&
-\frac{6 (-54+62 x^2+x^4+x^6)}{(9-x^2)(x^2-1)} \ln(x)
+\frac{-1161+251 x^2+61 x^4+9 x^6}{2 (9-x^2)(x^2-1)}\\
&& \hspace*{-2.5cm} \text{and the further equation} \hfill \nonumber
\\
\label{eq:f9bb}
f_{9a}(x) &=& -\frac{5}{8}(-13 - 16 x^2 + x^4) 
+  \frac{x^2}{2} (-24 + x^2) \ln(x)  + 3 x^2 \ln^2(x) - \frac{2}{3} f_{8a}(x)
\nonumber\\ &&
+  \frac{x}{6} \frac{d}{dx} f_{8a}(x).
\\
&& \hspace*{-2.5cm} \text{Likewise, one obtains for the second system} \hfill \nonumber
\\
\label{eq:two}
0 &=& 
\frac{d^2}{dx^2} f_{8b}(x)
- \frac{1 - 30 x^2 + 45 x^4}{x(9x^2-1)(x^2-1)} \frac{d}{dx} f_{8b}(x)
+ \frac{24 (-1 + 3 x^2)}{(9x^2-1)(x^2-1)} f_{8b}(x)
\nonumber\\ &&
-\frac{32 x^2 (-8 + 27 x^2)}{(9x^2-1)(x^2-1)} \ln^3(x)
+\frac{12 (1 - 13 x^2 - 216 x^4 + 162 x^6)}{(9x^2-1)(x^2-1)} \ln^2(x)
\nonumber\\ &&
-\frac{6 (6 - 46 x^2 - 399 x^4 + 81 x^6)}{(9x^2-1)(x^2-1)} \ln(x)
-\frac{61 - 415 x^2 + 2199 x^4 + 675 x^6}{2(9x^2-1)(x^2-1)}~,
\\
\label{eq:f9bc}
f_{9b}(x) &=& 
\label{eq:f9b}
 9 x^2 \left(1+6 x^2\right) \ln^2(x)
- \frac{9}{2} x^2 \left(8+15 x^2\right) \ln(x) 
+ \frac{15}{8} \left(-1+16 x^2+13 x^4\right)
            -2 f_{8b}(x) 
\nonumber\\ &&
+ \frac{1}{2} x \frac{d}{dx} f_{8b}(x)~.
\end{eqnarray}
The above differential equations of second order contain more than three singularities. We seek solutions in 
terms of Gau\ss{}' hypergeometric functions with rational arguments, following the algorithm described 
in Appendix~\ref{sec:8}.  It turns out that these differential equations have $_2F_1$ solutions.

Two more master integrals are obtained as integrals over the previous solutions. They obey the differential
equations
\begin{eqnarray}
\label{eq:10a}
\frac{d}{dx} f_{10a}(x) &=& 
\frac{6 x \big(x^2-6\big) H_{0}^2(x)}{(x^2-1)^2}
-\frac{4 x \big(x^2-3\big)}{3(x^2-1)^2} H_0^3(x)
+\frac{8}{x} \Biggr[H_{-1,-1,0}(x)- H_{-1,1,0}(x) - H_{1,-1,0}(x) 
\nonumber\\ &&
+ H_{1,1,0}(x) \Biggr]
-\frac{8 x}{x^2-1} \left[H_{0,-1,0}(x) - H_{0,1,0}(x) \right]
+\frac{x\big[342-51 x^2+2 \pi ^2 \big(x^2-1\big)\big]}{3(x^2-1)^2} 
\nonumber\\ &&
\times H_0(x)
+\frac{x^4 (165-176 \zeta_3)+8 x^2 (-105+22 \zeta_3) - 585}{12(x^2-1)^2 x}
-\frac{2 \pi ^2}{3x} \left[ H_{-1}(x) - H_1(x) \right]
\nonumber\\ &&
+\frac{4}{(x^2-1)^2 x} f_{8a}(x)
+\frac{2 \big(x^2+3\big)}{(x^2-1)^2 x} f_{9a}(x),
\\
\label{eq:10b}
\frac{d}{dx} f_{10b}(x) &=& 
-\frac{6 \big(15 x^2+2\big)}{(x^2-1)^2 x} H_0^2(x)
+\frac{4 \big(4 x^4+33 x^2+1\big)}{3 (x^2-1)^2 x} H_0^3(x)
+\frac{\big(8-16 x^2\big)}{(x^2-1) x} \Biggl[H_{0,-1,0}(x)
\nonumber\\ &&
-H_{0,1,0}(x)\Biggr]
+\frac{8}{x} \Biggl[ H_{-1,-1,0}(x)
- 2H_{-1,0,0}(x)
- H_{-1,1,0}(x)
- H_{1,-1,0}(x)
\nonumber\\ &&
+ 2 H_{1,0,0}(x)
+ H_{1,1,0}(x) \Biggr]
+\frac{3 \big(59 x^2+38\big)
        +\pi ^2 \big(4 x^4-6 x^2+2\big)
}{3(x^2-1)^2 x}  H_0(x) 
\nonumber\\ &&
+\frac{15
        -192 x^6 \zeta_3
        -8 x^2 (45+2 \zeta_3)
        +x^4 (-75+208 \zeta_3)
}{12 (x^2-1)^2 x^3} 
-\frac{2 \pi ^2}{3x} \left[H_{-1}(x) -  H_1(x)\right]
\nonumber\\ &&
+\frac{4}{3 (x^2-1)^2 x^3} f_{8b}(x)
+\frac{2 \big(
        3 x^2+1\big)}{3 (x^2-1)^2 x^3} f_{9b}(x)~,
\end{eqnarray}
with $\zeta_k, k \in \mathbb{N},~k \geq 2$ the values of the Riemann $\zeta$-function at integer argument and 
the harmonic polylogarithms $H_{\vec{a}}(x)$ are defined by \cite{Remiddi:1999ew}
\begin{eqnarray}
\label{eq:HPL}
H_{b,\vec{a}}(x) &=& \int_0^x dy f_b(y) H_{\vec{a}}(y);~~ f_b(x) \in \left\{f_0,f_1, f_{-1}\right\} 
\equiv \left\{\frac{1}{x},\frac{1}{1-x}, \frac{1}{1+x} \right\}; 
\nonumber\\ && H_{\tiny \underbrace{\tiny{0, ...,0}}_k}(x) =\frac{1}{k!} \ln^k(x); H_\emptyset(x) 
\equiv 1~. 
\end{eqnarray}
Subsequently, we will use the shorthand notation  $H_{\vec{a}}(x) \equiv H_{\vec{a}}$.
The harmonic polylogarithms occurring in the inhomogeneities of Eqs.~(\ref{eq:10a}, \ref{eq:10b}) can 
be rewritten as polynomials of 
\begin{eqnarray}
H_0, 
H_{1}, 
H_{-1}, 
H_{0,-1},  
H_{0,1}, 
H_{0,0,-1},
H_{0,0,1},
H_{0,-1,-1},  
H_{0,-1,1},
H_{0,1,-1}, 
H_{0,1,1},
\end{eqnarray}
cf.~\cite{Blumlein:2003gb}.

\section{Solution of the homogeneous equation}
\label{sec:3}

\vspace*{1mm}
\noindent
In the following we will derive the solution of the homogeneous part of Eqs.~(\ref{eq:one}, \ref{eq:two}) 
as examples in detail, using the algorithm outlined in Ref.~\cite{IVH}, see also Appendix~\ref{sec:8}.  

\begin{figure}[H]
\centering
\includegraphics[width=0.60\textwidth]{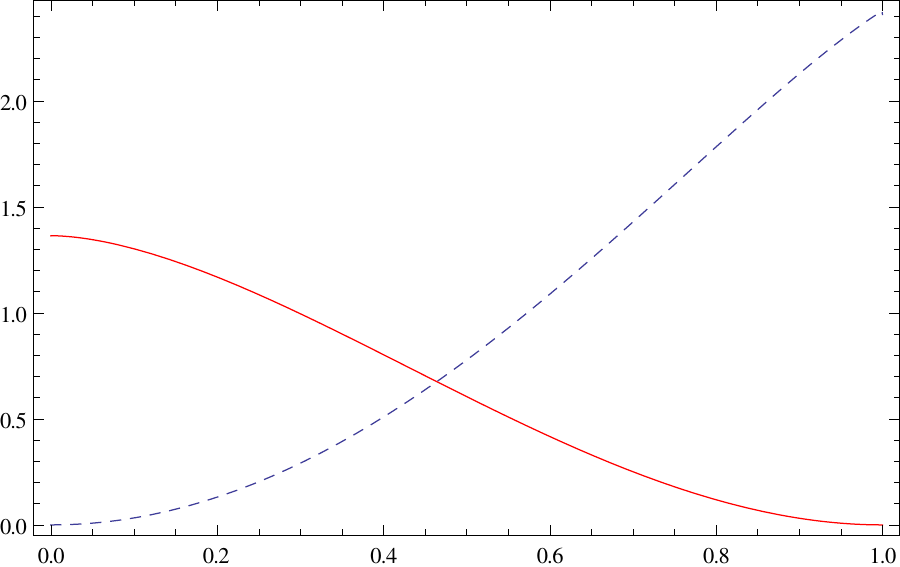}
\caption{\sf \small The homogeneous solutions (\ref{eq:ps1a}, \ref{eq:ps2a}) $\psi_{1a}^{(0)}$ (dashed line) and 
$\psi_{2a}^{(0)}$ (full line) as  functions of $x$.}
\label{fig:pic1}
\end{figure}

The homogeneous solutions of Eq.~(\ref{eq:one}) read
\begin{eqnarray}
\label{eq:ps1a}
\psi_{1a}^{(0)}(x) &=& \sqrt{2 \sqrt{3} \pi} 
\frac{x^2 (x^2-1)^2 (x^2-9)^2}{(x^2+3)^4}
\pFq{2}{1}{{\tfrac{4}{3}},\tfrac{5}{3}}{2}{z}
\\
\label{eq:ps2a}
\psi_{2a}^{(0)}(x) &=& \sqrt{2 \sqrt{3} \pi}
\frac{x^2 (x^2-1)^2 (x^2-9)^2}{(x^2+3)^4}
\pFq{2}{1}{{\tfrac{4}{3}},\tfrac{5}{3}}{2}{1-z},
\end{eqnarray}
with
\begin{eqnarray}
z = \frac{x^2(x^2-9)^2}{(x^2+3)^3}~.
\end{eqnarray}
The $_2F_1$ solutions (\ref{eq:ps1a}, \ref{eq:ps2a}) are close integer series \cite{VANH1} obeying
\begin{eqnarray}
\label{eq:CIS}
b \sum_{k=0}^\infty \tau_k (c \cdot z)^k  =  \sum_{k=0}^\infty m_k z^k,~\text{with}~~\tau_k, b \in
\mathbb{Q},~m_k \in \mathbb{Z},
\end{eqnarray}
with $c = 27$. The Wronskian for this system is
\begin{eqnarray}
\label{eq:W1}
W(x) = x (9 - x^2) (x^2-1).
\end{eqnarray}
The solutions are shown  in Figure~\ref{fig:pic1}.

\noindent
Equivalent solutions are found by applying relations due to triangle groups \cite{TAKEUCHI}, see 
Appendix~\ref{sec:8}, 
\begin{eqnarray}
\label{eq:ps1b}
\psi_{1b}^{(0)}(x) &=& \frac{\sqrt{\pi}}{4 \sqrt{6}} \Biggl\{
- (x-1)(x-3)(x+3)^2 \sqrt{\frac{x+1}{9-3x}} 
\pFq{2}{1}{{\tfrac{1}{2}},\tfrac{1}{2}}{1}{z}
\nonumber\\
&& + (x^2+3)(x-3)^2     \sqrt{\frac{x+1}{9-3x}}
\pFq{2}{1}{{\tfrac{1}{2}},-\tfrac{1}{2}}{1}{z} \Biggr\}
\\
\label{eq:ps2b}
\psi_{2b}^{(0)}(x) &=&
\frac{2\sqrt{\pi}}{\sqrt{6}}\Biggl\{x^2 \sqrt{(x+1)(9-3x)}
\pFq{2}{1}{{\tfrac{1}{2}},\tfrac{1}{2}}{1}{1-z}
\nonumber\\ && 
+\frac{1}{8} \sqrt{(x+1)(9-3x)} (x-3)(x^2+3)
\pFq{2}{1}{{\tfrac{1}{2}},-\tfrac{1}{2}}{1}{1-z}\Biggr\},
\end{eqnarray}
where
\begin{eqnarray}
\label{eq:z1}
z(x) = -\frac{16x^3}{(x+1)(x-3)^3}~.
\end{eqnarray}
These  solutions have the Wronskian (\ref{eq:W1}) up to a sign\footnote{The sign can be adjusted by $\psi_{1b}^{(0)} 
\leftrightarrow \psi_{2b}^{(0)}$.} but differ from those in (\ref{eq:ps1a}, \ref{eq:ps2a}).
The ratios of the homogeneous solutions are given by
\begin{eqnarray}
\frac{\psi_{1a}^{(0)}(x)}{\psi_{1b}^{(0)}(x)} &=& 3^{3/4} \sqrt{\frac{\pi}{2}} 
\\
\frac{\psi_{2a}^{(0)}(x)}{\psi_{2b}^{(0)}(x)} &=& - \frac{1}{3^{3/4}} \sqrt{\frac{2}{\pi}}~.
\end{eqnarray}

The hypergeometric functions appearing in (\ref{eq:ps1b}, \ref{eq:ps2b}) are given in terms of complete elliptic 
integrals \cite{TRICOMI}
\begin{eqnarray}
\label{ell:K}
\pFq{2}{1}{{\tfrac{1}{2}},\tfrac{1}{2}}{1}{z}  &=& \frac{2}{\pi} {\bf K}(z) \\
\label{ell:E}
\pFq{2}{1}{{\tfrac{1}{2}},-\tfrac{1}{2}}{1}{z} &=& \frac{2}{\pi} {\bf E}(z)~.
\end{eqnarray}
Both solutions obey (\ref{eq:CIS})  with $c=16$. 
\begin{figure}[H]
\centering
\includegraphics[width=0.60\textwidth]{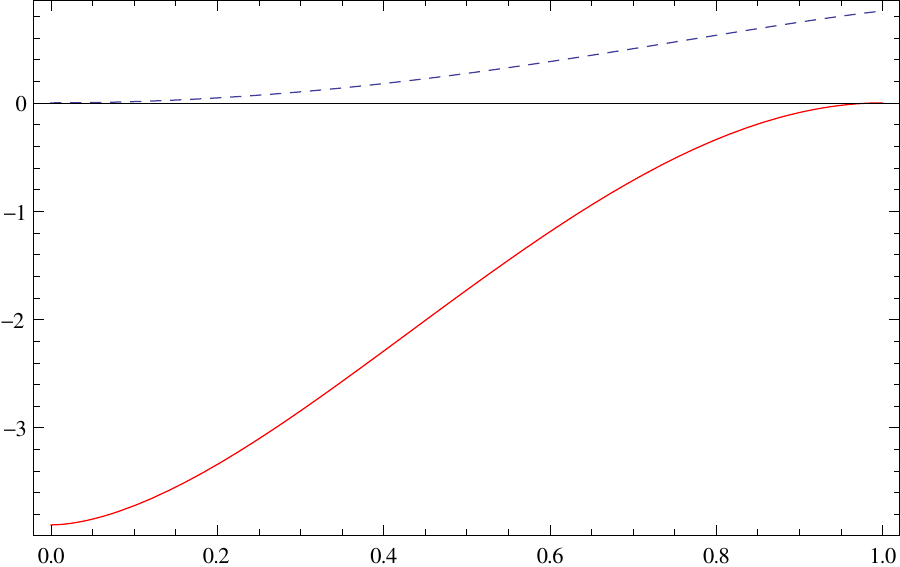}
\caption{\sf \small The homogeneous solutions (\ref{eq:ps1b}, \ref{eq:ps2b}) $\psi_{1b}^{(0)}$ (dashed line) and 
$\psi_{2b}^{(0)}$ (full line) as functions of $x$.}
\label{fig:pic2}
\end{figure}
\noindent
We also used the relation \cite{HYP1}
\begin{eqnarray}
\pFq{2}{1}{{\tfrac{3}{2}},\tfrac{3}{2}}{2}{z} = \frac{4}{\pi z(1-z)} \left[ {\bf E}(z) - (1-z) 
{\bf K}(z) \right],
\end{eqnarray}
noting that it is always possible to map a $_2F_1(a,b;c;x)$ function with $2a, 2b, c \in \mathbb{Z},  c > 
0$ into 
complete elliptic integrals. Their integral representations in Legendre's normal form \cite{LEGENDRE} read
\begin{eqnarray}
\label{eq:KL}
{\bf K}(z) &:=& \int_0^1 \frac{dt}{\sqrt{(1-t^2)(1-zt^2)}}\\
\label{eq:EL}
{\bf E}(z) &:=& \int_0^1 dt \sqrt{\frac{1-zt^2}{1-t^2}}~.
\end{eqnarray}
In going from (\ref{eq:ps1a}, \ref{eq:ps2a}) to (\ref{eq:ps1b}, \ref{eq:ps2b}) also a contiguous relation
had to be applied, leading to a linear combination of two hypergeometric functions.  The solutions are shown 
in Figure~\ref{fig:pic2}.

The ratio $\psi_{1b}^{(0)}/\psi_{2b}^{(0)}$ exhibits the interesting form
\begin{eqnarray}
\label{eq:rat1a}
\frac{\psi_{1b}^{(0)}(x)}{\psi_{2b}^{(0)}(x)} = - \frac{1}{3} \frac{{\bf E}(z)  - r_1(x) {\bf K}(z)}{{\bf 
E}(1-z)  
- (1-r_1(x)) {\bf K}(1-z)}
\end{eqnarray}
with
\begin{eqnarray}
\label{eq:rat1b}
r_1(x) = \frac{(x+3)^2 (x-1)}{(x^2+3)(x-3)},~~~~~\text{and}~~~~~\frac{r_1(x)}{r_1(-x)} = 1 - z(x)~.
\end{eqnarray}

Whether $_2F_1$ solutions emerging in single scale Feynman integral calculations as solutions of
differential equations for master integrals are always of the class to be reducible to complete elliptic integrals 
a priori is not known. However, one may use the algorithm given in Appendix~\ref{sec:8} to map a solution
to one represented by elliptic integrals, if the parameters of the respective $_2F_1$ solution match the required
pattern.

The homogeneous solutions of (\ref{eq:two}) read
\begin{eqnarray}
\label{eq:ps3}
\psi_{3}^{(0)}(x) &=&
-\frac{\sqrt{1-3x}\sqrt{x+1}}{2 \sqrt{2 \pi }} \Biggl[
   (x+1) \left(3 x^2+1\right)
   {\bf E}(z)
-  (x-1)^2 (3 x+1)
   {\bf K}(z)
\Biggr]
\\
\label{eq:ps4}
\psi_{4}^{(0)}(x) &=&
-\frac{\sqrt{1-3x}\sqrt{x+1}}{2\sqrt{2\pi}}
\Biggl[8 x^2  {\bf K}(1-z)
   - (x+1) \left(3
   x^2+1\right) {\bf E}(1-z)\Biggr],
\end{eqnarray}
with
\begin{eqnarray}
\label{eq:z2}
z  = \frac{16 x^3}{(x+1)^3 (3x-1)}~.
\end{eqnarray}
The argument $1-z$ appeared already in complete elliptic integrals by A.~Sabry in Ref.~\cite{SABRY},~Eq.~(68), 
with $x = -\lambda$, calculating the so-called kite-integral at 2 loops, 55 years ago; 
see also Ref.~\cite{Laporta:2004rb}, 
Eq.~(A.11), for the sunrise-diagram and  \cite{Remiddi:2016gno}, 
Eq.~(D.18), with $x = 1/\sqrt{u}$ for the kite-diagram. The latter aspect also shows the close relation between the 
elliptic structures appearing for both topologies, which has been mentioned in Ref.~\cite{Adams:2016xah}.

Using the Legendre identity \cite{LEGENDRE}
\begin{eqnarray}
\label{eq:LEGEND}
{\bf K}(z) {\bf E}(1-z) + {\bf E}(z) {\bf K}(1-z) - {\bf K}(z) {\bf K}(1-z) = \frac{\pi}{2}
\end{eqnarray}
one obtains the Wronskian of the system (\ref{eq:ps3}, \ref{eq:ps4})
\begin{eqnarray}
W(x) = x (9x^2-1) (x^2-1)~,
\end{eqnarray}
cf.~(\ref{eq:WR}). 

\begin{figure}[H]
\centering
\includegraphics[width=0.48\textwidth]{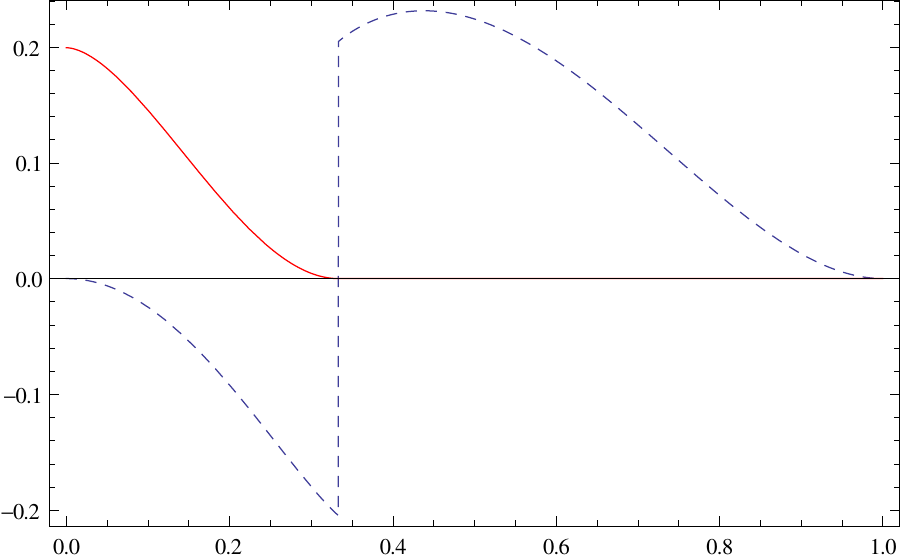}
\includegraphics[width=0.48\textwidth]{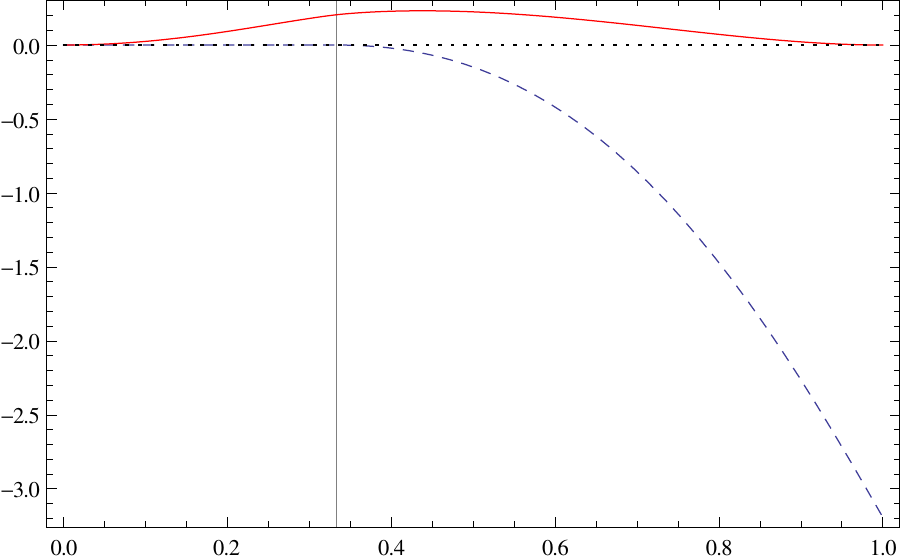}
\caption{\sf \small The homogeneous solutions (\ref{eq:ps3}, \ref{eq:ps3}) $\psi_{3}^{(0)}$ (dashed lines) and 
$\psi_{4}^{(0)}$ (full lines) as functions of $x$; left panel: real part, right panel: imaginary part.}
\label{fig:pic34}
\end{figure}

The homogeneous solutions (\ref{eq:ps3}, \ref{eq:ps4}), which are complex for $x \in [0,1]$, are shown in 
Figure~\ref{fig:pic34}.
The real part of $\psi_{3}^{(0)}(x)$ has a discontinuity at $x = 1/3$ moving from $- (4/9) \sqrt{2/(3 \pi)}$ to 
$(4/9) \sqrt{2/(3 \pi)}$, while {\sf Re}$(\psi_{4}^{(0)}(x))$ vanishes for $x > 1/3$. Likewise, {\sf Im}$(\psi_{4}^{(0)}(x))$ 
vanishes for $0\leq x \leq 1/3$. 

\noindent
We finally consider the ratio $\psi_3^{(0)}/\psi_4^{(0)}$
\begin{eqnarray}
\frac{\psi_3^{(0)}(x)}{\psi_4^{(0)}(x)}
&=& -\frac{{\bf E}(z) - r_2(x){\bf K}(z)}{{\bf E}(1-z) - (1-r_2(x)){\bf K}(1-z)},
\end{eqnarray}
where
\begin{eqnarray}
r_2(x) = \frac{(x-1)^2 (3x+1)}{(x+1)(3x^2+1)}~~~~~\text{and}~~~~~\frac{r_2(x)}{r_2(-x)} = 1 - z(x)~.
\end{eqnarray}
This structure is the same as in (\ref{eq:rat1a}, \ref{eq:rat1b}) up to the pre-factor.

The solution of the inhomogeneous equations (\ref{eq:one}, \ref{eq:two}) are obtained form (\ref{eq:INHOM})
specifying the constants $C_{1,2}$ by physical requirements.
The previous calculation of the corresponding master integrals in \cite{Grigo:2012ji} used expansions of the 
propagators \cite{Q2EXP,Steinhauser:2000ry}, obtaining series representations around $x = 0$ and $x = 1$.
The first expansion coefficients of these will be used to determine $C_1$ and $C_2$. The 
inhomogeneous solutions are given by
\begin{eqnarray}
\label{eq:PS2}
\psi(x) = \psi_1^{(0)}(x) \left[C_1 - I_2(x)\right]
        + \psi_2^{(0)}(x) \left[C_2 + I_1(x)\right],
\end{eqnarray}
with 
\begin{eqnarray}
\label{eq:PS2aA}
I_{1(2)}(x) = \int dx \psi_{1(2)}(x) \frac{N(x)}{W(x)}~. 
\end{eqnarray}
Eq.~(\ref{eq:PS2}) is an integral which cannot be represented within the class of iterative integrals. It
therefore requires a generalization. We present this in Section~\ref{sec:4}. Efficient numerical 
representations using series expansions are given in Section~\ref{sec:42}.

\section{Iterated Integrals over Definite Integrals}
\label{sec:4}

\vspace*{1mm}
\noindent
The elliptic integrals (\ref{eq:KL}, \ref{eq:EL}) cannot be rewritten as integrals in which their argument $x$
only appears in one of their integral boundaries.\footnote{Iterative non-iterative integrals have been 
introduced by the 2nd author in a talk on the 5th International Congress on Mathematical Software, held at 
FU Berlin, July 11-14, 2016, with a series of colleagues present, cf. \cite{ICMS16}.} 
Therefore, the integrals of the type of  Eq.~(\ref{eq:PS2}) 
do not belong to the iterative integrals of the type given in 
Refs.~\cite{Remiddi:1999ew,Ablinger:2011te,Ablinger:2013cf,Ablinger:2014bra} and generalizations thereof to
general alphabets, which have the form
\begin{eqnarray}
\label{eq:ITI}
H_{b,\vec{a}}(x) = \int_0^x dy f_b(y) H_{\vec{a}}(y)~.
\end{eqnarray}
For a given difference equation, associated to a corresponding differential equation,
the algorithms of~\cite{SIG1,SIG2} based on \cite{SIGBASE} allow to decide whether or 
not the recurrence is first order factorizable.
In the first case the corresponding nested sum-product structure is returned. In the case the problem is not
first order factorizable, integrals will be introduced whose integrands depend on variables that cannot be 
moved to the integration boundaries and over which one will integrate by later integrals.
This is the case if the corresponding quantity obeys a differential equation of order $m \geq 2$, 
not being reducible to lower orders. Examples of this kind are irreducible Gau\ss{}' $_2F_1$ functions, to 
which also the complete elliptic integrals ${\bf E}(z)$ and ${\bf K}(z)$ belong.

The new iterative integrals are given by  
\begin{eqnarray}
\label{eq:ITNEW}
\mathbb{H}_{a_1,..., a_{m-1};\{ a_m; F_m(r(y_m))\},a_{m+1},...,a_q}(x) &=& \int_0^x dy_1 
f_{a_1}(y_1) \int_0^{y_1} dy_2 ... \int_0^{y_{m-1}} dy_m f_{a_m}(y_m) F_m[r(y_m)] 
\nonumber\\ &&
\times H_{a_{m+1},...,a_q}(y_{m+1}),
\end{eqnarray}
and cases in which more than one definite integral $F_m$ appears. Here the $f_{a_i}(y)$ are the 
usual letters of the different classes considered in 
\cite{Remiddi:1999ew,Ablinger:2011te,Ablinger:2013cf,Ablinger:2014bra} multiplied by 
hyperexponential pre-factors 
\begin{eqnarray}
r(y) y^{r_1}(1-y)^{r_2},~~~~r_i \in \mathbb{Q},~r(y) \in \mathbb{Q}[y]
\end{eqnarray}
and $F[r(y)]$ is given by
\begin{eqnarray}
F[r(y)] = \int_0^1 dz g(z,r(y)),~~~r(y) \in \mathbb{Q}[y], 
\end{eqnarray}
such that the $y$-dependence cannot be transformed into one of the integration boundaries completely.
We have chosen here $r(y)$ as a rational function because of concrete examples in this paper, which, 
however, is not necessary. Specifically we have
\begin{eqnarray}
\label{eq:F1}
F[r(y)] &=& \pFq{2}{1}{a,b}{c}{r(y)} = \frac{\Gamma(c)}{\Gamma(b) \Gamma(c-b)} 
\int_0^1 dz z^{b} (1-z)^{c-b-1} \left(1-r(y) z\right)^{-a},
\nonumber\\ &&  \hspace*{8cm}
r(y) \in \mathbb{Q}[y], a,b,c 
\in 
\mathbb{Q}.
\end{eqnarray}

The new iterated integral (\ref{eq:ITNEW}) is not limited to the emergence of the functions (\ref{eq:F1}).
Multiple definite integrals are allowed as well. They emerge e.g. in the case of Appell-functions \cite{HYP,SLATER} 
and even more involved higher functions. These integrals also obey relations of the shuffle type w.r.t. their letters 
$f_{a_m}(y_m) (F_m[r(y_m)])$,~cf. e.g.~\cite{SHUF,Blumlein:2003gb}.

Within the analyticity region of the problem one may derive series expansions of the corresponding solutions
around special values, e.g. $x=0, x=1$ and other values to map out the function for its whole argument range.
In many cases, one will even find convergent, widely overlapping representations, which are highly accurate and 
provide a numerical solution in terms of a finite number of analytic expansion coefficients.
We apply this method to the solution of the differential equations in Section~\ref{sec:2} in the following 
section and return to the construction of a closed form analytic representation using $q$-series 
and Dedekind $\eta$ functions in Section~\ref{sec:5}.
\section{The Solution of the Inhomogeneous Equation by Series Expansion}
\label{sec:42}

\vspace*{1mm}
\noindent
The inhomogeneous solutions of type (\ref{eq:PS2}) can be expanded into series around 
$x=0$ and $x=1$ analytically using computer algebra packages like {\tt Mathematica} or {\tt maple}.
One either obtains Taylor series or superpositions of Taylor series times a factor $\ln^k(x), k \in \mathbb{N}$. For all 
solutions both expansions have a wide overlap\footnote{This technique has also been used in Ref.~\cite{TLSR2a}.}
and one may obtain in this way a highly accurate representation of all solutions in the complete region $x \in [0,1]$. 

In the following we present the first terms of the series expansion for the 
functions $f_{8(9,10),a(b)}(x)$ around $x=0$ and $x=1$. 

For $f_{8a}$ we obtain 
\begin{eqnarray}
f_{8a}(x) &=& 
-\sqrt{3} \Biggl[
        \pi ^3 \Biggl(
 \frac{35 x^2}{108}
-\frac{35 x^4}{486}
-\frac{35 x^6}{4374}
-\frac{35 x^8}{13122}
-\frac{70x^{10}}{59049}
-\frac{665 x^{12}}{1062882}
\Biggr)
+\Biggl(
 12 x^2
-\frac{8 x^4}{3}
\nonumber\\ && 
-\frac{8 x^6}{27}
-\frac{8 x^8}{81}
-\frac{32 x^{10}}{729}
-\frac{152 x^{12}}{6561}
\Biggr) 
{\sf Im}\Biggl[\text{Li}_3\left(\frac{e^{-\frac{i \pi}{6}}}{\sqrt{3}}\right)\Biggr]
\Biggr]
-\pi ^2 \Biggl(
1
+\frac{x^4}{9}
-\frac{4 x^6}{243}
-\frac{46 x^8}{6561}
\nonumber\\ && 
-\frac{214 x^{10}}{59049}
-\frac{5546 x^{12}}{2657205}
\Biggr)
-\Biggl(
-\frac{3}{2}
-\frac{x^4}{6}
+\frac{2 x^6}{81}
+\frac{23 x^8}{2187}
+\frac{107 x^{10}}{19683}
+\frac{2773 x^{12}}{885735}
\Biggr) 
\psi ^{(1)}\left(\frac{1}{3}\right)
\nonumber\\ && 
-\sqrt{3} \pi  \Biggl(
 \frac{x^2}{4}
-\frac{x^4}{18}
-\frac{x^6}{162}
-\frac{x^8}{486}
-\frac{2 x^{10}}{2187}
-\frac{19 x^{12}}{39366}
\Biggr) \ln^2(3)
-\Biggl[
        33 x^2
        -\frac{5 x^4}{4}
        -\frac{11 x^6}{54}
\nonumber\\ &&
        -\frac{19 x^8}{324}
        -\frac{751 x^{10}}{29160}
        -\frac{2227 x^{12}}{164025}
        +\pi ^2 \Biggl(
 \frac{4 x^2}{3}
-\frac{8 x^4}{27}
-\frac{8 x^6}{243}
-\frac{8 x^8}{729}
-\frac{32 x^{10}}{6561}
-\frac{152 x^{12}}{59049}
        \Biggr)
\nonumber\\ &&
        +\Biggl(
-2 x^2
+\frac{4 x^4}{9}
+\frac{4 x^6}{81}
+\frac{4 x^8}{243}
+\frac{16 x^{10}}{2187}
+\frac{76 x^{12}}{19683}
\Biggr) 
\psi^{(1)}\left(\frac{1}{3}\right)
\Biggr] \ln(x)
+\frac{135}{16}
+19 x^2
\nonumber\\ &&
-\frac{43 x^4}{48}
-\frac{89 x^6}{324}
-\frac{1493 x^8}{23328}
-\frac{132503 x^{10}}{5248800}
-\frac{2924131 x^{12}}{236196000}
-\Biggl(
        \frac{x^4}{2}-12 x^2\Biggr) \ln^2(x)
\nonumber\\ &&
-2 x^2 \ln^3(x)
+ O\left(x^{14} \ln(x)\right)
\end{eqnarray}
around $x=0$. Here we also applied a series of relations for $\psi^{(k)}$-functions at rational 
argument, cf.~Ref.~\cite{Ablinger:2011te}.

Likewise, one may expand around $y= 1-x = 0$. In this case, we can rewrite the inhomogeneous solution
given in (\ref{eq:PS2}) as
\begin{eqnarray}
\label{eq:PS2a}
\psi(y) = \psi_1^{(0)}(y) \left[\overline{C}_1 - \bar{I}_2(y)\right]
        + \psi_2^{(0)}(y) \left[\overline{C}_2 + \bar{I}_1(y)\right],
\end{eqnarray}
with 
\begin{eqnarray}
\bar{I}_{1(2)}(x) &=& \int dy~\psi_{1(2)}^{(0)}(y) \frac{N(y)}{W(y)},
\\
W(y) &=& 
  \psi_1^{(0)}(y) \frac{d}{dy} \psi_2^{(0)}(y) 
- \psi_2^{(0)}(y) \frac{d}{dy} \psi_1^{(0)}(y).
\end{eqnarray}
One obtains
\begin{eqnarray}
f_{8a}(x) &=& 
\frac{275}{12} + \frac{10}{3} y  - 25 y^2 + \frac{4}{3}  y^3  + \frac{11}{12} y^4 
+ y^5 + \frac{47}{96} y^6 + \frac{307}{960} y^7  + \frac{19541}{80640 } y^8 
+ \frac{22133}{120960} y^9 
\nonumber\\ &&
+ \frac{1107443}{7741440} y^{10} + 
 \frac{96653063}{851558400} y^{11} + \frac{3127748803}{34062336000} y^{12} 
\nonumber\\ &&
+ 7 \Biggl(2 y^2 
- y^3 - \frac{1}{8} y^4  - \frac{1}{64} y^6 - \frac{1}{128} y^7  - \frac{3}{512} y^8 - 
\frac{1}{256} y^9 - \frac{47}{16384} y^{10} 
\nonumber\\ &&
- \frac{69}{32768} y^{11} - \frac{421}{262144} y^{12}
\Biggr) \zeta_3 + O(y^{13})~.
\end{eqnarray}

The solution of Eq.~(\ref{eq:f9bb}) around $x=0$ reads
\begin{eqnarray}
f_{9a}(x) &=& 
\sqrt{3} \Biggl(
 4 x^2
+\frac{8 x^6}{81}
+\frac{16 x^8}{243}
+\frac{32 x^{10}}{729}
+\frac{608 x^{12}}{19683}
\Biggr) {\sf Im}\Biggl(\text{Li}_3\Biggl(\frac{e^{-\frac{i \pi }{6}}}{\sqrt{3}}\Biggr)\Biggr]
+\sqrt{3} \pi ^3 \Biggl(
 \frac{35 x^2}{324}
\nonumber\\ &&
+\frac{35 x^6}{13122}
+\frac{35 x^8}{19683}
+\frac{70 x^{10}}{59049}
+\frac{1330 x^{12}}{1594323}
\Biggr)
+\sqrt{3} \pi  \Biggl(
 \frac{x^2}{12}
+\frac{x^6}{486}
+\frac{x^8}{729}
+\frac{2 x^{10}}{2187}
+\frac{38 x^{12}}{59049}
\Biggr) 
\nonumber\\ && \times
\ln^2(3)
+\pi ^2 \Biggl(
 \frac{2}{3}
-\frac{2 x^2}{9}
+\frac{4 x^4}{81}
+\frac{8 x^6}{729}
+\frac{128 x^8}{19683}
+\frac{262 x^{10}}{59049}
+\frac{25604 x^{12}}{7971615}
\Biggr)
+\Biggl(
-1
+\frac{x^2}{3}
\nonumber\\ &&
-\frac{2 x^4}{27}
-\frac{4 x^6}{243}
-\frac{64 x^8}{6561}
-\frac{131 x^{10}}{19683}
-\frac{12802 x^{12}}{2657205}
\Biggr) \psi ^{(1)}\left(\frac{1}{3}\right)
+\Biggl(
        3 x^2
        +\frac{x^4}{3}
        +\frac{11 x^6}{162}
        +\frac{19 x^8}{486}
\nonumber\\ &&
        +\frac{751 x^{10}}{29160}
        +\frac{8908 x^{12}}{492075}
        +\pi ^2 \Biggl(
 \frac{4 x^2}{9}
+\frac{8 x^6}{729}
+\frac{16 x^8}{2187}
+\frac{32 x^{10}}{6561}
+\frac{608 x^{12}}{177147}
\Biggr)
+\Biggl[
-\frac{2 x^2}{3}
-\frac{4 x^6}{243}
\nonumber\\ &&
-\frac{8 x^8}{729}
-\frac{16 x^{10}}{2187}
-\frac{304 x^{12}}{59049}
\Biggr) \psi ^{(1)}\left(\frac{1}{3}\right)
\Biggr] \log (x)
+\frac{5}{2}
-\frac{11 x^2}{6}
-\frac{5 x^4}{12}
-\frac{14 x^6}{243}
-\frac{1151 x^8}{34992}
\nonumber\\ &&
-\frac{109973 x^{10}}{5248800}
-\frac{2523271 x^{12}}{177147000}
-2 x^2 \log ^2(x)
+\frac{2}{3} x^2 \log ^3(x) + O\left(x^{14} \ln(x)\right)~.
\end{eqnarray}
The corresponding expansion around $x=1$ is given by
\begin{eqnarray}
f_{9a}(x) &=& 
\frac{5}{3} + \frac{2}{3} y  + \frac{2}{3} y^2 + \frac{1}{2} y^3 + \frac{1}{3} y^4 - 
\frac{11}{480} y^5 + 
\frac{13}{1920} y^6  - \frac{2461}{120960} y^7 - \frac{3701}{241920} y^8 
\nonumber\\ &&
- 
\frac{76627}{4644864} y^9 - \frac{1289527}{92897280} y^{10} - 
\frac{635723359}{51093504000} y^{11} - \frac{13482517}{1261568000} y^{12} 
\nonumber\\ &&
+ 
 7 \Biggl(
- \frac{2}{3} y  - \frac{1}{6} y^2 + \frac{1}{4} y^3 + \frac{1}{64} y^5 + \frac{1}{256} y^6 + 
\frac{1}{256} y^7 + \frac{1}{512} y^8
+ \frac{25}{16384} y^9 + \frac{65}{65536} y^{10} 
\nonumber\\ &&
+ \frac{99}{131072} y^{11} + 
\frac{145}{262144} y^{12} \Biggr) \zeta_3 + O\left(y^{13}\right)~.
\end{eqnarray}
Here the integration constants $C_{1,2}$ and $\overline{C}_{1,2}$ are\footnote{We thank P.~Marquard for having 
provided all the necessary constants and a series of expansion parameters for the solutions given in 
Ref.~\cite{Grigo:2012ji} in computer readable form.}
\begin{eqnarray}
C_1 &=&  \frac{35 \pi^3}{72} 
+ 18 {\sf Im}\left[\Li_3\left(
-\frac{e^{5i\pi/6}}{\sqrt{3}}\right)\right] 
+ \frac{2 \pi^2  \ln(3)}{\sqrt{3}} 
+ \frac{3}{8} \pi \ln^2(3)
\nonumber\\ &&
+   \frac{\sqrt{3}}{16} \left[25 - 2 \ln(3) \left(45 + 8 \psi^{(1)}
\left(\frac{1}{3}\right)\right)\right]
\\
C_2 &=& -\left[
\frac{135}{16} - \pi^2 + \frac{3}{2} \psi^{(1)}\left(\frac{1}{3}\right)\right]
\left(-\frac{2 \pi}{9 \sqrt{3}}\right)
\\
\overline{C}_1 &=& \frac{275}{32} \pi 
\\
\overline{C}_2 &=& \frac{275}{64} - \frac{275}{48} \ln(2) - \frac{7}{3} \zeta_3~.
\end{eqnarray}

The solution of (\ref{eq:10a}) is an integral containing the functions $f_{8a}(x)$ and $f_{9a}(x)$.
Its series around $x=0$ reads 
\begin{eqnarray}
f_{10a}(x) &=& 
\sqrt{3} \Biggl(
-12 x^2
-\frac{22 x^4}{3}
-\frac{148 x^6}{27}
-\frac{359 x^8}{81}
-\frac{13652 x^{10}}{3645}
-\frac{21370 x^{12}}{6561}
\Biggr) 
\nonumber\\ && \times
{\sf Im}\Biggl[
        \text{Li}_3\left(
                -\frac{(-1)^{5/6}}{\sqrt{3}}\right)\Biggr]
+\Biggl(
6
+\frac{22 x^2}{3}
+\frac{11 x^4}{3}
+\frac{22 x^6}{9}
+\frac{11 x^8}{6}
+\frac{22 x^{10}}{15}
+\frac{11 x^{12}}{9}
\Biggr) \zeta_3
\nonumber\\ &&
+\pi ^2 \Biggl(
 \frac{7 x^2}{6}
+\frac{13 x^4}{72}
-\frac{x^6}{486}
-\frac{12739 x^8}{209952}
-\frac{245263 x^{10}}{2952450}
-\frac{1950047 x^{12}}{21257640}
\Biggr)
\nonumber\\ &&
+\sqrt{3} \pi ^3 \Biggl(
-\frac{35 x^2}{108}
-\frac{385 x^4}{1944}
-\frac{1295 x^6}{8748}
-\frac{12565 x^8}{104976}
-\frac{23891 x^{10}}{236196}
-\frac{373975 x^{12}}{4251528}
\Biggr)
\nonumber\\ &&
+\sqrt{3} \pi  \Biggl(
-\frac{x^2}{4}
-\frac{11 x^4}{72}
-\frac{37 x^6}{324}
-\frac{359 x^8}{3888}
-\frac{3413 x^{10}}{43740}
-\frac{10685 x^{12}}{157464}
\Biggr) \ln^2(3)
\nonumber\\ &&
+\Biggl(
 \frac{2 \pi ^2}{3}
-x^2
+\frac{7 x^6}{81}
+\frac{4825 x^8}{34992}
+\frac{76078 x^{10}}{492075}
-\frac{x^4}{12}
+\frac{561323 x^{12}}{3542940}
\Biggr) \psi ^{(1)}\left(\frac{1}{3}\right)
\nonumber\\ &&
+\Biggl[
-x^4
-\frac{32 x^6}{27}
-\frac{761 x^8}{648}
-\frac{3251 x^{10}}{2916}
-\frac{27455 x^{12}}{26244}
+\pi ^2 \Biggl(
-\frac{5 x^2}{3}
-\frac{53 x^4}{54}
-\frac{175 x^6}{243}
\nonumber\\ &&
-\frac{1679 x^8}{2916}
-\frac{15839 x^{10}}{32805}
-\frac{49301 x^{12}}{118098}
\Biggr)
+\Biggl(
 2 x^2
+\frac{11 x^4}{9}
+\frac{74 x^6}{81}
+\frac{359 x^8}{486}
+\frac{6826 x^{10}}{10935}
\nonumber\\ &&
+\frac{10685 x^{12}}{19683}
\Biggr) \psi ^{(1)}\left(\frac{1}{3}\right)
\Biggr] \ln(x)
-\frac{19 \pi ^4}{72}
-\frac{1}{2} \psi^{(1)}\left(\frac{1}{3}\right)^2
+\frac{3 x^4}{2}
+\frac{245 x^6}{162}
+\frac{31723 x^8}{23328}
\nonumber\\ &&
+\frac{634597 x^{10}}{524880}
+\frac{10219913 x^{12}}{9447840}
+ O\left(x^{14} \ln(x)\right)~.
\end{eqnarray}
Likewise, one obtains the expansion around $x=1$, which is given by
\begin{eqnarray}
f_{10a}(x) &=&
-\frac{11 \pi ^4}{45}
+\frac{4 \ln^4(2)}{3}
-\frac{4}{3} \pi ^2 \ln^2(2)
+32 \text{Li}_4\left(\frac{1}{2}\right)
+\Bigg[
6
+3 y
-2 y^2
-\frac{13 y^3}{8}
-\frac{163 y^4}{128}
\nonumber\\ &&
-\frac{631 y^5}{640}
-\frac{1213 y^6}{1536}
-\frac{2335 y^7}{3584}
-\frac{36247 y^8}{65536}
-\frac{47221 y^9}{98304}
-\frac{69631 y^{10}}{163840}
-\frac{1100145 y^{11}}{2883584}
\nonumber\\ &&
-\frac{544987 y^{12}}{1572864}
-\frac{1082435 y^{13}}{3407872}
\Biggr] \zeta_3
+\frac{5 y^2}{2}
+\frac{7 y^3}{4}
+\frac{2363 y^4}{1728}
+\frac{1867 y^5}{1728}
+\frac{2293073 y^6}{2592000}
\nonumber\\ &&
+\frac{71317 y^7}{96000}
+\frac{8080140871 y^8}{12644352000}
+\frac{31879816079 y^9}{56899584000}
+\frac{255571071379 y^{10}}{512096256000}
\nonumber\\ &&
+\frac{1844349403987 y^{11}}{4096770048000}
+\frac{13424123319977921 y^{12}}{32716805603328000}
+\frac{2056360866308893 y^{13}}{5452800933888000}
+O(y^{13})
\nonumber\\ 
\end{eqnarray}
with
\begin{eqnarray}
C_3 &=& -\frac{19}{72} \pi^4  + \frac{2}{3} \pi^2 
\psi^{(1)}\left(\frac{1}{3}\right) - \frac{1}{2} 
\psi^{(1)}\left(\frac{1}{3}\right)^2 + 6 \zeta_3
\\
\overline{C}_3 &=& 
9 \zeta_4 - 6 \zeta_3 - 2 {\sf B}_4, 
\end{eqnarray}
with \cite{Broadhurst:1991fi}
\begin{eqnarray}
{\sf B}_4 &=& - 4 \zeta_2 \ln^2(2) + \frac{2}{3} \ln^4(2) - \frac{13}{2} \zeta_4 
+ 16 \Li_4\left(\frac{1}{2}\right),
\end{eqnarray}
as integration constants in this case.

The series expansion of the solution of Eq.~(\ref{eq:two}) is given by
\begin{eqnarray}
f_{8b}(x) &=& - \Biggl\{
\frac{145}{48} - 19 x^2 - \frac{261}{16} x^4  + \frac{19}{12} x^6  + \frac{4157}{288} x^8 + 
\frac{510593}{7200} x^{10} + \frac{13208647}{36000} x^{12}
\nonumber\\ &&
+\left(
\frac{1}{2}
+\frac{9}{2} x^4
-6 x^6
-23 x^8
-107 x^{10}
-\frac{2773}{5} x^{12}
\right) \zeta_2
\nonumber\\ &&
+2 x^2 \left(
-1
+2 x^2
+2 x^4
+6 x^6
+24 x^8
+114 x^{10}
\right) \zeta_3
\nonumber\\ &&
-2 x^2 
\left(
-1
-14 x^2
+4 x^4
+12 x^6
+48 x^8
+228 x^{10}
\right) \ln^3(x)
\nonumber\\ &&
-\frac{1}{10} x^2 \left(
120
+ 585 x^2
+ 120 x^4
+ 460 x^6
+ 2140 x^8
+ 11092 x^{10}
\right) \ln^2(x)
\nonumber\\ &&
+\Biggl[(
        33 x^2
        +\frac{201}{4} x^4
        +\frac{29}{2} x^6
        +\frac{307}{12} x^8
        +\frac{7927}{120} x^{10}
        +\frac{14107}{75} x^{12}
\nonumber\\ &&
        -6 x^2 \Biggl(
-1
+2 x^2
+2 x^4
+6 x^6
+24 x^8
+114 x^{10}
\Biggr) \zeta_2
\Biggr] \ln(x) + O\left(x^{14} \ln^3(x)\right)\Biggr\}~.
\nonumber\\ &&
\end{eqnarray}

The solution of Eq.~(\ref{eq:f9bc}) reads
\begin{eqnarray}
f_{9b}(x) &=&
- \frac{95}{12} + \frac{131}{2} x^2 + \frac{99}{2} x^4 + \frac{53}{6} x^6 
+ \frac{5999}{144} x^8 + \frac{196621}{800} x^{10} + \frac{14055067}{9000} x^{12}
\nonumber\\ &&
 + \left(-1 + 3 x^2 - 6 x^4 - 12 x^6 - 64 x^8 - 393 x^{10} - \frac{12802}{5} x^{12} \right) 
\zeta_2
\nonumber\\ &&
+ 2 \left(x^2 + 2 x^6 + 12 x^8 + 72 x^{10} + 456 x^{12}\right) \zeta_3
\nonumber\\ &&
- 2 \left(x^2 + 4 x^6 + 24 x^8 + 144 x^{10} + 912 x^{12}\right) \ln^3(x)  
\nonumber\\ &&
+ \left(24 x^2 + 96 x^4 - 24 x^6 - 128 x^8 - 786 x^{10} - \frac{25604}{5} x^{12}
    \right) \ln^2(x) 
\nonumber\\ &&
+ \Biggl[-81 x^2 - 126 x^4 + \frac{5}{2} x^6 + \frac{31}{6} x^8
- \frac{633}{40} x^{10} - \frac{26762}{75} x^{12} 
\nonumber\\ &&
- 6 \left(x^2 + 2 x^6 + 12 x^8 + 72 x^{10} + 456 x^{12}\right) \zeta_2 \Biggr] \ln(x) + 
O\left(x^{14} \ln^3(x)\right)~.
\end{eqnarray}
Here the constants in (\ref{eq:PS2}) have been fixed by comparing to the first 
expansion coefficients in \cite{Grigo:2012ji}
\begin{eqnarray}
C_1 &=& - \frac{1}{24}\left[2 i \pi (145  + 4 \pi^2) + 3 (165 + 16 \zeta_3)\right]\\
C_2 &=& -\frac{1}{12} \pi (145 + 4 \pi^2) = {\sf Im}(C_1)~.
\end{eqnarray}

The solution of (\ref{eq:10b}) is given as an integral containing $f_{8(9)b}(x)$ with the constant 
\begin{eqnarray}
C_3 &=& 3 \zeta_4 + 6 \zeta_3
\end{eqnarray}
and reads
\begin{eqnarray}
f_{10b}(x) &=& 
3 \zeta_4
-4 x^2+\frac{7}{4} x^4-\frac{553}{81} x^6 - \frac{87587}{1728} x^8 - \frac{9136091}{33750} x^{10}
-\frac{236649223}{162000} x^{12}
\nonumber\\ &&
+\left(-6 x^2-\frac{1}{2} x^4+ \frac{46}{3} x^6 
+\frac{1957}{24} x^8 +\frac{30907}{75} x^{10} + \frac{40103}{18} x^{12}\right) \zeta_2 
\nonumber\\ &&
+\Biggl[
12 x^2- \frac{15}{2} x^4 -\frac{257}{9} x^6 - \frac{3613}{48} x^8 - \frac{103577}{500} 
x^{10} - \frac{1039019}{1800} x^{12}
\nonumber\\ &&
+ \left(8 x^2+16 x^4+ \frac{116}{3} x^6 
 + 128 x^8 + \frac{2708}{5} x^{10} + \frac{8062}{3} x^{12} \right) \zeta_2 \Biggr] \ln(x)
\nonumber\\ &&
+ \left(-12 x^2-x^4+ \frac{92}{3} x^6 +\frac{1957}{12} x^8 
+ \frac{61814}{75} x^{10} + \frac{40103}{9} x^{12} \right) \ln^2(x)
\nonumber\\ &&
+\left( \frac{16}{3} x^2 + \frac{32}{3} x^4 
+ \frac{232}{9} x^6 +\frac{256}{3} x^8 
+ \frac{5416}{15} x^{10} +\frac{16124}{9} x^{12} \right) \ln^3(x)
\nonumber\\ &&
+\left(6+4 x^2-2 x^4-\frac{32}{3} x^6 - 41 x^8-\frac{896}{5} x^{10}-\frac{2684}{3} 
x^{12}\right) \zeta_3 + O\left(x^{14} \ln^3(x)\right)~.
\nonumber\\ &&
\end{eqnarray}

The corresponding solutions around $x=1$ have the expansions
\begin{eqnarray}
f_{8b}(x) &=& \frac{275}{12} + 10 y - 71 y^2 + 12 y^3 + \frac{57}{4} y^4 + 18 y^5 
- \frac{1079}{160} y^6 - \frac{621}{320} y^7 - \frac{30967}{80640} y^8 
+ \frac{3449}{24192} y^9 
\nonumber\\ &&
+ \frac{13850687}{38707200} y^{10} + 
\frac{81562673}{170311680} y^{11} + \frac{6586514681}{11354112000} y^{12} 
\nonumber\\ &&
+ 
 7 \Biggl(2 y^2 - 3 y^3 + \frac{7}{8} y^4 - \frac{1}{64} y^6 - \frac{3}{128} y^7 
- \frac{15}{512} y^8 - \frac{9}{256} y^9 - \frac{687}{16384} y^{10} - \frac{1647}{32768} y^{11} 
\nonumber\\ &&
- \frac{15933}{262144} y^{12}\Biggr) \zeta_3 + O\left(y^{13}\right)
\end{eqnarray}
\begin{eqnarray}
f_{9b}(x) &=& 
\frac{5}{3} + 2 y + 2 y^2 + \frac{3}{2} y^3 - \frac{3}{2} y^4 - \frac{171}{160} y^5 
- \frac{577}{640} y^6  - \frac{35851}{40320} y^7 - \frac{77957}{80640} y^8 
- \frac{1726163}{1548288} y^9 
\nonumber\\ &&
- \frac{41342669}{30965760} y^{10} 
- \frac{27949201859}{17031168000} y^{11} + \frac{6932053241}{2838528000} y^{12} 
\nonumber\\ &&
+ 
 7 \Biggl(-2 y + \frac{5}{2} y^2  - \frac{1}{4} y^3 + \frac{(3}{64} y^5 + \frac{(17}{256} y^6 
+ \frac{21}{256} y^7 + \frac{5}{512} y^8  + \frac{1995}{16384} y^9 + \frac{9873}{65536} y^{10} 
\nonumber\\ &&
+ \frac{24741}{131072} y^{11} - \frac{15933}{65536} y^{12}\Biggr) \zeta_3 + O\left(y^{13}\right).
\end{eqnarray}

\begin{eqnarray}
f_{10b}(x) &=&      
2 {\sf B}_4 - 9 \zeta_4
+ \frac{5}{2}  y^2 
+ \frac{13}{4} y^3 
+ \frac{6251}{1728} y^4 
+ \frac{6721}{1728} y^5 
+ \frac{10775573}{2592000} y^6 
+ \frac{142659}{32000} y^7 
\nonumber\\ &&
+ \frac{60860651591}{12644352000} y^8 
+ \frac{298199146349}{56899584000} y^9 
+ \frac{1475031521177}{256048128000} y^{10}  
+ \frac{26211821446117}{4096770048000} y^{11} \nonumber\\ &&
+ \frac{235080972861513791}{32716805603328000} y^{12}
+ \Biggl[6 + 9 y + y^2 - \frac{11}{8} y^3 - \frac{307}{128} y^4 
- \frac{1893}{640} y^5  - \frac{5137}{1536} y^6 
\nonumber\\ &&
- \frac{13179}{3584} y^7 - \frac{263063}{65536}
y^8 - \frac{431519}{98304} y^9 - \frac{395741}{81920} y^{10} - \frac{15466743}{2883584} y^{11} 
\nonumber\\ && 
- \frac{9465637}{1572864} y^{12}\Biggr] \zeta_3 + O\left(y^{13}\right),
\end{eqnarray}
with the integration constants 
\begin{eqnarray}
\overline{C}_1 &=& - i \pi \frac{275}{48}
\\
\overline{C}_2 &=&  i \left[\overline{C}_1 - \frac{297}{32} + \frac{275}{8} \ln(2)  + 14 
\zeta_3\right]
\\
\overline{C}_3 &=& 
\frac{5}{2} - 9 \zeta_4 - \frac{466638231901}{12595494912} \zeta_3 + 2 {\sf B}_4,
\end{eqnarray}
obtained by comparing again to the first expansion coefficients in \cite{Grigo:2012ji}. The constants are complex here. 

\begin{figure}[H]
\centering
\includegraphics[width=0.47\textwidth]{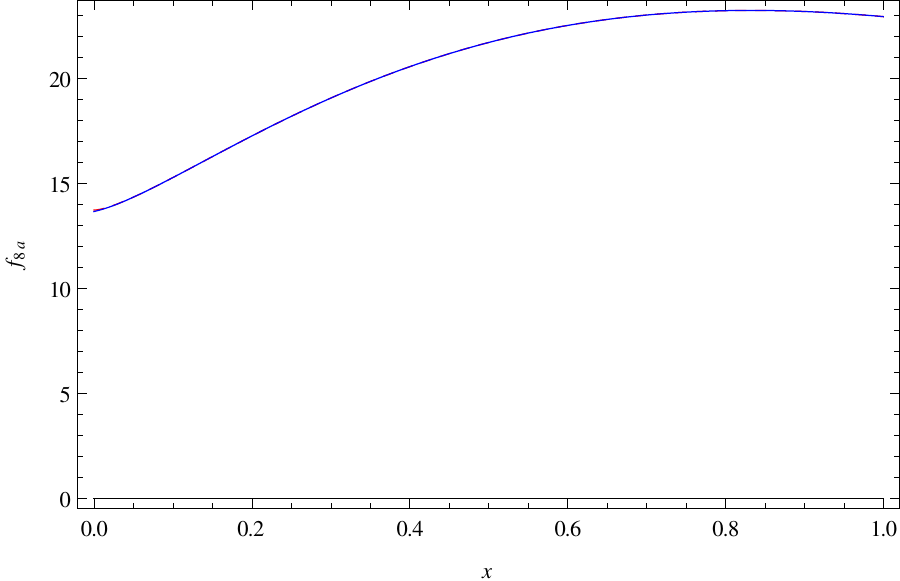}
\includegraphics[width=0.52\textwidth]{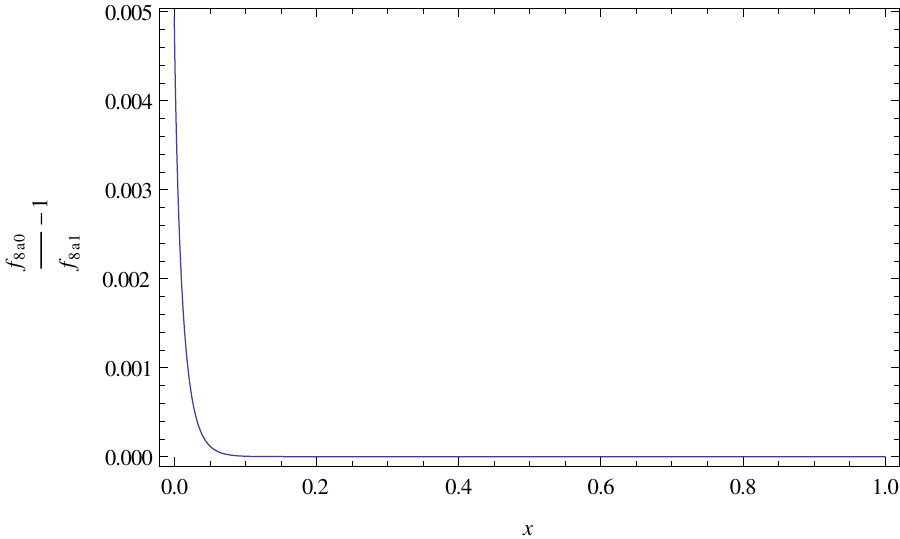}
\caption{\sf \small The inhomogeneous solution of 
Eq.~(\ref{eq:one}) as a function of $x$. Left panel: Red dashed line: expansion 
around $x=0$; blue line: expansion around $x=1$. Right panel: illustration of the 
relative accuracy and overlap of the two solutions $f_{8a}(x)$ around 0 and 1.
\label{fig:f8a}}
\end{figure}

\begin{figure}[H]
\centering
\includegraphics[width=0.47\textwidth]{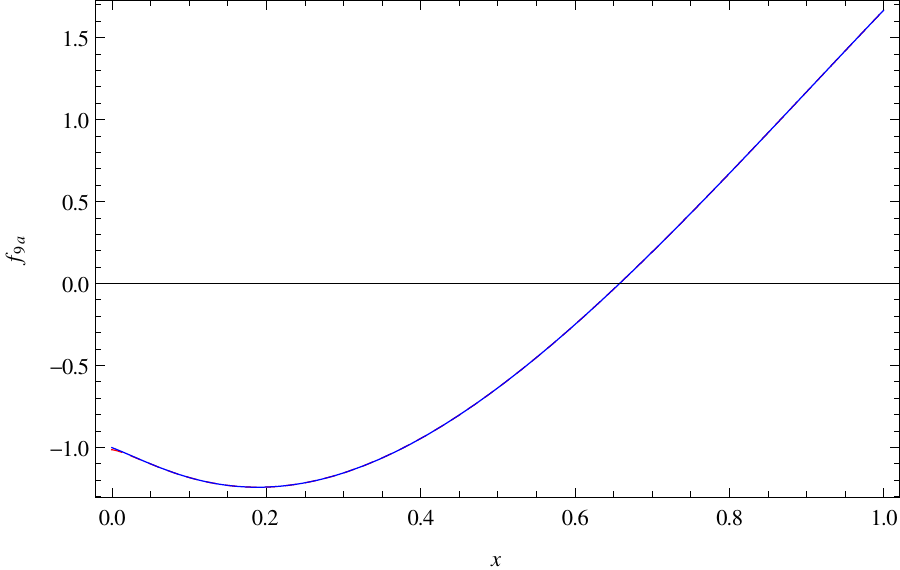}
\includegraphics[width=0.52\textwidth]{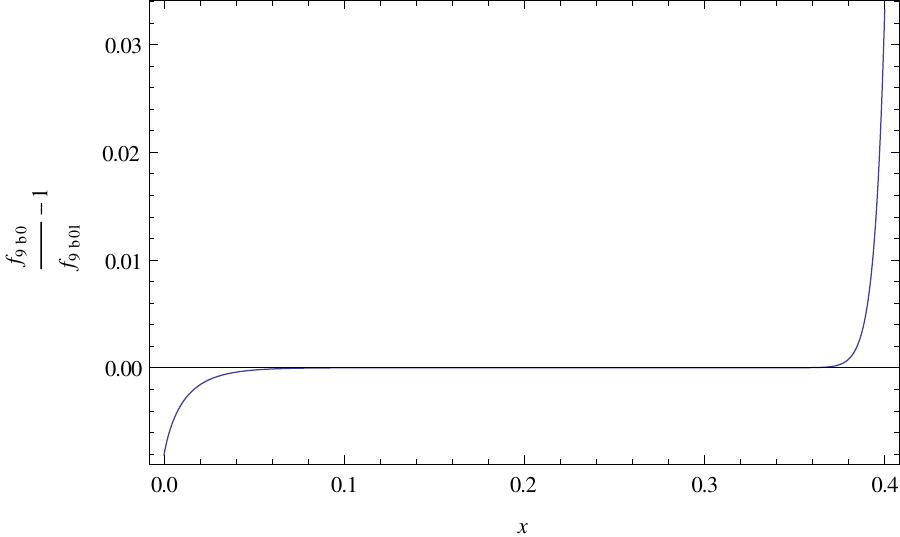}
\caption{\sf \small The inhomogeneous solution of 
Eq.~(\ref{eq:f9bb}) as a function of $x$. Left panel: Red dashed line: expansion 
around $x=0$; blue line: expansion around $x=1$. Right panel: illustration of the 
relative accuracy and overlap of the two solutions $f_{9a}(x)$ around 0 and 1.
\label{fig:f9a}}
\end{figure}

\begin{figure}[H]
\centering
\includegraphics[width=0.47\textwidth]{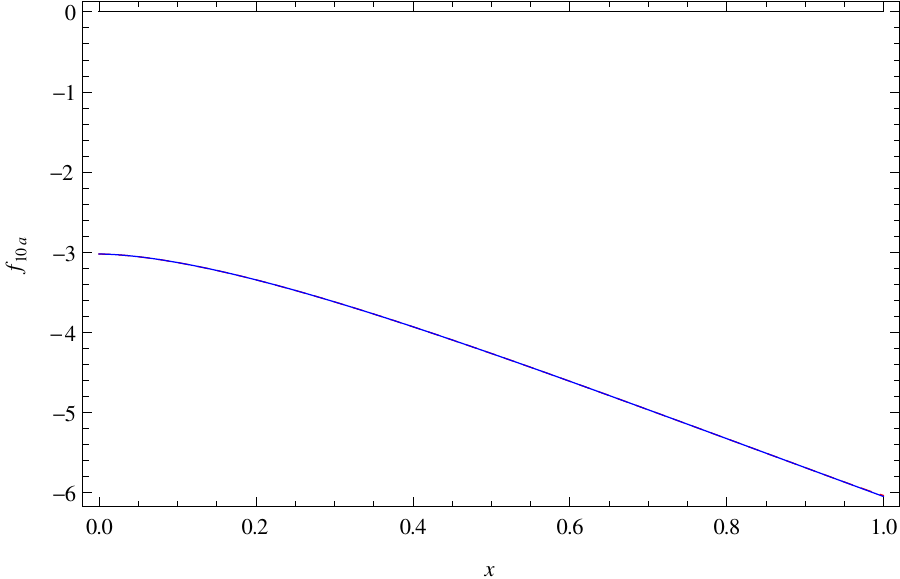}
\includegraphics[width=0.52\textwidth]{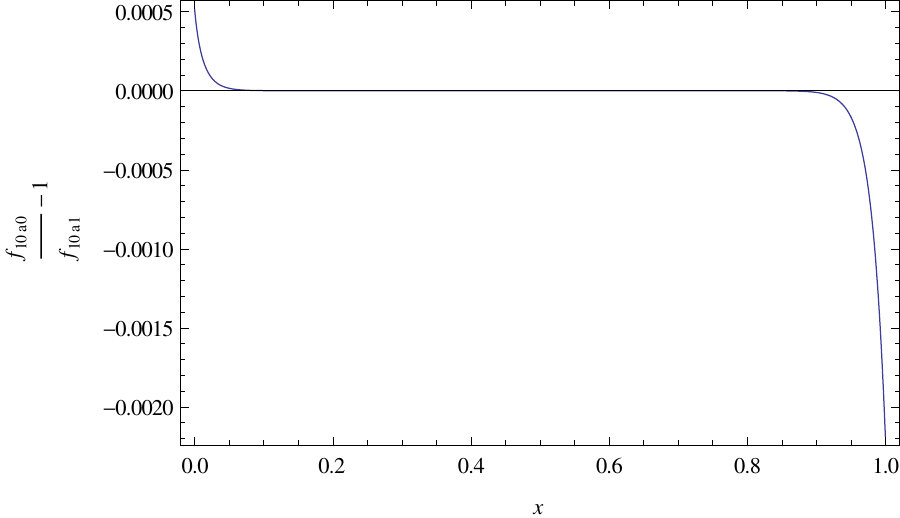}
\caption{\sf \small The inhomogeneous solution of 
Eq.~(\ref{eq:10a}) as a function of $x$. Left panel: Red dashed line: expansion 
around $x=0$; blue line: expansion around $x=1$. Right panel: illustration of the 
relative accuracy and overlap of the two solutions $f_{10a}(x)$ around 0 and 1.
\label{fig:f10a}}
\end{figure}

\begin{figure}[H]
\centering
\includegraphics[width=0.47\textwidth]{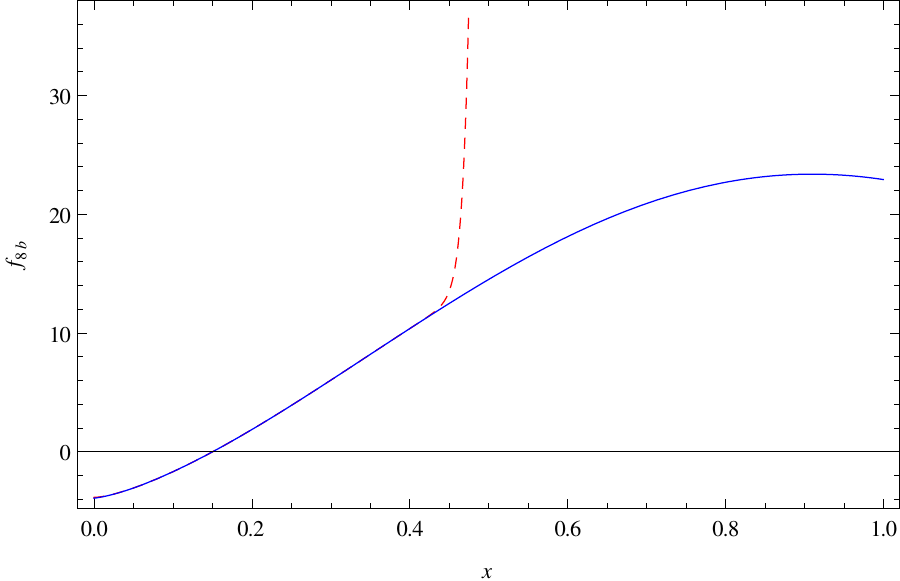}
\includegraphics[width=0.52\textwidth]{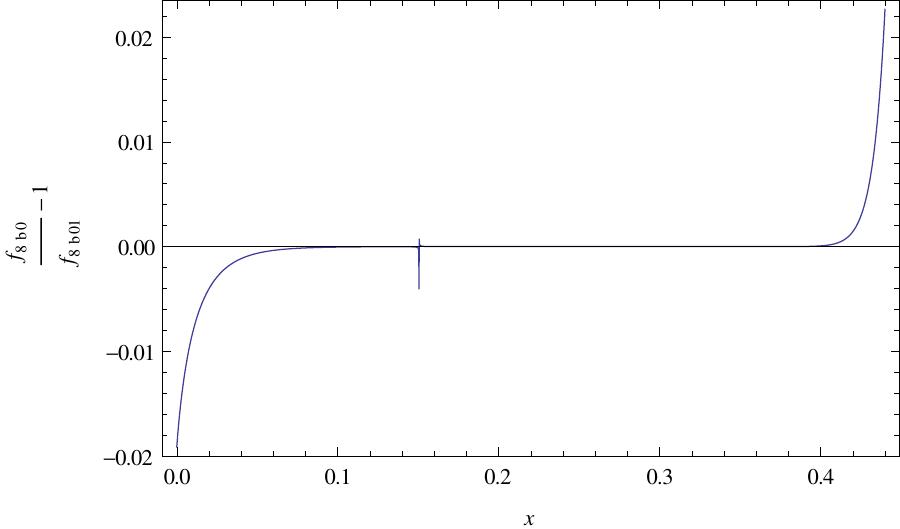}
\caption{\sf \small The inhomogeneous solution of 
Eq.~(\ref{eq:two}) as a function of $x$. Left panel: Red dashed line: expansion 
around $x=0$; blue line: expansion around $x=1$. Right panel: illustration of the 
relative accuracy and overlap of the two solutions $f_{8b}(x)$ around 0 and 1.
\label{fig:f8b}}
\end{figure}

\begin{figure}[H]
\centering
\includegraphics[width=0.47\textwidth]{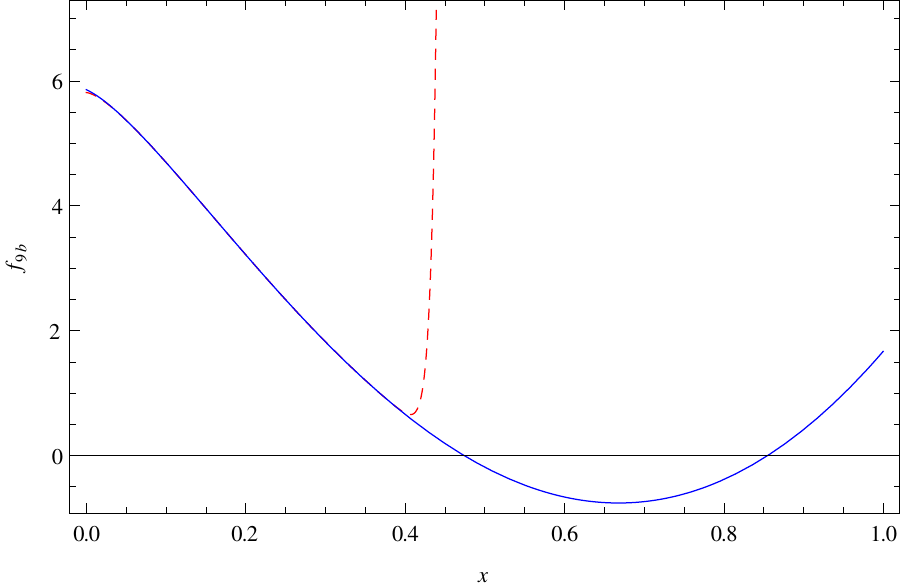}
\includegraphics[width=0.52\textwidth]{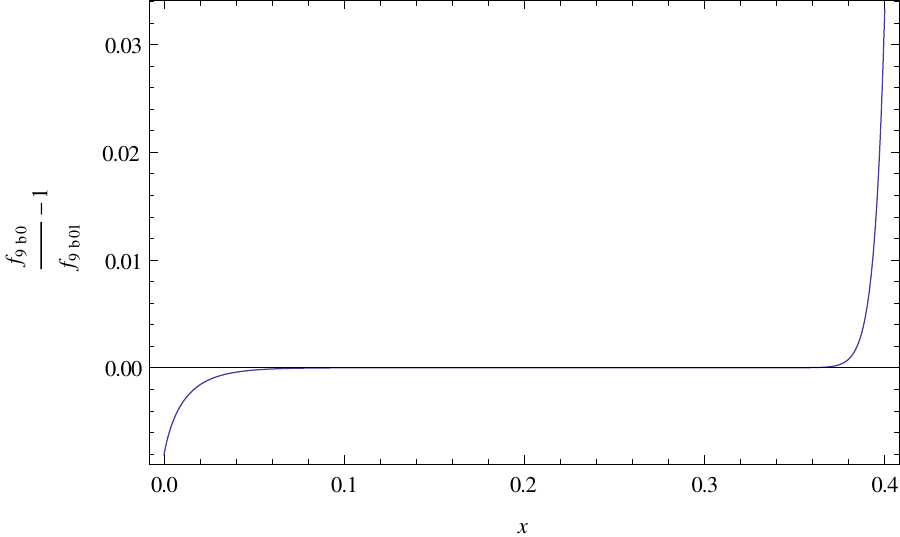}
\caption{\sf \small The inhomogeneous solution of 
Eq.~(\ref{eq:f9bc}) as a function of $x$. Left panel: Red dashed line: expansion 
around $x=0$; blue line: expansion around $x=1$. Right panel: illustration of the 
relative accuracy and overlap of the two solutions $f_{9b}(x)$ around 0 and 1.
\label{fig:f9b}}
\end{figure}

\begin{figure}[H]
\centering
\includegraphics[width=0.47\textwidth]{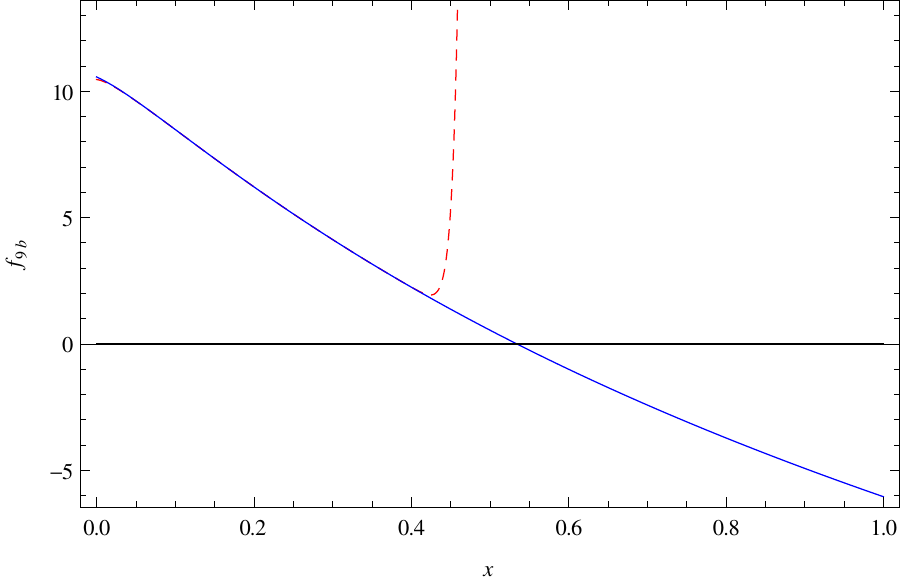}
\includegraphics[width=0.52\textwidth]{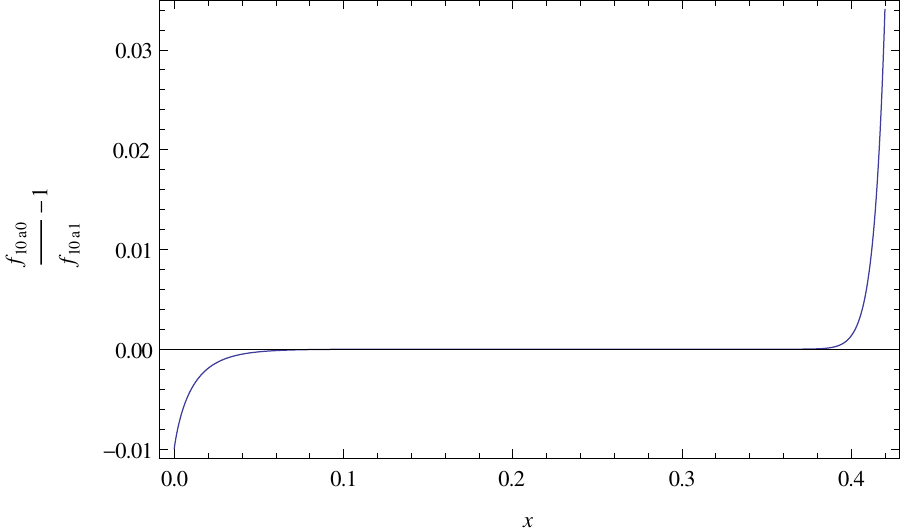}
\caption{\sf \small The inhomogeneous solution of 
Eq.~(\ref{eq:10b}) as a function of $x$. Left panel: Red dashed line: expansion 
around $x=0$; blue line: expansion around $x=1$. Right panel: illustration of the 
relative accuracy and overlap of the two solutions $f_{10b}(x)$ around 0 and 1.
\label{fig:f10b}}
\end{figure}

The solutions are illustrated in Figures~\ref{fig:f8a}--\ref{fig:f10b}. The expansions around $x=0$ and $x=1$ have 
wide overlapping regions in all cases. We use expansions up to $O(x^{50})$ and $O(y^{50})$, respectively. Due to the 
constants $C_i (\overline{C}_i), i = 1,2,3,$ which are imposed by the physical case studied, all solutions are 
real
in the region $x \in [0,1]$. The fact, that the homogeneous solutions in the cases $b$, have a branch point, has, 
however, consequences for the solutions around $x=0$, as will be shown below.

The function $f_{8a}(x)$ is shown in Figure~\ref{fig:f8a}. Its boundary values at $x=0,1$ read
\begin{eqnarray}
f_{8a}(0) = \frac{135}{16} - \pi^2 +\frac{3}{2} \psi^{(1)}\left(\frac{1}{3}\right)
~~~~~\text{and}~~~~~f_{8a}(1) = \frac{275}{12}~.
\end{eqnarray}
\noindent
At very small $x$, the expansion around $x=1$ delivers too small values, while at large $x$ the small $x$ 
expansion evaluates to somewhat larger values, however, well below double precision.
\noindent
$f_{9a}(x)$ is shown in Figure~\ref{fig:f9a} with the values 
\begin{eqnarray}
f_{9a}(0) = \frac{5}{2} + \frac{2}{3} \pi^2  - \psi^{(1)}\left(\frac{1}{3}\right)
~~~~~\text{and}~~~~~f_{9a}(1) = \frac{5}{3}~,
\end{eqnarray}
\noindent
\noindent
at $x=0,1$ and a very similar behaviour for the approximation around $x=0$ and 1 as in the case of $f_{8a}$.
Figure~\ref{fig:f10a} shows the function $f_{10a}$, for which the boundaries are 
\begin{eqnarray}
f_{10a}(0) = 
-\frac{19}{72} \pi^4 + \frac{2}{3} \pi^2 \psi^{(1)}\left(\frac{1}{3}\right)
- \psi^{(1)}\left(\frac{1}{3}\right)^2 + 6 \zeta_3 
~~~~~\text{and}~~~~~f_{10a}(1) = 
2 {\sf B}_4 -9 \zeta_4~.
\end{eqnarray}
Here somewhat larger deviations of the series solutions around $x=0$ at 1 and $x=1$ at 0 are visible.
\noindent

In Figure~\ref{fig:f8b} the behaviour of $f_{8b}(x)$ is illustrated.
\noindent
The series expansion around $x=0$ starts to diverge at $x \sim 0.4$, while the expansion around $x=1$ still holds at 
$x \sim 0.1$. The boundary values of $f_{8b}$ at $x=0,1$ are
\begin{eqnarray}
f_{8b}(0) = - \frac{145}{48} - \frac{1}{2} \zeta_2 
~~~~~\text{and}~~~~~
f_{8b}(1) = \frac{275}{12}~.
\end{eqnarray}
There is a numerical artefact in Figure~\ref{fig:f8b}b at $x \sim 0.14$ implied by the zero-transition of
$f_{8b}$ in this region.
\noindent

A similar behaviour to that of $f_{8b}$ is exhibited by $f_{9b}(x)$, shown in Figure~\ref{fig:f9b}. Again the 
series-solution around $x=0$ starts to diverge for $x \sim 0.4$. However, the one around $x=1$ holds even below $x \sim 
0.1$.
The boundary values of $f_{9b}$ at $x=0,1$ are
\begin{eqnarray}
f_{9b}(0) = \frac{25}{6} + \zeta_2
~~~~~\text{and}~~~~~
f_{9b}(1) = \frac{5}{3}~.
\end{eqnarray}

$f_{10b}(x)$ is shown in Figure~\ref{fig:f10b}. The validity of the serial expansions around $x=0$ and 1 are very 
similar to the cases of $f_{8(9)b}(x)$, discussed above.
\noindent
The boundary values at $x=0,1$ are
\begin{eqnarray}
f_{10b}(0) = 3 \zeta_4 + 6 \zeta_3
~~~~~\text{and}~~~~~
f_{10b}(1) = 2 {\sf B}_4 - 9 \zeta_4~.
\end{eqnarray}

Notice that the representations (\ref{eq:two}, \ref{eq:PS2a}) allow for the 
analytic determination of the $N$th expansion coefficient of the corresponding series around $x = 0~ 
(y=0)$ using the techniques of the package {\tt 
HarmonicSums.m}~\cite{HARMONICSUMS,Ablinger:PhDThesis,Ablinger:2011te,Ablinger:2013cf,Ablinger:2014bra}.

The series expansions agree with those  obtained by solving the differential equations
through series Ans\"atze in \cite{Grigo:2012ji}. In an attachment to this paper, we present the expansion
of the solutions around $x=0$ and $x=1$ up to terms of $O(x^{50})$ for further use. The solutions
are well overlapping in wider ranges in $x$. In the case of the functions $f_{8(9,10)a}(x)$ the power series 
expansion around $x=1$ reflects the branch point at $x = 1/3$ in the homogeneous solution. Our general 
expressions easily allow expansions around other fixed values of $x$, which may be useful in special 
numerical applications.

The above representations constitute a practical analytic solution in the case of iterative non-iterative 
integrals. Indeed it applies to the whole class of these functions within their analyticity regions.
Thus the method is not limited to cases in which elliptic integrals contribute. Since, however, the case
in which $_2F_1$ solutions may be related in a non-trivial manner, see ii) and iii) in Section~\ref{sec:5}, to 
solutions through elliptic integrals with rational argument is very frequent, we turn now to a more detailed 
discussion of this case.
\section{Elliptic Solutions}
\label{sec:5}

\vspace*{1mm}
\noindent
As we have seen, in special cases the solutions of a second order differential equation having a $_2F_1$ 
solution may be expressed in 
terms of the complete elliptic integrals ${\bf E}(r(z))$ and ${\bf K}(r(z))$. Our general goal is to 
represent the emerging structures in terms of $q$-series with explicit predicted expansion 
coefficients in closed form as far as possible, if not even simpler representations can be found.

Different levels of complexity can be distinguished, depending on the structure of $r(z)$ and whether only 
elliptic integrals of the first kind or also of the second kind necessarily contribute. Furthermore, there 
are requirements to other building blocks emerging in the solutions, which we will discuss below.
 
\begin{enumerate}[label=(\roman*)]
\item If the complete elliptic integrals are given by ${\bf K}(z)$ or ${\bf K}(1-z)$, choosing the case $z \in 
[0,1]$, and similarly for ${\bf E}$, one may solve the difference equation, obtained from the differential equation
by a Mellin transform. It turns out that this difference equation factorizes to first order, unlike the differential 
equation in $x$-space; see \cite{vonManteuffel:2017hms} for an example. The Mellin transforms (\ref{eq:MELL}) are 
given by
\begin{eqnarray}
\label{eq:Ksimp}
\Mvec[{\bf K}(1-z)](N)  &=& \frac{2^{4N+1}}{\displaystyle (1+2N)^2 \binom{2N}{N}^2} 
\\
\label{eq:Esimp}
\Mvec[{\bf E}(1-z)](N)  &=& \frac{2^{4N+2}}{\displaystyle (1+2N)^2 (3+2N) \binom{2N}{N}^2},
\end{eqnarray}
since
\begin{eqnarray}
{\bf K}(1-z) &=& \frac{1}{2} \frac{1}{\sqrt{1-z}} \otimes \frac{1}{\sqrt{1-z}} 
\\
{\bf E}(1-z) &=& \frac{1}{2} \frac{z}{\sqrt{1-z}} \otimes \frac{1}{\sqrt{1-z}}~. 
\end{eqnarray}
Here the Mellin convolution is  defined by
\begin{eqnarray}
A(z) \otimes B(z) &=& \int_0^1 dz_1 \int_0^1 dz_2 \delta(z-z_1 z_2) A(z_1) B(z_2).
\end{eqnarray}
Eqs.~(\ref{eq:Ksimp}) and (\ref{eq:Esimp}) are hypergeometric terms in $N$, which has been shown already in 
Ref.~\cite{Ablinger:2013eba} for ${\bf K}(1-z)$, see also \cite{Ablinger:2014bra}. As we outlined in 
Ref.~\cite{Ablinger:2015tua} the solution of systems of differential equations or difference equations can 
always be obtained algorithmically in the case either of those factorizes to first order. The transition to 
$z$-space is then straightforward. In $z$-space also the analytic continuation to the other kinematic regions is 
performed.

\item In a second set of cases, only the elliptic integrals ${\bf K}(r(z))$ and ${\bf K}'(r(z))$ contribute, with 
$r(z)$ a rational function. In transforming from $z$- to $q$-space, furthermore, no terms in the solution emerge 
which cannot be expressed in terms of modular forms 
\cite{KF,KOECHER,MILNE,RADEMACHER,DIAMOND,SCHOENENBERG,APOSTOL,KOEHLER,ONO,MIYAKE,SERRE,MART1}, except terms 
$\propto \ln^k(q), k~\in~\mathbb{N}$. This is 
the situation e.g. in Refs.~\cite{BLOCH2,Adams:2014vja,Adams:2016xah}. We will show below that here 
both the homogenous solution and the integrand of the inhomogeneous solution can be expressed by Lambert--Eisenstein 
series \cite{LAMBERT,EISENSTEIN}, also known as elliptic polylogarithms, modulo eventual terms $\ln^k(q)$. The remaining 
$q$-integral in the inhomogeneous term can be carried out in the class of elliptic polylogarithms \cite{ELLPOL}, 
see \cite{Adams:2016xah}.

\item
In the cases presented in Section~\ref{sec:3}, the solutions depend both on the elliptic 
integrals ${\bf K}(r(z))$, ${\bf E}(r(z))$ and ${\bf K}'(r(z))$, ${\bf E}'(r(z))$, see also Section~\ref{sec:6.2}. 
Both ${\bf E}(r(z))$ and ${\bf E}'(r(z))$ can be mapped to modular forms representing them by the nome $q$ 
according to Eqs.~(\ref{eq:nome}, \ref{qeqa}), powers of $\ln(q)$, and polylogarithms, like $\Li_0(q)$ 
\cite{GH}, and the $\eta$-factor given in Eq.~(\ref{eq:etaF}). These aspects lead to a generalization 
w.r.t. the cases covered by ii), since in a series of building blocks the factor $1/\eta^k(\tau)$ has to be
split off to obtain a suitable modular form. This factor is a $q$-Pochhammer symbol and also emerges in the 
$q$-integral in the inhomogeneous solution.
\end{enumerate}

Since the topic of  analytic $q$-series representations is a very recent one and it is only on the way to be
algorithmized and automated for the application to a larger number of cases appearing in Feynman parameter 
integrals, we are going to summarize the necessary definitions and central properties for a wider audience 
in Section~\ref{sec:6.1}. Then we will show in Section~\ref{sec:6.2} that in the case of the differential equations 
(\ref{eq:one}, \ref{eq:two}) both the elliptic integrals {\bf K} and {\bf E} are contributing, which implies the 
appearance of the additional $\eta$-factor (\ref{eq:etaF}). In Section~\ref{sec-52} we will then 
construct the building blocks for the homogeneous and inhomogeneous solutions of all terms through polynomials of 
$\eta$-weighted Lambert--Eisenstein series, referring to the examples (\ref{eq:ps3}, \ref{eq:ps4}). Here we use 
methods of the theory of modular functions and modular forms.
\subsection{From Elliptic Integrals to Lambert--Eisenstein Series}
\label{sec:6.1}

\vspace{1mm}
\noindent 
There are various sets of functions which can be used to express the complete elliptic integrals and their 
inverse, the elliptic functions, which have been worked out starting with Euler \cite{EUL3}, Legendre 
\cite{LEGENDRE} and Abel \cite{ABEL}, followed by Jacobi's seminal work \cite{JAC1,JAC2} and the final 
generalization by Weierstra\ss{} \cite{WEIER}\footnote{For $q$-expansions starting with the Weierstra\ss{}' 
$\wp$ and $\sigma$ functions see e.g. \cite{LANG2}.}.
We first present a collection of relations out of the theory of elliptic integrals, their related functions 
and modular forms \cite{KF,KOECHER,MILNE} for the convenience of the reader. They are essential to derive 
integrals over complete elliptic integrals at rational arguments, which can be represented in terms of
elliptic polylogarithms. Later, we will consider the different steps for a representation of the inhomogeneous 
solution based on the homogeneous solutions $\psi_3$ and $\psi_4$ given before.

We first summarize a series of properties of Jacobi $\vartheta_i$ and the Dedekind $\eta$ functions in 
Section~\ref{sec:6.1.1}, followed by the representation of the complete elliptic integrals 
of the first and second kind by the parameters of the elliptic curve and by the Jacobi $\vartheta_i$ and the Dedekind 
$\eta$ functions in Section~\ref{sec:6.1.2}. Basic facts about modular functions and modular forms are
summarized in Section~\ref{sec:6.1.3} for the later representation of the building blocks of the  homogeneous and 
inhomogeneous solutions of the second order differential equations of Section~\ref{sec:2}. In Section~\ref{sec:6.1.5}
we collect some relations on elliptic polylogarithms and give representations of $\eta$-ratios in terms of
modular forms weighted by a factor $1/\eta^k(\tau)$ in Section~\ref{sec:6.1.4}. The modular forms are expressed
over bases formed by Lambert--Eisenstein series and products thereof. 
\subsubsection{\boldmath The Jacobi $\vartheta_i$ and Dedekind $\eta$ Functions}
\label{sec:6.1.1}

\vspace*{1mm}
\noindent 
As entrance point we use Jacobi's $\vartheta_i$ functions \cite{JAC2}.
The $\vartheta$ functions possess $q$-series and product representations\footnote{In the literature 
different definitions of the Jacobi $\vartheta$-functions are given, cf.~\cite{TRICOMI},~p.~305. We follow the 
one used by {\tt Mathematica}.}

\newpage
\begin{eqnarray}
\label{eq:J1}
\vartheta_1(q,z) &=& \sum_{k = -\infty}^\infty  (-1)^{\left(k-\tfrac{1}{2}\right)}q^{(k+1/2)^2} \exp[(2k+1) iz] 
=  2 q^{\tfrac{1}{4}} \sum_{k=0}^\infty (-1)^n q^{n(n+1)} \sin[(2n+1) z]
\nonumber\\
\\
\label{eq:J2}
\vartheta_2(q,0) &\equiv&  \vartheta_2(q) = \sum_{k = -\infty}^\infty  q^{(k-1/2)^2} 
\\
\label{eq:J3}
\vartheta_3(q,0) &\equiv&  \vartheta_3(q) = \sum_{k = -\infty}^\infty  q^{k^2} 
\\
\label{eq:J4}
\vartheta_4(q,0) &\equiv&  \vartheta_4(q) = \sum_{k = -\infty}^\infty  (-1)^k q^{k^2}.
\end{eqnarray}
The elliptic polylogarithms, introduced in (\ref{eq:ELP1}, \ref{eq:ELP2}) below are also $q$-series, 
containing a specific parameter pattern which allows to accommodate certain classes of $q$-series emerging in Feynman 
integral calculations.
The product representations associated to (\ref{eq:J2}--\ref{eq:J4}) read
\begin{eqnarray}
\vartheta_2(q) &=& 2 q^{\frac{1}{4}} \prod_{k=1}^\infty 
\left(1-q^{2k}\right)\left(1+q^{2k}\right)^2
\\
\vartheta_3(q) &=& \prod_{k=1}^\infty \left(1-q^{2k}\right)\left(1+q^{2k-1}\right)^2
\\
\vartheta_4(q) &=& \prod_{k=1}^\infty \left(1-q^{2k}\right)\left(1-q^{2k-1}\right)^2.
\end{eqnarray}
They are closely related to Euler's totient function \cite{EUL4} 
\begin{eqnarray}
\phi(q) = \prod_{k=1}^\infty \frac{1}{1-q^k},
\end{eqnarray}
the first emergence of $q$-products, and to Dedekind's $\eta$ function \cite{DEDEKINDeta}.\footnote{The $\vartheta$ and $\eta$ 
functions, as well as their $q$-series, play also an important role in other branches of physics, as e.g. in lattice 
models in statistical physics in form of Rogers-Ramanujan identities, see~e.g.~\cite{BAXTER,MCCOY,BOSTAN},  percolation 
theory \cite{KZ}, and other applications, e.g. in attempting to describe properties of deep-inelastic structure 
functions~\cite{Scott:1999nm}. In the latter case, the asymptotic behavior of Dedekind's $\eta$ function at $x \sim 1$ 
seems to resemble the structure function for a wide range down to $x~\sim~10^{-5}$. It has a surprisingly similar form as 
the small-$x$ asymptotic wave equation solution \cite{WAVE}, however, with a rising power of the soft 
pomeron \cite{Donnachie:1996gq}.}
\begin{eqnarray}
\label{eq:DEDmain}
\eta(\tau) = \frac{q^{\frac{1}{12}}}{\phi(q^2)}.
\end{eqnarray}
One has\footnote{It is usually desirable to work with $\eta$-functions depending on integer
multiples of $\tau$ only, cf.~\cite{KOEHLER}, which can be achieved by rescaling the power of $q$.}
\begin{eqnarray}
\vartheta_2(q) &=& \frac{2 \eta^2(2\tau)}{\eta(\tau)}
\\
\vartheta_3(q) &=& \frac{\eta^5(\tau)}{\eta^2\left(\frac{1}{2}\tau\right) \eta^2(2 \tau)}
\\
\vartheta_4(q) &=& \frac{\eta^2\left(\frac{1}{2}\tau\right)}{\eta(\tau)}.
\end{eqnarray}

In the following we will make use of series representations of both Jacobi $\vartheta$- and
Dedekind $\eta$-functions.  We list a series of important relations for convenience :
\begin{align}
\eta(\tau + n) &= e^{i \frac{\pi n}{12}} \eta(\tau),~~~n \in \mathbb{N}         
\\
\eta(\tau)     &= q^{\frac{1}{12}} \sum_{k = -\infty}^\infty (-1)^k q^{3k^2+k}  
& \text{%
\cite{EULER1}
}
\\
\eta^3(\tau)   &= q^{\frac{1}{4}}  \sum_{k = -\infty}^\infty (4k+1) q^{4k^2+2k} 
& \text{%
\cite{JAC1}
}
\\
\frac{\eta^2(\tau)}{\eta(2\tau)} 
               &= \sum_{k = -\infty}^{\infty} (-1)^k q^{2 k^2}                  & \text{%
\cite{GAUSS2}   
}
\\ 
\frac{\eta^2(2\tau)}{\eta(\tau)} 
               &= q^{\frac{1}{4}} \sum_{k = -\infty}^{\infty} q^{4k^2+2k}       & \text{%
\cite{GAUSS2}   
}
\\
\frac{\eta(2\tau)^5}{\eta(\tau)^2} &= 
q^{\frac{2}{3}} \sum_{k = -\infty}^{\infty} (-1)^k (3k+1) q^{6k^2+4k}           & \text{%
\cite{GORDON1}   
}
\\
\frac{\eta(\tau)^5}{\eta(2\tau)^2} &= 
q^{\frac{1}{12}} \sum_{k = -\infty}^{\infty} (6k+1) q^{3k^2+k}                  & \text{%
\cite{GORDON1}    
}
\\
\frac{\eta(\tau) \eta(6\tau)^2}{\eta(2\tau) \eta(3\tau)} &= 
q^{\frac{2}{3}} \sum_{k = -\infty}^{\infty} (-1)^k q^{6k^2+4k}                                  & \text{%
\cite{KAC}   
}
\\
\frac{\eta(2\tau) \eta(3\tau)^2}{\eta(\tau) \eta(6\tau)} &= 
q^{\frac{1}{12}} \sum_{k = -\infty}^{\infty} q^{3k^2 + k}                       & \text{%
\cite{KAC}   
}
\\
\frac{\eta(\tau)^2 \eta(6\tau)}{\eta(2\tau) \eta(3\tau)} &= 
q^{\frac{1}{4}} \sum_{k = -\infty}^{\infty} \left(q^{9k^2+3k} -q^{(3k+1)(3k+2)} \right).                      & 
\text{%
\cite{KAC}   
}
\end{align}
Many other identities hold and can be found e.g. 
in~Refs.~\cite{KOEHLER,MACDONALD,KAC,LEPOWSKY,ZUCKER,MARTIN,WILLIAMS,KENDRIL,ONO}.

\subsubsection{\boldmath Representations of the Modulus and the Elliptic Integrals}
\label{sec:6.1.2}

\vspace*{1mm}
\noindent 
For later use we consider also the structure of the differential equation of the  Weierstra\ss{}' function 
$\wp(z)$ \cite{WEIER},
\begin{eqnarray}
\label{eq:WEI1}
\wp'^2(z) = 4 \wp^3(z) - g_2 \wp(z) -g_3 = 4 
(\wp(z) - e_1)
(\wp(z) - e_2)
(\wp(z) - e_3)~.
\end{eqnarray}
The functions $g_2, g_3, e_1, e_2$ and $e_3$ are given by
\begin{eqnarray}
g_2 &=& -4 [e_2 e_3 + e_3 e_1 + e_1 e_2] = 2[e_1^2 + e_2^2 + e_3^2]
\\
g_3 &=& 4 e_1 e_2 e_3 = \frac{4}{3} [e_1^3 + e_2^3 + e_3^3]
\\
\label{eq:eI}
e_1 + e_2 + e_3
&=&  0,
\end{eqnarray}
\begin{figure}[H]
\centering
\includegraphics[width=0.60\textwidth]{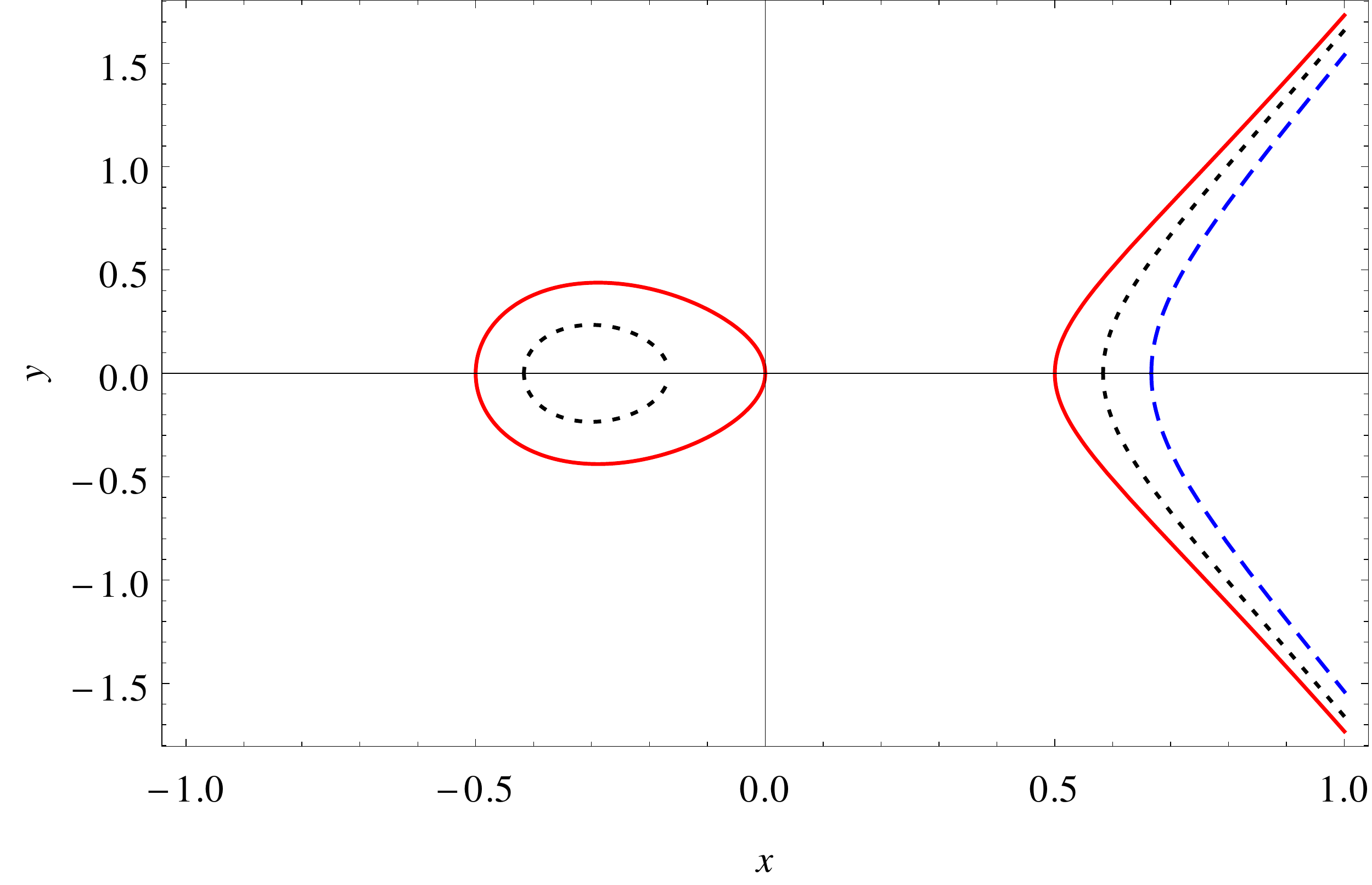}
\caption{\sf \small The elliptic curve for $k = 0$ (dashed blue line), $k = 1/2$ (dotted black lines), and 
$k = 1/\sqrt{2}$ full red lines.}
\label{fig:ELLC}
\end{figure}

\noindent
and the following representation in terms  of Jacobi $\vartheta$ functions holds:
\begin{eqnarray}
e_1 &=& ~~\frac{\pi^2}{12 \omega^2} \left[\vartheta_3^4(q) + \vartheta_4^4(q)\right]
\\ 
e_2 &=& ~~\frac{\pi^2}{12 \omega^2} \left[\vartheta_2^4(q) + \vartheta_4^4(q)\right]
\\ 
e_3 &=& -\frac{\pi^2}{12 \omega^2} \left[\vartheta_2^4(q) + \vartheta_3^4(q)\right].
\end{eqnarray}
Here Jacobi's identity is implied by (\ref{eq:eI}) with
\begin{eqnarray}
\vartheta_3^4(q) &=& \vartheta_2^4(q) + \vartheta_4^4(q)~.
\end{eqnarray}
The r.h.s. of (\ref{eq:WEI1}) parameterizes the elliptic curve 
\begin{eqnarray}
\label{eq:ELLC}
y^2 = 4  (x - e_1)(x - e_2)(x - e_3)
\end{eqnarray}
of the corresponding problem.
Setting $e_1 - e_2 = 1$ for the purpose of illustration, the elliptic curves corresponding to the module $k$ is 
shown in Figure~\ref{fig:ELLC}, choosing specific values.

The modulus $k$ can be represented in terms of the functions $e_i$ by
\begin{eqnarray}
k^2 = z(x),
\end{eqnarray}
cf.~(\ref{eq:z1}, \ref{eq:z2}). $k$ and $k'= \sqrt{1-k^2}$ are given by
\begin{eqnarray}
k  &=& \sqrt{\frac{e_3 - e_2}{e_1 - e_2}} 
= \frac{\vartheta_2^2(q)}{\vartheta_3^2(q)} 
\equiv \frac{4 \eta^{8}(2\tau) \eta^4\left(\frac{\tau}{2}\right)}{\eta^{12}(\tau)} 
\\
k' &=& \sqrt{\frac{e_1 - e_3}{e_1 - e_2}}
=
\frac{\vartheta_4^2(q)}{\vartheta_3^2(q)} \equiv \frac{\eta^{4}(2\tau) 
\eta^{8}\left(\frac{\tau}{2}\right)}{\eta^{12}(\tau)}, 
\end{eqnarray}
cf.~(\ref{eq:WEI1}), which  implies the following relation for $\eta$ functions 
\begin{eqnarray}
\label{eq:ID4}
1 = \frac{\eta^8\left(\frac{\tau}{2}\right) \eta^8(2\tau)}{\eta^{24}(\tau)} \left[16 \eta^8(2\tau) 
+ \eta^8\left(\frac{\tau}{2}\right) \right]~.
\end{eqnarray}
Further, one may express the elliptic integral of the first kind {\bf K} by
\begin{eqnarray}
\label{eq:K}
{\bf K}(k^2) &=& \omega \sqrt{e_1 - e_3},~~~~\text{with}~~~~\sqrt{e_1 - e_3} = \frac{\pi}{2 \omega} 
\vartheta_3^2(q) \equiv  \frac{\pi}{2} \frac{\eta^{10}(\tau)}{\eta^4\left(\frac{\tau}{2}\right)   
\eta^4(2\tau)},
\\
\label{eq:Kp}
{\bf K'}(k^2) &=& - \frac{1}{\pi} {\bf K}(k^2)~ \ln(q)~.
\end{eqnarray}
Sometimes one also introduces the Jacobi functions $\omega, \omega', \eta$ and $\eta'$, which are defined by
\begin{eqnarray}
\omega    &=& \frac{{\bf K}}{\sqrt{e_1 - e_3}}
\\
\omega'   &=& i\frac{{\bf K'}}{\sqrt{e_1 - e_3}} = \omega \tau = \frac{\omega}{i \pi} \ln(q)
\\
\eta  &=& - \frac{1}{12 \omega} \frac{\vartheta_1'''(q)}{\vartheta_1'(q)},
\end{eqnarray}
with
\begin{eqnarray}
\vartheta_1^{(k)}(q) = \lim_{z \rightarrow 0} \frac{d^k}{dz^k} \vartheta_1(q,z).
\end{eqnarray}
The function
$\eta'$ can be obtained using Legendre's identity (\ref{eq:LEGEND}) in the form
\begin{eqnarray}
\eta \omega' - \eta' \omega = i \frac{\pi}{2}.
\end{eqnarray}
One obtains the following representations of the elliptic integrals of the second kind by
\begin{eqnarray}
{\bf E}(k^2)   &=& \frac{e_1 \omega + \eta}{\sqrt{e_1 - e_3}}
\\
{\bf E'}(k^2)  &=& i \frac{e_3 \omega' + \eta'}{\sqrt{e_1 - e_3}}.
\end{eqnarray}
Later on we will use the relation \cite{ERDELYI2,ABRSTE} for {\bf E}
\begin{eqnarray}
{\bf E}(k^2)   &=& {\bf K}(k^2) + \frac{\pi^2 q}{{\bf K}(k^2)} \frac{d}{dq} \ln\left[\vartheta_4(q)\right] 
\label{eq:EMain}
\end{eqnarray}
and the Legendre identity (\ref{eq:LEGEND}) to express {\bf E}$'$,
\begin{eqnarray}
{\bf E}'(k^2)   &=& \frac{\pi}{2{\bf K}(k^2)} \left[1 + 2 \ln(q)~q\frac{d}{dq} 
\ln\left[\vartheta_4(q)\right]\right].
\label{eq:EMainPR}
\end{eqnarray}
\subsubsection{\boldmath Modular Forms and Modular Functions}
\label{sec:6.1.3}

\vspace*{1mm}
\noindent 
All building blocks forming the homogeneous solutions and the integrand of the inhomogeneous solutions
of the second order differential equations considered above can be expressed in terms of $\eta$-ratios.
They are defined as follows.
\definition
\label{def:6.1}
\noindent
Let $r = \left(r_\delta\right)_{\delta|N}$ be a finite sequence of integers indexed by the divisors 
$\delta$ of $N \in \mathbb{N} \backslash \{0\}$. The function $f_r(\tau)$ 
\begin{eqnarray}
f_r(\tau) := \prod_{d|N} \eta(d \tau)^{r_d},~~~d,N \in \mathbb{N} \backslash\{0\},~~r_d \in \mathbb{Z},
\label{eq:etaR}
\end{eqnarray}
is called {\sf $\eta$-ratio}.
\rm

\noindent
These are modular functions or modular forms; the former ones can be obtained as the ratio of two
modular forms. In the following we summarize a series of basic facts on these quantities in a series 
of definitions and theorems needed in the calculation of the present paper, cf. also Refs.~\cite{KF,KOECHER,
MILNE,RADEMACHER,DIAMOND,SCHOENENBERG,APOSTOL,KOEHLER,ONO,MIYAKE,SERRE,MART1}. 

\definition 
\label{def:6.2}
Let
\begin{eqnarray}
{\rm SL}_2(\mathbb{Z}) = \Biggl\{ M = \left(
\begin{array}{cc} a & b \\ c & d \end{array}
\right),~a,b,c,d \in \mathbb{Z},~\det(M) = 1 \Biggr\}.
\nonumber
\end{eqnarray}
${\rm SL}_2(\mathbb{Z})$ is the {\sf modular group}.
\rm

\noindent	
For $g = \left(\begin{array}{cc} a & b \\ c & d \end{array} \right) \in {\rm SL}_2(\mathbb{Z})$ and $z \in \mathbb{C} \cup 
\infty$ one defines the M\"obius transformation
\begin{eqnarray}
g z \mapsto \frac{az + b}{cz + d}.
\nonumber
\end{eqnarray}
Let 
\begin{eqnarray}
S = \left(\begin{array}{rr} 0 & -1 \\ 1 & 0 \end{array}\right),~~~~\text{and}~~~~
T = \left(\begin{array}{rr} 1 &  1 \\ 0 & 1 \end{array}\right),~~~S, T \in {\rm SL}_2(\mathbb{Z}).
\nonumber
\end{eqnarray}
These elements generate ${\rm SL}_2(\mathbb{Z})$ and one has
\begin{align*}
Sz      &\mapsto -\frac{1}{z}, &
Tz      &\mapsto z+1, &
S^2z    &\mapsto z, &
(ST)^3z &\mapsto z.
\end{align*}

\definition
\label{def:6.3}
For $N \in \mathbb{N} \backslash \{0\}$ one considers the {\sf congruence subgroups} of 
${\rm SL}_2(\mathbb{Z})$, $\Gamma_0(N)$,
$\Gamma_1(N)$ and $\Gamma(N)$, defined by
\begin{align*}
\Gamma_0(N) &:= \left\{ \left(\begin{array}{cc} a & b \\ c & d \end{array} \right) \in {\rm SL}_2(\mathbb{Z}), 
c \equiv 0~~(\modd~N) \right\},
\\
\Gamma_1(N) &:= \left\{ \left(\begin{array}{cc} a & b \\ c & d \end{array} \right) \in {\rm SL}_2(\mathbb{Z}), 
a \equiv d \equiv 1~~(\modd~N),~~c \equiv 0~(\modd~N) \right\},
\\
\Gamma(N) &:= \left\{ \left(\begin{array}{cc} a & b \\ c & d \end{array} \right) \in {\rm SL}_2(\mathbb{Z}), 
a \equiv d \equiv 1~~(\modd~N),~~b \equiv c \equiv 0~~(\modd~N) \right\},
\end{align*}
with ${\rm SL}_2(\mathbb{Z}) 
\supseteq \Gamma_0(N) 
\supseteq \Gamma_1(N) 
\supseteq \Gamma(N)$ and $\Gamma_0(N) \subseteq \Gamma_0(M),~M|N$.
\rm

\noindent

\proposition
\label{prop:6.4}
If $N \in \mathbb{N} \backslash \{0\}$, then the {\sf index} of $\Gamma_0(N)$ in $\Gamma_0(1)$ is
\begin{align*}
\mu_0(N) = [\Gamma_0(1) : \Gamma_0(N)] = N \prod_{p|N} \left(1 + \frac{1}{p}\right).
\end{align*}
The product is over the prime divisors $p$ of $N$.
\rm

\definition
\label{def:6.6}
For any congruence subgroup $G$ of {\rm SL}$_2(\mathbb{Z}$) a {\sf cusp} of $G$ is an equivalence class in
$\mathbb{Q} \cup \infty$ under the action of $G$, cf.~{\rm \cite{ONO}}.  
\rm

\definition
\label{def:6.5}
Let $x \in \mathbb{Z} \backslash \{0\}$. The analytic function $f: \mathbb{H} \rightarrow \mathbb{C}$ is a 
{\sf modular form} of {\sf weight} \newline
{\sf w} = $k$ for $\Gamma_0(N)$ and {\sf character} $a \mapsto \left(\tfrac{x}{a}\right)$ if 
\begin{enumerate}[label=(\roman*)]
\item \[
f\left(\frac{az +b}{cz +d}\right) = \left(\frac{x}{a}\right) (cz+d)^k f(z),~~~\forall z \in 
\mathbb{H},~\forall \left( \begin{array}{cc} a & b \\ c &d \end{array} \right) \in \Gamma_0(N).\]
\item $f(z)$ is holomorphic in $\mathbb{H}$
\item $f(z)$ is holomorphic at the cusps of $\Gamma_0(N)$, cf.~{\rm \cite{PR1}, p.~532}.
\end{enumerate}
Here $\left(\tfrac{x}{a}\right)$ denotes the Jacobi symbol {\rm \cite{JAC6}}.\footnote{For its efficient 
evaluation
see e.g. \cite{LEUT}.} A modular form is called a {\sf cusp form} if it vanishes at the cusps.
\rm

\definition
\label{def:6.7}
A {\sf modular function} $f$ for $\Gamma_0(N)$ and weight {\sf w = $k$} obeys
\begin{enumerate}[label=(\roman*)]
\item $f(\gamma z) = (cz+d)^{k} f(z),~~~\forall z \in \mathbb{H}~~\text{and}~~\forall \gamma \in \Gamma_0(N)$ 
\item $f$ is meromorphic in $\mathbb{H}$
\item $f$ is meromorphic at the cusps of $\Gamma_0(N)$.
\end{enumerate}
\rm

\vspace*{1mm}\noindent
The $q$ expansion of a modular function has the form
\begin{equation}
f^*(q) = \sum_{k=-N_0}^\infty a_k q^k,~~~\text{for~some}~~N_0 \in \mathbb{N}.
\nonumber
\end{equation}

\lemma
\label{lem:6.8}
The set of functions ${\cal M}(k; N; x)$ for $\Gamma_0(N)$ and character $x$ obeying Definition~{\rm \ref{def:6.5}}
forms a finite dimensional vector space over $\mathbb{C}$. In particular, for
any non-zero function $f\in {\cal M}(k; N; x)$ we have
\begin{eqnarray}
{\rm ord}(f) \leq b = \frac{k}{12} \mu_0(N),
\label{eq:DIMV}
\end{eqnarray}
cf. e.g.~{\rm \cite{OGG,KOECHER,KOEHLER}}. 
\rm

\noindent
The bound (\ref{eq:DIMV}) on the dimension has been refined, cf. 
e.g.~\cite{HECKE,MILNE,SCHOENENBERG,MIYAKE}\footnote{
The dimension of the corresponding vector space can be also calculated using the {\tt Sage} program
by W.~Stein \cite{STEIN}.}.
The number of independent modular forms $f \in {\cal M}(k; N; x)$ is $\leq b$, allowing for a basis 
representation in finite terms.

For any $\eta$-ratio $f_r$ (\ref{eq:etaR}) one can prove that there exists a minimal integer $l \in \mathbb{N}$,
an integer $N \in \mathbb{N}$ and a character $x$ such that
\begin{eqnarray}
\label{eq:ETAF}
\bar{f}_r(\tau) = \eta^l(\tau) f_r(\tau) \in {\cal M}(k; N; x)
\end{eqnarray}
is a modular form. All quantities which are expanded in $q$-series below will be first brought into the form
(\ref{eq:ETAF}). In some cases one has $l =0$. The form (\ref{eq:ETAF}) is of importance to obtain
Lambert-Eisenstein series (Section~\ref{sec:6.1.4}), which can be rewritten in terms of elliptic polylogarithms 
(Section~\ref{sec:6.1.5}).

Applying the following Theorem, one can find the  $\eta$-ratios belonging to ${\cal M}({\sf w}; N; 1)$. 
\theorem {\it (Paule, Radu, Newman); {\rm \cite{PR,NEWMAN}}.\\
\label{thm:6.9}
Let $f_r$ be an $\eta$-ratio of weight {\sf w} = $\tfrac{1}{2} \sum_{d|N} r_d$. 
$f_r \in {\cal M}({\sf w}; N; 1)$ if the following conditions are satisfied
\begin{enumerate}[label=(\roman*)]
\item $\sum_{d|N} d r_d \equiv 0~({\rm mod}~24)$
\item $\sum_{d|N} N r_d/d \equiv 0~({\rm mod}~24)$
\item $\prod_{d|N} d^{r_d}$ is the square of a rational number
\item $\sum_{d|N} r_d \equiv 0 ~({\rm mod}~4)$
\item $\sum_{d|N} {\gcd}^2(d,\delta) r_d/d \geq 0,~~~\forall \delta|N$.
\end{enumerate}}
\rm
If we refer to modular forms they are thought to be those of ${\rm SL}_2(\mathbb{Z})$, if not specified
otherwise.
\subsubsection{\boldmath Elliptic Polylogarithms}
\label{sec:6.1.5}

\vspace*{1mm}
\noindent 
The elliptic polylogarithm is defined by \cite{ELLPOL}\footnote{For a recent numerical representation of
elliptic polylogartithms see \cite{Passarino:2016zcd}.} 
\begin{eqnarray}
\label{eq:ELP1}
{\rm ELi}_{n;m}(x;y;q) = 
\sum_{k = 1}^\infty
\sum_{l = 1}^\infty \frac{x^k}{k^n} \frac{y^l}{l^m} q^{kl}.
\end{eqnarray} 
It appears in the present context, because it is a function which allows to represent
the different Lambert--Eisenstein series, cf. Section~\ref{sec:6.1.4}, spanning the $\eta$-ratios 
$\bar{f}_r(\tau)$. In the following we briefly describe a few of its properties, which will 
be applied later on.

Sometimes it appears useful, cf.~\cite{Adams:2016xah}, to refer also to 
\begin{eqnarray}
\label{eq:ELP3}
\overline{E}_{n;m}(x;y;q) = \Biggl\{
\renewcommand{\arraystretch}{1.5}
\begin{array}{ll}
\tfrac{1}{i} [\ELI_{n;m}(x;y;q) - \ELI_{n;m}(x^{-1};y^{-1};q)], & n+m~~\text{even}\\
~~\ELI_{n;m}(x;y;q) + \ELI_{n;m}(x^{-1};y^{-1};q), & n+m~~\text{odd}.
\\
\end{array} 
\renewcommand{\arraystretch}{1}
\end{eqnarray}

The multiplication relation of elliptic polylogarithms is given by \cite{ELLPOL}
\begin{align}
\label{eq:ELP4a}
&{\ELI}_{n_1,...,n_l;m_1,...,m_l;0,2o_2,...,2o_{l-1}}(x_1,...,x_l;y_1,...,y_l;q) = 
\nonumber\\
& 
\hspace*{4cm}
{\ELI}_{n_1;m_1}(x_1;y_1;q)
{\ELI}_{n_2,...,n_l;m_2,...,m_l;2o_2,...,2o_{l-1}}(x_2,...,x_l;y_2,...,y_l;q), 
\end{align}
with
\begin{eqnarray}
\label{eq:ELP2}
\ELI_{n_,...,n_l;m_1,...,m_l;2o_1,...,2o_{l-1}}(x_1,...,x_l;y_1,...y_l;q)
&=& \sum_{j_1=1}^\infty ...
\sum_{j_l=1}^\infty 
\sum_{k_1=1}^\infty ...
\sum_{k_l=1}^\infty 
\frac{x_1^{j_1}}{j_1^{n_1}} ...
\frac{x_l^{j_l}}{j_l^{n_l}} 
\frac{y_1^{k_1}}{k_1^{m_1}} 
\frac{y_l^{k_l}}{k_l^{m_l}} 
\nonumber\\ && \times
\frac{q^{j_1 k_1+...+q_l k_l}}{\prod_{i=1}^{l-1}(j_ik_i+...+j_lk_l)^{o_{i}}},l > 0.
\end{eqnarray}
For the synchronization of different elliptic polylogarithms w.r.t. the argument $q$, also the 
relation
\begin{eqnarray}
\label{eq:ELP2S}
&&\hspace*{-1cm}
\ELI_{n_,...,n_l;m_1,...,m_l;2o_1,...,2o_{l-1}}(x_1,...,x_l;y_1,...y_l;-q)
\nonumber\\ 
&&\hspace*{2.8cm}
= \ELI_{n_,...,n_l;m_1,...,m_l;2o_1,...,2o_{l-1}}(-x_1,...,-x_l;-y_1,...-y_l;q)
\end{eqnarray}
is used. In deriving representations in terms of Lambert--Eisenstein series, it often occurs that
the variable is not $q$ but $q^m, m > 1, m \in \mathbb{N}$. Its synchronization to $q$
is shown in Section~\ref{sec:6.1.4}.

The logarithmic integral of an elliptic polylogarithm is given by
\begin{align}
\label{eq:ELP5}
&\ELI_{n_1,...,n_l;m_1,...,m_l;2(o_1+1),2o_2,...,2o_{l-1}}(x_1,...,x_l;y_1,...,y_l;q) = 
\nonumber\\
& 
\hspace*{4cm}
\int_0^q 
\frac{dq'}{q'} 
\ELI_{n_1,...,n_l;m_1,...,m_l;2o_1,...,2o_{l-1}}(x_1,...,x_l;y_1,...,y_l;q'). 
\end{align}
Similarly, cf.~\cite{Adams:2016xah}
\begin{align}
&\overline{E}_{n_1,...,n_l;m_1,...,m_l;0,2o_2,...,2o_{l-1}}(x_1,...,x_l;y_1,...,y_l;q) =
\nonumber\\
& \hspace*{5cm}
\overline{E}_{n_1;m_1}(x_1;y_1;q)
\overline{E}_{n_2,...,n_l;m_2,...,m_l;2o_2,...,2o_{l-1}}(x_1,...,x_l;y_1,...,y_l;q)
\\
&\overline{E}_{n_1,...,n_l;m_1,...,m_l;2(o_1+1),2o_2,...,2o_{l-1}}(x_1,...,x_l;y_1,...,y_l;q) =
\nonumber\\
& \hspace*{5cm}
\int_0^q \frac{dq'}{q'}\overline{E}_{n_1,...,n_l;m_1,...,m_l;2o_1,...,2o_{l-1}}(x_1,...,x_l;y_1,...,y_l;q') 
\end{align}
holds.

The integral over the product of two more general elliptic polylogarithms is given by
\begin{eqnarray}
\int_0^q \frac{d\bar{q}}{\bar{q}} \ELI_{m,n}(x,q^a,q^b) \ELI_{m',n'}(x',q^{a'},q^{b'})
&=& 
\sum_{k=1}^\infty
\sum_{l=1}^\infty
\sum_{k'=1}^\infty
\sum_{l'=1}^\infty
\frac{x^k}{k^m}
\frac{x'^k}{k'^{m'}}
\frac{{q}^{al}}{l^{n}}
\frac{{q}^{a'l'}}{l'^{n}}
\nonumber\\ && \times
\frac{{q}^{bkl + b'k'l'}}{al + a'l' + bkl +bk'l'}.
\end{eqnarray}
Integrals over other products are obtained accordingly. 
\subsubsection{\boldmath Representations in terms of $\eta$-Weighted Lambert--Eisenstein Series}
\label{sec:6.1.4}

\vspace*{1mm}
\noindent 
We turn now to the basis representation of the modular forms of ${\cal M}(k; N; x)$, cf.~Lemma~\ref{lem:6.8}.
It is given by the Eisenstein series \cite{LAMBERT,EISENSTEIN} for weight {\sf w} $= k$  and products of 
Eisenstein series of total weight $k$. In the cases dealt with below products of two Eisenstein series turned 
out to be sufficient. In more involved cases also products of more Eisenstein series might appear.

The Eisenstein series are defined by
\begin{eqnarray}
\label{eq:EISEN1}
G_{2k}(z)  =  \sum_{{m,n} \in \mathbb{Z}^2 \backslash \{0,0\}} \frac{1}{(m+nz)^{2k}},
\end{eqnarray} 
which can be  rewritten in normalized form by 
\begin{eqnarray}
\label{eq:EISEN2}
E_{2k}(q) = \frac{G_{2k}(q)}{2 \zeta_{2k}}  =
1 - \frac{4k}{B_{2k}} \sum_{n=1}^\infty \frac{n^{2k-1} 
q^n}{1-q^n},
\end{eqnarray} 
with $B_{2k}$ the Bernoulli numbers. Eisenstein series are modular forms for $k \geq 2$.
\begin{eqnarray}
\label{eq:EISEN2a}
E_{2}(q) = 
1 - 24 \sum_{n=1}^\infty \frac{n q^n}{1-q^n}
\end{eqnarray}
is not a modular form but one has 

\lemma
\label{lem:E2}
The function $E_2(\tau) - N E_2(N\tau)$ is a modular form of weight {\sf w = 2} for the
group $\Gamma_0(N)$ with the trivial character $x=1$.
\rm

The Eisenstein series are associated to the earlier Lambert series 
\cite{LAMBERT,BB,AGRAWAL,ARNDT,ALZ} which are defined by
\begin{eqnarray}
\label{eq:LAM1}
\sum_{k=1}^\infty \frac{k^\alpha q^k}{1 - q^k} = \sum_{k=1}^\infty \sigma_{\alpha}(k) 
q^k,~~~\sigma_\alpha(k) = 
\sum_{d|k} d^\alpha,~~~\alpha \in \mathbb{N}.
\end{eqnarray} 
Eq.~(\ref{eq:EISEN2}) can be obtained from (\ref{eq:EISEN1}) by applying the Lipschitz summation 
formula \cite{LIPS}. Finally, Eq.~(\ref{eq:LAM1}) can be rewritten in terms of elliptic polylogarithms, 
cf.~Eq.~(\ref{eq:ELP1}), by
\begin{eqnarray}
\sum_{k=1}^\infty \frac{k^\alpha q^k}{1 - q^k} = \sum_{k=1}^\infty k^\alpha \Li_0(q^k) = \sum_{k,l = 
1}^\infty k^\alpha q^{kl} = {\rm ELi}_{-\alpha;0}(1;1;q).
\end{eqnarray} 

\noindent
In the derivation often the argument $q^m,~~m \in \mathbb{N}, m > 0$, appears, which 
shall be mapped to the variable $q$. We do this for the Lambert series using the
replacement
\begin{eqnarray}
\label{eq:xm}
\Li_0(x^m) =
\frac{x^m}{1-x^m} = \frac{1}{m} \sum_{k=1}^m \frac{\rho_m^k x}{1 - \rho_m^k x} = \frac{1}{m} \sum_{k=1}^m \Li_0(\rho_m^k 
x),
\end{eqnarray} 
with 
\begin{eqnarray}
\rho_m = \exp\left(\frac{2\pi i}{m}\right).
\end{eqnarray} 
One has
\begin{eqnarray}
\label{eq:CYC}
\sum_{k=1}^\infty \frac{k^\alpha q^{mk}}{1 - q^{mk}} = \ELI_{-\alpha;0}(1;1;q^m) =
\frac{1}{m^{\alpha+1}} \sum_{n=1}^m 
{\rm ELi}_{-\alpha;0}(\rho_m^n;1;q))~.
\end{eqnarray}
Relations like (\ref{eq:xm}, \ref{eq:CYC}) and similar ones are the sources of the $m$th roots 
of unity, which correspondingly appear in the parameters of the elliptic polylogarithms  through the 
Lambert series. 

Furthermore, the following sums occur
\begin{eqnarray}
\label{eq:extELP1}
\sum_{m=1}^\infty \frac{(am+b)^l q^{am+b}}{1 -  q^{am+b}} &=& \sum_{n=1}^l \binom{l}{n} a^{n} b^{l-n}
\sum_{m=1}^\infty \frac{m^n q^{am+b}}{1-q^{am+b}},~~~a, l \in \mathbb{N},~~~b \in \mathbb{Z}
\end{eqnarray}
and 
\begin{eqnarray}
\label{eq:extELP2}
\sum_{m=1}^\infty \frac{m^n q^{am+b}}{1-q^{am+b}} = \ELI_{-n;0}(1;q^b;q^a) = \frac{1}{a^{n+1}} \sum_{\nu = 
1}^a \ELI_{-n;0}(\rho_a^\nu;q^b;q)~.
\end{eqnarray}

Likewise, one has
\begin{eqnarray}
\label{eq:extELP2a}
\sum_{m=1}^\infty \frac{(-1)^m m^n q^{am+b}}{1-q^{am+b}} &=& 
\ELI_{-n;0}(-1; q^b; q^a) 
\nonumber\\ &=&
\frac{1}{a^{n+1}} \left\{
\sum_{\nu = 1}^{2a} \ELI_{-n;0}(\rho_{2a}^\nu;q^b;q)
-\sum_{\nu = 1}^{a} \ELI_{-n;0}(\rho_{a}^\nu;q^b;q)\right\}
\end{eqnarray}
In intermediate representations also Jacobi symbols appear, obeying the identities
\begin{eqnarray}
\label{eq:JACs}
\left(\frac{-1}{(2k)\cdot n + (2l+1)}\right) = (-1)^{k+l};~~~~
\left(\frac{-1}{a b}\right) =
\left(\frac{-1}{a}\right)
\left(\frac{-1}{b}\right).
\end{eqnarray}
In the case of an even value of the denominator one may factor $\left(\tfrac{-1}{2}\right) = 1$ and 
consider the case of the remaining odd-valued denominator.  

We found also Lambert series of the kind 
\begin{eqnarray}
\label{eq:extELP2b}
\sum_{m=1}^\infty \frac{q^{(c-a)m}}{1-q^{cm}} &=& \ELI_{0;0}(1;q^{-a};q^c) = \frac{1}{c} \sum_{n=1}^c
\ELI_{0;0}(\rho_c^n;q^{-a};q)  
\\
\sum_{m=1}^\infty (-1)^m \frac{q^{(c-a)m}}{1-q^{cm}} &=& \ELI_{0;0}(1;-q^{-a};q^c) = \frac{1}{c} \sum_{n=1}^c
\ELI_{0;0}(\rho_c^n;-q^{-a};q),~~~a,c \in \mathbb{N} \backslash \{0\}  
\nonumber\\
\end{eqnarray}
in intermediate steps of the calculation.

For later use we also introduce the functions
\begin{eqnarray}
\label{eq:Y}
Y_{m,n,l} &:=& \sum_{k=0}^\infty \frac{(mk+n)^{l-1} q^{mk+n}}{1-q^{mk+n}} =
n^{l-1} \Li_0(q^n) + \sum_{j=0}^{l-1} \binom{l-1}{j} n^{l-1-j} m^j \ELI_{-j;0}(1;q^n;q^m)
\nonumber\\
\\
\label{eq:Z}
Z_{m,n,l} &:=& \sum_{k=1}^\infty \frac{k^{m-1} q^{nk}}{1-q^{lk}}  = \ELI_{0;-(m-1)}(1;q^{n-l};q^l)
\\
T_{m,n,l,a,b} &:=& \sum_{k=0}^\infty \frac{(mk + n)^{l - 1} q^{a (mk + n)}}{1 - q^{b (mk + n)}} \nonumber\\ &=& 
n^{l-1} q^{n(a-b)} 
\Li_0\left(q^{nb}\right) 
+ q^{n(a-b)} \sum_{j=0}^{l-1} \binom{l-1}{j} m^j n^{l-1-j} \ELI_{-j;0}\left(q^{m(a-b)};q^{nb};q^{mb}\right),
\nonumber\\
\label{eq:T}
\end{eqnarray}
keeping the $q$-dependence implicit. The functions $Y,Z$ and $T$ allow for more compact representations 
for a series of building blocks given below. Note that (part of) the parameters $(x;y)$ of 
the elliptic polylogarithms can become $q$-dependent.
\subsection{\boldmath The Emergence of ${\rm \bf E}(r(z))$}
\label{sec:6.2}

\vspace{1mm}
\noindent
The solutions of the homogeneous part of the equations (\ref{eq:one}) and (\ref{eq:two}) needed the elliptic 
integrals of the first and second kind. The question arises, whether one would also find solutions based on the
elliptic integral of the first kind only, as it was possible e.g. in the case of Ref.~\cite{Adams:2014vja}. 
There the reason is that the sunrise integral can be written as an integral over $1/\sqrt{y^2}$ where $y^2$ defines an 
elliptic curve. Let us first transform (\ref{eq:one}) and (\ref{eq:two}) into Heun equations with four singularities 
setting $t = x^2$
\begin{eqnarray}
\frac{d^2}{dr^2} F_{8a}(t) - \left(\frac{1}{t-1} + \frac{1}{t-9}\right) \frac{d}{dt} F_{8a}(t) + 
\frac{2(t-3)}{t(t-1)(t-9)} F_{8a}(t) &=& 0,
\\
\frac{d^2}{dr^2} F_{8b}(t) - \left(\frac{1}{t-1} + \frac{1}{t-\tfrac{1}{9}}\right) \frac{d}{dt} F_{8b}(t) +
\frac{2\left(t-\frac{1}{3}\right)}{t(t-1)\left(t-\tfrac{1}{9}\right)} F_{8b}(t) &=& 0.
\end{eqnarray}
One may now consult Refs.~\cite{RHEUN1,RHEUN2}\footnote{J.B. would like to thank S. Weinzierl for having pointed 
out these references to him.}. In this form, both equations do not belong to the cases for which the solution can 
be found as an integral over an algebraic curve, as one finds inspecting the tables given in 
\cite{RHEUN1,RHEUN2}.  However, one may investigate the solution of differential equations associated
to (\ref{eq:one}, \ref{eq:two}), which obey the conditions of \cite{RHEUN1,RHEUN2}. We found equations of this 
type, but needed an additional differential operator to map them back to the original equations. 
The differentiation of an elliptic integral of the first kind will now imply that an elliptic integral
of the second kind is present, as already the well-known relation \cite{TRICOMI}
\begin{eqnarray}
{\bf E}(k^2)   &=&  2 k^2 (1-k^2) \frac{d{\bf K}(k^2)}{dk^2} + (1-k^2){\bf K}(k^2)
\end{eqnarray}
shows. In general, the derivative is for $x$, where $k = k(x)$. One retains nonetheless the dependence on {\bf E},
which has no representation in terms of Lambert--Eisenstein series only, as we show in Section~\ref{sec-52}.

We remark that in the case of the equal-mass sunrise and kite diagrams \cite{BLOCH2,Adams:2014vja,Adams:2016xah} 
one obtains elliptic integrals of the first kind only. The reason for this consists in the direct dispersive 
integral representation of the former and similarly for the kite integral. The solution obeys a corresponding 
second order differential equation in accordance with \cite{RHEUN1,RHEUN2}. 
\subsection{\boldmath The $q$-Series of the $\eta$-Ratio Representations of the Basic Building Blocks}
\label{sec-52}

\vspace*{1mm}
\noindent
In the following we seek a series representations in the nome $q$ (\ref{eq:nome}) of the different 
building blocks of the solutions (\ref{eq:one}--\ref{eq:10b}). We will as widely as possible apply
an algorithmic approach, which is applicable to a wide class of systems emerging in calculations of 
Feynman integrals of a similar type, i.e. being solutions of second order differential equations 
leading to solutions in terms of (complete) elliptic integrals. In this context the theory of modular 
forms and modular functions 
\cite{KF,KOECHER,MILNE,OGG,RADEMACHER,DIAMOND,SCHOENENBERG,APOSTOL,MIYAKE,SERRE,MART1} plays a central 
role.

The different building blocks depend on the kinematic variable $x$, which we discussed first. All 
contributing functions are mapped 
to modular forms $\bar{f}_r$, splitting off a factor $1/\eta^k(\tau)$ if necessary. They are obtained as 
polynomials of Lambert-Eisenstein series, and are mapped to elliptic polylogarithms following 
Sections~\ref{sec:6.1.4}, \ref{sec:6.1.5}. 
\subsubsection{\boldmath The kinematic variable $x$} \label{sec-521} 

\vspace*{1mm}
\noindent
We consider the representation of one of the sets of homogeneous solutions $\psi_{3,4}(z(x))$, with 
$z(x)$ given by (\ref{eq:z2}) and set $\overline{x} = -x$.
\begin{eqnarray}
\overline{x} = -\frac{1}{3y}
\end{eqnarray}
maps the modulus
\begin{eqnarray}
k^2 = z(x) = \frac{16 \overline{x}^3}{(1-\overline{x})^3 (1+3\overline{x})},
\end{eqnarray}
into
\begin{eqnarray}
l^2 = z(y) = \frac{16 y}{(1-y) (1+3y)^3},
\end{eqnarray}
obeying Legendre's modular equation, cf.~\cite{KF},
\begin{eqnarray}
\sqrt{kl} + \sqrt{k'l'} =1,
\end{eqnarray}
cf.~\cite{BORWEIN1,Bailey:2008ib,Broadhurst:2008mx,JOYCE}. The nome $q_k =\exp(-\pi {\bf K}(k'^2)/{\bf K}(k^2))$
is the cube of the nome $q_l \equiv q =\ \exp(-\pi {\bf K}(l'^2)/{\bf K}(l^2))$ and is obtained by
a cubic Legendre-Jacobi transformation \cite{LEGENDRE1,JAC5}\footnote{This is, besides the well-know Landen 
transformation \cite{TRICOMI,LANDENGAUSS}, the next higher modular transformation. There exist even higher order
transformations, which were derived in Refs.~\cite{JAC5,SOHNKE,JOUBERT,KOENIGSBERGER,CAYLEY}. Also for the 
hypergeometric function $\pFq{2}{1}{\tfrac{1}{r},1-\tfrac{1}{r}}{1}{z(x)}$ there are rational modular transformations 
\cite{MAIER}.}. 

According to \cite{BORWEIN1,Bailey:2008ib,Broadhurst:2008mx,JOYCE}
\begin{eqnarray}
\label{eq:cubic1}
\frac{16 y}{(1-y) (1+3y)^3} = \frac{\vartheta_2^4(q)}{\vartheta_3^4(q)}
\end{eqnarray}
is solved by
\begin{eqnarray}
\label{eq:cubic2}
y = \frac{\vartheta_2^2(q^3)}{\vartheta_2^2(q)} \equiv - \frac{1}{3\overline{x}} = \frac{1}{3x}.
\end{eqnarray}
Both the expressions (\ref{eq:cubic1}, \ref{eq:cubic2}) are modular functions. For definiteness, 
we consider the range in $q$
\begin{eqnarray}
\label{eq:kincond}
q \in [-1,1]~~~\text{which~corresponds~to}~~~
y \in \left[0, \tfrac{1}{3}\right],~~~~x \in \left[1, +\infty\right[
\end{eqnarray}
in the following. Here the variable $x$ lies in the unphysical region. However, the nome $q$ has to obey the
condition (\ref{eq:kincond}). Other kinematic regions can be reached performing analytic 
continuations.\footnote{Representations of this kind are frequently used working 
first in a region which is free of singularities, see e.g. \cite{Gluza:2009yy}.}

We would like to make use of the method of proving conjectured $\eta$-ratios by knowing a finite number of 
terms in their $q$-series expansion. For this purpose, we refer to modular forms. In general
it will be therefore necessary to split off an $\eta$-factor from the respective quantity, such that the 
$\eta$-ratio is analytic at the cusps, cf. {\it v)} in Theorem~\ref{thm:6.9}. 
We can achieve this by separating off a common factor of 
\begin{eqnarray}
\frac{1}{\eta^{12}(\tau)}.
\end{eqnarray}
A basis in the corresponding spaces ${\cal M}(k; N; x)$ is used to represent the corresponding 
quantities.

To give a first example we consider the $\eta$-representation for $x$. The associated $\eta$-ratio can  be 
represented in terms of Lambert--Eisenstein series at different powers of $q$ as follows 
\begin{eqnarray}
\label{eq:x1}
x = \frac{1}{3} 
\frac{\eta^{4}(2\tau) 
\eta^{2}(3\tau)}{\eta^2(\tau) \eta^4(6\tau)}
&=&
\frac{1}{\displaystyle 3\eta^{12}(\tau)} \Biggl\{\frac{1}{16} \Biggl(
1-8 \sum_{n=1}^\infty \frac{n^5 (-1)^n q^{2n}}{1-q^{2n}}\Biggr)
+ \frac{3}{16} \Biggl(1 + 16 \sum_{n=1}^\infty \frac{n^3 (-1)^n q^{2n}}{1-q^{2n}}\Biggr)
\nonumber\\ && \times
\left[5 + 24 \sum_{n=1}^\infty \left(\frac{n  q^{2n}}{1-q^{2n}} - \frac{6n 
q^{12n}}{1-q^{12n}}\right)\right]\Biggr\}.
\end{eqnarray}
Following (\ref{eq:CYC}) one obtains
\begin{eqnarray}
x &=& \frac{1}{\displaystyle 3 \eta^{12}(\tau)}\Biggl\{
\frac{1}{16} \Biggl(1 - 8 \ELI_{-5;0}(-1;1;q^2) \Biggr)
+ \frac{3}{16}\left[1 + 16 \ELI_{-3;0}(-1;1;q^2)\right]
\nonumber\\ && 
\times \left\{5 + 24 \ELI_{-1;0}(1;1;q^2) 
- 144 \ELI_{-1;0}(1;1;q^{12}) \right\}\Biggr\}.
\label{eq:xELLP}
\end{eqnarray}
One may synchronize the arguments to $q$ using Eq.~(\ref{eq:extELP2}) and
the products in (\ref{eq:xELLP})  may be formally collected using Eq.~(\ref{eq:ELP4a}). 
For
\begin{eqnarray}
\tilde{x} = x \eta^{12}(\tau),
\end{eqnarray}
both sides are modular forms, and the r.h.s. is expressed as a polynomial of Lambert series.
According to Lemma~\ref{lem:6.8} they agree if the first $b$ non-vanishing expansion coefficients of 
their $q$-series agree. Here we have extracted the power of $1/\eta^{12}(\tau)$, to choose a factor 
often appearing. It is also the minimal factor necessary.
\subsubsection{\boldmath How to find the complete $q$-series of the building blocks?}
\label{sec-htf}

\vspace*{1mm}
\noindent
After having found an exact representation of the kinematic variable $x$ in terms of an $\eta$-ratio  
in Section~\ref{sec-521}, we are in the position to perform the variable transformation from $x$- to 
$q$-space by series expansion at any depth. However, we still have to find the associated $\eta$-ratios
for the corresponding building blocks. An empiric way to derive the $\eta$-ratio would consist in 
systematically enlarging an Ansatz using larger and larger structures (\ref{eq:etaR}) and to compare their 
$q$-series to the one required for a sufficiently large number of terms according to Lemma~\ref{lem:6.8},
after having projected on to a suitable modular form, cf.~Section~\ref{sec:6.1.3}. This is a possible, 
but time-consuming way. Quite a series of $q$-series expressions of $\eta$-ratios are, however, contained 
in Sloan's Online Encyclopedia of Integer Sequences \cite{OEIS}, often with detailed references to the literature,
which one therefore should consult first. Lemma~\ref{lem:6.8} will then allow to prove the corresponding
equality of the two modular forms comparing their $q$-series up to the necessary number of non-vanishing
expansion coefficients. All the relevant $q$-series needed in the following could be found in this way.
\subsubsection{\boldmath The Ingredients of the Homogeneous Solution}
\label{sec-522}

\vspace*{1mm}
\noindent
Let us now construct the individual $q$-series of the further building blocks. The representation of
the elliptic integral of the first kind ${\bf K}(z)$ using (\ref{eq:K}, \ref{eq:LAM1}) is well-known
\begin{eqnarray}
\label{eq:Kmod}
{\bf K}(z) = \frac{\pi}{2}\sum_{k=1}^\infty \frac{q^k}{1+q^{2k}} 
=  \frac{\pi}{i}\sum_{k=1}^\infty \left[
\Li_0\left(i q^k\right) - \Li_0\left(-i q^k\right) \right] 
= \frac{\pi}{4} \overline{E}_{0;0}(i;1;q),
\end{eqnarray}
cf.~\cite{OEIS} {\tt A002654} by M.~Somos, \cite{BYFR}, and Ref.~\cite{KOEHLER}, Eq.~(13.10). ${\bf K}'(z)$ 
is given by (\ref{eq:Kp}). Another quantity, which enters the representation of {\bf E}($k^2$) 
(\ref{eq:EMain}), can also be obtained in terms of Lambert-series directly 
\begin{eqnarray}
q \frac{\vartheta_4'(q)}{\vartheta_4(q)} &=& - \frac{1}{2}\left[\ELI_{-1;0}(1;1;q) + 
\ELI_{-1;0}(-1;1;q)\right]
+ \left[\ELI_{0;0}(1;q^{-1};q) + \ELI_{0;0}(-1;q^{-1};q)\right]
\nonumber\\ &&
- \left[\ELI_{-1;0}(1;q^{-1};q) + \ELI_{-1;0}(-1;q^{-1};q)\right].
\end{eqnarray}
We still need the following $\eta$-weighted $q$-series 
\begin{eqnarray}
\frac{1}{{\bf K}(k^2)} &=& \frac{2}{\pi \eta^{12}(\tau)}
\Biggl\{
\frac{5}{48}\Biggl\{
1 - 24 \ELI_{-1; 0}( 1; 1; q)
  - 4 \Biggl[
    1 - 24 \ELI_{-1; 0}( 1; 1; q^4)\Biggr]\Biggr\} 
\nonumber\\ &&
\times
\Biggl\{
    -1 - 4 \Biggl[
    \ELI_{0; 0}( -1; 1/q; q^2) - 4 \ELI_{-1; 0}( -1; 1/q; q^2)
    + 4 \ELI_{-2; 0}( -1; 1/q; q^2)\Biggr]\Biggr\}
\nonumber\\ &&
- \frac{1}{16}\Biggl\{
  5 - 4 \Biggl[\ELI_{0; 0}( -1; 1/q; q^2) - 8 \ELI_{-1; 0}( -1; 1/q; q^2) +
       24 \ELI_{-2; 0}( -1; 1/q; q^2) 
\nonumber\\ &&
- 32 \ELI_{-3; 0}( -1; 1/q; q^2) +
       16 \ELI_{-4; 0}( -1; 1/q; q^2)
       \Biggr]\Biggr\}
\label{eq:ONEoK}
\end{eqnarray}
to express
${\bf E}(k^2)$, Eq.~(\ref{eq:EMain}). ${\bf E}'(k^2$) is then obtained by (\ref{eq:LEGEND}, 
\ref{eq:EMainPR}).

Next we express the square root factor appearing in (\ref{eq:ps3}, \ref{eq:ps4}), for which the following 
representation in an $\eta$-ratio holds \cite{OEIS} {\tt A256637}
\begin{eqnarray}
\sqrt{(1-3x)(1+x)}     &=& \frac{i}{\sqrt{3}} \left. 
\frac{
\eta\left(\tfrac{\tau}{2}\right)
\eta\left(\tfrac{3\tau}{2}\right) \eta(2\tau) \eta(3\tau)} 
{\eta(\tau)\eta^3(6\tau)}\right|_{q\rightarrow -q}.
\end{eqnarray}
The corresponding $q$-series is given by
\begin{eqnarray}
\sqrt{(1-3x)(1+x)}     &=& \frac{i}{\sqrt{3}} \Biggl\{1+\frac{54}{7} \left[T_{2,1,3,1,12}
+ T_{2,1,3,5,12} - T_{2,1,3,7,12} - T_{2,1,3,11,12} 
- T_{2,2,3,1,12}
\right.
\nonumber\\ && \left. 
- T_{2,2,3,5,12} 
+ T_{2,2,3,7,12} + T_{2,2,3,11,12}\right]
-\frac{26}{7} \left[T_{6,1,3,1,4} - T_{6,1,3,3,4} + T_{6,2,3,3,4}  \right.
\nonumber\\ && \left. 
- 2 T_{6,3,3,1,4} + 2 T_{6,3,3,3,4} - T_{6,4,3,1,4} + T_{6,4,3,3,4} + T_{6,5,3,1,4} - T_{6,5,3,3,4} 
\right. \nonumber\\ && \left. 
+2 T_{6,6,3,1,4} 
- 2 T_{6,6,3,3,4}\right] - 8 \left[Y_{8,2,3}-Y_{8,6,3}\right]
+5 \left[-Y_{12,1,3} - 2Y_{12,3,3}
\right. \nonumber\\ && \left. 
- Y_{12,5,3}+Y_{12,7,3}\right]
-\frac{35}{4} Y_{12,8,3}
+\frac{27}{14} \left[T_{2,1,3,1,12} +T_{2,1,3,5,12} -T_{2,1,3,7,12} 
\right. \nonumber\\ && \left.  
- T_{2,1,3,11,12} 
- T_{2,2,3,1,12} - T_{2,2,3,5,12} + T_{2,2,3,7,12} + T_{2,2,3,11,12} \right] Y_{12,8,3}
\nonumber\\ &&
- \frac{13}{14} \left[
T_{6,1,3,1,4}- T_{6,1,3,3,4}+T_{6,2,3,3,4}-2 T_{6,3,3,1,4}+2 
T_{6,3,3,3,4}-T_{6,4,3,1,4}
\right.
\nonumber\\ && \left.
+T_{6,4,3,3,4}
+T_{6,5,3,1,4}-T_{6,5,3,3,4}+2 T_{6,6,3,1,4}-2 T_{6,6,3,3,4} 
\right] Y_{12,8,3} 
+\left[-8 Y_{4,3,3} \right. 
\nonumber\\ && \left. 
-2 Y_{8,2,3} +2 Y_{8,6,3}\right] Y_{12,8,3}
+\frac{1}{4} \left[-Y_{12,1,3} -2 Y_{12,3,3} -Y_{12,5,3} 
+Y_{12,7,3} \right. 
\nonumber\\ && \left. 
+2 Y_{12,9,3}\right] Y_{12,8,3}
+10 Y_{12,9,3}
+T_{6,2,3,1,4} \left[\frac{26}{7} + \frac{13}{14} \left( -Y_{12,4,3}+Y_{12,8,3}
\right. \right.
\nonumber\\ &&
\left. \left. 
+ Y_{12,10,3}\right)\right]
+\frac{3}{2} Y_{24,2,3}
+\frac{3}{8} Y_{12,8,3} Y_{24,2,3}
-\frac{9}{4} Y_{24,4,3}
-2 Y_{4,3,3} Y_{24,4,3}
\nonumber\\ && 
+\frac{1}{4} \left[Y_{12,1,3} + 2 Y_{12,3,3} +Y_{12,5,3}-Y_{12,7,3} -2 Y_{12,9,3}\right] Y_{24,4,3}
+3 Y_{24,6,3}
\nonumber\\ && 
+\frac{3}{4} Y_{12,8,3} Y_{24,6,3}
-\frac{9}{4} Y_{24,8,3}
-2 Y_{4,3,3} Y_{24,8,3}
+\frac{1}{4} \left[Y_{12,1,3} + 2 Y_{12,3,3} 
\right.
\nonumber\\ && \left. 
+ Y_{12,5,3} - Y_{12,7,3} -2 Y_{12,9,3}\right]Y_{24,8,3}
+\frac{3}{2} Y_{24,10,3}
+\frac{3}{8} Y_{12,8,3} Y_{24,10,3}
-\frac{3}{2} Y_{24,14,3}
\nonumber\\ && 
-\frac{3}{8} Y_{12,8,3} Y_{24,14,3}
+\frac{9}{4} Y_{24,16,3}
+2 Y_{4,3,3} Y_{24,16,3}
-\frac{1}{4} \left[Y_{12,1,3} + 2 Y_{12,3,3} \right. 
\nonumber\\ && \left.
+Y_{12,5,3} - Y_{12,7,3} -2 Y_{12,9,3}\right] Y_{24,16,3}
-3 Y_{24,18,3}
-\frac{3}{4} Y_{12,8,3} Y_{24,18,3}
\nonumber\\ && 
+Y_{4,1,3} \left[8 Y_{12,8,3}+2 Y_{24,4,3}+2 Y_{24,8,3}-2 Y_{24,16,3}-2 Y_{24,20,3}\right]
\nonumber\\ && 
+Y_{12,11,3} \left[5+\frac{1}{4} \left(-Y_{12,4,3}+Y_{12,8,3}-Y_{24,4,3}- Y_{24,8,3}+ Y_{24,16,3}
\right. \right.
\nonumber\\ && \left. \left. 
+ 
Y_{24,20,3}\right)\right]
+\frac{9}{4} Y_{24,20,3}
+2 Y_{4,3,3} Y_{24,20,3}
-\frac{1}{4} \left[
    Y_{12,1,3} 
+ 2 Y_{12,3,3} 
+   Y_{12,5,3} \right.  
\nonumber\\ && \left.
-   Y_{12,7,3} 
- 2 Y_{12,9,3} 
\right] Y_{24,20,3}
+Y_{12,10,3} \left[-\frac{35}{4}
+\frac{27}{14} \left(T_{2,1,3,1,12}
                    +T_{2,1,3,5,12} \right. \right.
\nonumber\\ && \left. \left. 
                    -T_{2,1,3,7,12} 
                    -T_{2,1,3,11,12}
                    -T_{2,2,3,1,12}
                    -T_{2,2,3,5,12}
                    +T_{2,2,3,7,12}
                    +T_{2,2,3,11,12}\right)
\right.
\nonumber\\ && \left. 
+\frac{13}{14} \left(-  T_{6,1,3,1,4}
             +  T_{6,1,3,3,4}
             -  T_{6,2,3,3,4}
             +2 T_{6,3,3,1,4}
             -2 T_{6,3,3,3,4}
             +  T_{6,4,3,1,4}
\right. \right. \nonumber\\ && \left. \left.
             -  T_{6,4,3,3,4}
             -  T_{6,5,3,1,4}
             +  T_{6,5,3,3,4}
             -2 T_{6,6,3,1,4}
             +2 T_{6,6,3,3,4}\right)
+8 Y_{4,1,3}-8 Y_{4,3,3} \right.
\nonumber\\ && \left. 
-2 Y_{8,2,3}+2 Y_{8,6,3}
+\frac{1}{4} \left(
-Y_{12,1,3}- 2Y_{12,3,3}- Y_{12,5,3}+Y_{12,7,3}+2 Y_{12,9,3}
\right. \right.
\nonumber\\ && \left. \left. 
+ Y_{12,11,3}\right)
+\frac{3}{8} \left(Y_{24,2,3}+2 Y_{24,6,3}+ Y_{24,10,3}- Y_{24,14,3}-2 Y_{24,18,3}- Y_{24,22,3}\right)\right]
\nonumber\\ && 
+Y_{12,4,3} \left[\frac{35}{4}+\frac{27}{14}\left( -T_{2,1,3,1,12}- T_{2,1,3,5,12}+
T_{2,1,3,7,12}+T_{2,1,3,11,12}
\right. \right. \nonumber\\ && \left. \left. 
+ T_{2,2,3,1,12}+ T_{2,2,3,5,12}- T_{2,2,3,7,12}- T_{2,2,3,11,12}\right)
+\frac{13}{14} \left( T_{6,1,3,1,4}
\right. \right. 
\nonumber\\ && 
\left. \left.  
-T_{6,1,3,3,4}+ T_{6,2,3,3,4}-2 T_{6,3,3,1,4}+2 T_{6,3,3,3,4}- 
T_{6,4,3,1,4}+ 
T_{6,4,3,3,4}
\right. \right.
\nonumber\\ && \left. \left. 
+ T_{6,5,3,1,4}- T_{6,5,3,3,4}+2 T_{6,6,3,1,4}-2 T_{6,6,3,3,4}\right)
-8 Y_{4,1,3}+8 Y_{4,3,3}+2 Y_{8,2,3}
\right. \nonumber\\ && \left. 
-2 Y_{8,6,3}
+\frac{1}{4} \left(Y_{12,1,3}+2 Y_{12,3,3}+ Y_{12,5,3}- Y_{12,7,3}-2 Y_{12,9,3}\right)
+\frac{3}{8} \left(-Y_{24,2,3} \right. \right.
\nonumber\\ && \left. \left. 
-2 Y_{24,6,3} - Y_{24,10,3}+ Y_{24,14,3}+2 Y_{24,18,3}+ Y_{24,22,3}\right)\right]
+Y_{12,2,3} \left[\frac{35}{4} \right.
\nonumber\\ && \left. 
+\frac{27}{14} \left(-T_{2,1,3,1,12}- T_{2,1,3,5,12}+ 
T_{2,1,3,7,12}+ T_{2,1,3,11,12}+ T_{2,2,3,1,12} \right. \right.
\nonumber\\ && \left. \left. 
+ T_{2,2,3,5,12}- T_{2,2,3,7,12}- T_{2,2,3,11,12}\right)
+\frac{13}{14}\left(T_{6,1,3,1,4}-T_{6,1,3,3,4}-T_{6,2,3,1,4} \right. \right.
\nonumber\\ && \left. \left. 
+ T_{6,2,3,3,4}-2 T_{6,3,3,1,4}+2 T_{6,3,3,3,4}- 
T_{6,4,3,1,4}+ T_{6,4,3,3,4}+ T_{6,5,3,1,4}
\right. \right.
\nonumber\\ && \left. \left. 
- T_{6,5,3,3,4}+2 T_{6,6,3,1,4}-2 T_{6,6,3,3,4}\right)
-8 Y_{4,1,3}+8 Y_{4,3,3}+2 Y_{8,2,3} \right.
\nonumber\\ && \left. 
-2 Y_{8,6,3}
+\frac{1}{4} \left(Y_{12,1,3}+2 Y_{12,3,3}+ Y_{12,5,3}- Y_{12,7,3}-2 Y_{12,9,3}- Y_{12,11,3}\right)
\right.
\nonumber\\ && \left. 
+\frac{3}{8} \left( 
-Y_{24,2,3}-2 Y_{24,6,3}- Y_{24,10,3}+ Y_{24,14,3}+2 Y_{24,18,3} + Y_{24,22,3}\right)\right]
\nonumber\\ && \left.
-\frac{3}{2} Y_{24,22,3}-\frac{3}{8} Y_{12,8,3} Y_{24,22,3}\Biggr\}\right|_{q \rightarrow -q}.
\label{eq:sqrt}
\end{eqnarray}

The polynomials $(x+1)(3 x^2+1)$ and  $(x-1)^2 (3x+1)$ can be assembled using (\ref{eq:xELLP}).

Let us  also list the $q$-series for Jacobi's  $\eta$-function, cf. also~\cite{OEIS} {\tt A000203} 
\begin{eqnarray}
\eta = - \frac{1}{12 \omega}\left[-1 + 24 \sum_{k=1}^\infty k \Li_0\left(q^{2k}\right)\right]
= 
-\frac{1}{12 \omega} \frac{\vartheta_1^{(3)}(q)}{\vartheta_1^{(1)}(q)},
\end{eqnarray}
which is related to
\begin{eqnarray}
\frac{\vartheta_1^{(3)}(q)}{\vartheta_1^{(1)}(q)} &=& 
= -1 + 12\left[ \ELI_{0;-1}(1;1;q) + \ELI_{0;-1}(-1;1;q)\right].
\end{eqnarray}
\subsubsection{\boldmath The Inhomogeneity}
\label{sec-523}

\vspace*{1mm}
\noindent
The integral over the inhomogeneity (\ref{eq:PS2aA}) in the case of the homogeneous solutions
$\psi_{3,4}(x)$ has the following structure
\begin{eqnarray}
\label{eq:INHOM34}
I = \sum_{m=1}^8 c_m \int \frac{dx}{x} H_0^n(x) \hat{f}_m(x) \psi_{3,4}(x),~~~n \in \{0,1,2,3\},~~
c_m \in \mathbb{Q}
\end{eqnarray}
and 
\begin{eqnarray}
\hat{f}_m \in \left\{
\frac{1}{1 \pm x}, 
\frac{1}{(1 \pm x)^2}, 
\frac{1}{1 \pm 3x},
\frac{1}{(1 \pm 3x)^2}\right\}.
\end{eqnarray}
For the functions $f_a^k$
\renewcommand{\arraystretch}{2}
\begin{eqnarray}
f_a^k(x)    &=&\displaystyle \frac{1}{(1- ax)^k},~~~k \geq 1,~~k,a \in \mathbb{N}, 
\label{mir:1}
\end{eqnarray}
\renewcommand{\arraystretch}{1.}

\noindent
the structure of $x$,~Eq.~~(\ref{eq:cubic2}), leads to the following symmetry
\renewcommand{\arraystretch}{2}
\begin{eqnarray}
f_a^k(-x)    &=& \left.f_a^k(x)\right|_{q \rightarrow -q}. 
\label{mir:2}
\end{eqnarray}
\renewcommand{\arraystretch}{1.}

\noindent
For convenience we introduce the variable $\xi$,
\begin{eqnarray}
\xi =  \frac{1}{x} = 3 y,~~~~~~\xi \in ]0,1] \leftrightarrow q \in [0,1].
\end{eqnarray}
Under this change of variables the harmonic polylogarithms $H_{\vec{a}}(x)$ can be transformed using
the command {\tt TransformH} of the package 
{\tt HarmonicSums}~\cite{HARMONICSUMS,Ablinger:PhDThesis,Ablinger:2011te,Ablinger:2013cf,Ablinger:2014bra}.

One obtains the following $\eta$-ratios, cf.~{\tt  A187100, A187153} \cite{OEIS} 
\begin{eqnarray}
\frac{1}{1-x}    &=& -\frac{\xi}{1-\xi} = -3 \frac{\eta^2(\tau) \eta\left(\tfrac{3}{2}\tau\right) 
\eta^3(6\tau)}{\eta^3\left(\tfrac{1}{2}\tau\right) \eta(2\tau) \eta^2(3\tau)}
\\
\frac{1}{1 - 3x} &=& -\frac{\xi}{3 - \xi} = -\frac{\left[\eta(\tau) \eta\left(\tfrac{3}{2}\tau\right) 
\eta^2(6\tau)\right]^3}{\eta\left(\tfrac{1}{2}\tau\right) \eta^2(2\tau) \eta^9(3\tau)},
\end{eqnarray}
for which we get the representation in terms of an $\eta$-factor and elliptic polylogarithms using the
relations to Lambert--series given in Section~\ref{sec:6.1.4}.
\begin{eqnarray}
\label{eq:den1}
\frac{1}{1-x}    &=& 
\frac{1}{\eta^{12}(\tau)} \Biggl\{
-\frac{637}{51840} \left[Y_{2,1,6} - Y_{2,2,6}\right]
-\frac{49}{46080} [Y_{3,1,6} + Y_{3,2,6}] 
+\frac{49}{23040} Y_{3,3,6}
+\frac{91}{360} Y_{4,1,3}
\nonumber\\ &&
-\frac{721}{1620} Y^2_{4,1,3}
+\frac{721}{103680} Y_{4,2,6}
-\frac{91}{360} Y_{4,3,3}
+\frac{721}{810} Y_{4,1,3} Y_{4,3,3}
-\frac{721}{1620} Y^2_{4,3,3}
-\frac{721}{103680} Y_{4,4,6}
\nonumber\\ &&
-\frac{7}{414720} \left[Y_{6,1,6} - Y_{6,2,6} - 2 Y_{6,3,6} - Y_{6,4,6} + Y_{6,5,6} + 2 Y_{6,6,6}\right]     
-\frac{119}{144} Y_{12,1,3}
\nonumber\\ &&
-\frac{7}{1620} \left[Y_{4,1,3} - Y_{4,3,3}\right] Y_{12,1,3}
+\frac{383}{6480} Y^2_{12,1,3}
-\frac{7}{16} Y_{12,2,3}
+\frac{7}{128} Y^2_{12,2,3}
+\frac{67}{51840} Y_{12,2,6}
\nonumber\\ &&
-\frac{119}{72} Y_{12,3,3}
-\frac{7}{810} \left[Y_{4,1,3} - Y_{4,3,3} \right] Y_{12,3,3}
+\frac{383}{1620} Y_{12,1,3} Y_{12,3,3} 
+\frac{383}{1620} Y^2_{12,3,3} 
-\frac{7}{16} Y_{12,4,3}
\nonumber\\ &&
+\frac{7}{64} Y_{12,2,3} Y_{12,4,3}
+\frac{7}{128} Y^2_{12,4,3}
-\frac{67}{51840} Y_{12,4,6}
-\frac{119}{144} Y_{12,5,3}
-\frac{7}{1620}\left[Y_{4,1,3} - Y_{4,3,3} \right] Y_{12,5,3}
\nonumber\\ &&
+\frac{383}{3240} \left[Y_{12,1,3} + 2Y_{12,3,3} -Y_{12,7,3}\right] Y_{12,5,3}
+\frac{383}{6480} Y^2_{12,5,3}
-\frac{67}{25920} Y_{12,6,6}
\nonumber\\ &&
+\frac{119}{144} Y_{12,7,3}
+\frac{7}{1620}\left[Y_{4,1,3}-Y_{4,3,3}\right] Y_{12,7,3}
-\frac{383}{3240} \left[Y_{12,1,3} 
+ 2 Y_{12,3,3}\right] Y_{12,7,3}
\nonumber\\ && 
+\frac{383}{6480} Y^2_{12,7,3}
+\frac{7}{16} Y_{12,8,3}
-\frac{7}{64} \left[Y_{12,2,3} + Y_{12,4,3}\right] Y_{12,8,3}
+\frac{7}{128} Y^2_{12,8,3}
\nonumber\\ &&
-\frac{67}{51840} Y_{12,8,6}
+\frac{119}{72} Y_{12,9,3}
+\frac{7}{810} \left[Y_{4,1,3}-Y_{4,1,3}\right] Y_{12,9,3}
-\frac{383}{1620}\left[ Y_{12,1,3} + 2 Y_{12,3,3} 
\right.
\nonumber\\ && \left.
+Y_{12,5,3} - Y_{12,7,3} \right] 
Y_{12,9,3} 
+\frac{383}{1620} Y^2_{12,9,3}
+\frac{7}{16} Y_{12,10,3}
-\frac{7}{64} \left[Y_{12,2,3}  
+ Y_{12,4,3} 
\right. 
\nonumber\\ && \left.
- Y_{12,8,3} \right] Y_{12,10,3}
+\frac{7}{128} Y^2_{12,10,3}
+\frac{67}{51840} Y_{12,10,6}
+\frac{119}{144} Y_{12,11,3}
+\frac{7}{1620} \left[
Y_{4,1,3} -Y_{4,3,3} \right] 
\nonumber\\ &&
\times Y_{12,11,3} 
-\frac{383}{3240} \left[ Y_{12,1,3} + 2 Y_{12,3,3} + Y_{12,5,3} - Y_{12,7,3} - 2 Y_{12,9,3}\right]  Y_{12,11,3}
\nonumber\\ &&
+\frac{383}{6480} Y^2_{12,11,3}
+\frac{67}{25920} Y_{12,12,6}
-\frac{259}{18} Z_{3,1,4}
+\frac{37}{18} [Y_{12,1,3} + 2Y_{12,3,3} +Y_{12,5,3} -Y_{12,7,3} 
\nonumber\\&& 
-2Y_{12,9,3}-Y_{12,11,3}] Z_{3,1,4}
+\frac{63}{4} Z_{3,1,12}
-\frac{9}{4} \left[Y_{12,1,3} + 2 Y_{12,3,3} +  Y_{12,5,3} -Y_{12,7,3} 
-2Y_{12,9,3} \right.
\nonumber\\ && 
\left.
-Y_{12,11,3}\right] Z_{3,1,12}
+\frac{117}{5} Z^2_{3,1,12}
+\frac{259}{18} Z_{3,3,4}
-\frac{37}{18} \left[
Y_{12,1,3} + 2Y_{12,3,3} + Y_{12,5,3}
\right.
\nonumber\\ &&  \left.
-Y_{12,7,3} - 2Y_{12,9,3} - Y_{12,11,3}\right] Z_{3,3,4}
+\frac{63}{4} Z_{3,5,12}
-\frac{9}{4} \left[Y_{12,1,3} + 2Y_{12,3,3} + Y_{12,5,3}
\right.
\nonumber\\ &&
\left. -Y_{12,7,3} - 2Y_{12,9,3} - Y_{12,11,3}\right] Z_{3,5,12}
+\frac{234}{5} Z_{3,1,12} Z_{3,5,12}
+\frac{117}{5} Z^2_{3,5,12}
-\frac{63}{4} Z_{3,7,12}
\nonumber\\ &&
+\frac{9}{4} \left[Y_{12,1,3} + 2Y_{12,3,3} + Y_{12,5,3}
-Y_{12,7,3} - 2Y_{12,9,3} - Y_{12,11,3}\right]
Z_{3,7,12}
\nonumber\\ && 
-\frac{234}{5} \left[Z_{3,1,12} + Z_{3,5,12}\right] Z_{3,7,12}
+\frac{117}{5} Z^2_{3,7,12}
-\frac{63}{4} Z_{3,11,12}
\nonumber\\ &&
+\frac{9}{4} \left[
Y_{12,1,3} 
+ 2Y_{12,3,3} 
+ Y_{12,5,3} 
- Y_{12,7,3} 
- 2Y_{12,9,3} 
- Y_{12,11,3} \right] Z_{3,11,12}
\nonumber\\ &&
-\frac{234}{5} \left[
 Z_{3,1,12} 
+Z_{3,5,12} 
-Z_{3,7,12}\right]  Z_{3,11,12}
+\frac{117}{5} Z^2_{3,11,12}
-\frac{4459}{5760} Z_{6,1,1}
\nonumber\\ &&
+ \frac{7}{810} Y_{12,9,3} \left[Y_{4,1,3} - Y_{4,3,3}\right]
\Biggr\},
\\
\label{eq:den2}
\frac{1}{1 - 3x} &=&
\frac{1}{\eta^{12}(\tau)} \Biggl\{
-\frac{2071}{1170} \left[Y_{4,1,3}^2
- 2 Y_{4,3,3} Y_{4,1,3} \right]
+\frac{41}{8190}\left[
 Y_{12,1,3} 
+2Y_{12,3,3} 
+Y_{12,5,3} 
-Y_{12,7,3}
\right. 
\nonumber\\ &&
\left. 
-2Y_{12,9,3} 
-Y_{12,11,3}\right] Y_{4,1,3} 
+\frac{17}{20} Y_{4,1,3}
-\frac{2071}{1170} Y^2_{4,3,3}
+\frac{5651}{32760} Y_{12,1,3}^2
+\frac{9}{64} \left[Y^2_{12,2,3} 
\right. \nonumber\\ && \left.
+ Y^2_{12,4,3}
+ Y^2_{12,8,3}
+ Y^2_{12,10,3}
\right]
+\frac{5651}{32760}\left[ 
  4Y^2_{12,3,3} 
+  Y^2_{12,5,3}
+  Y^2_{12,7,3}
+ Y^2_{12,11,3}
\right]
\nonumber\\ && 
+\frac{162}{5} \left[
   Z_{3,1,12}^2
+   Z_{3,5,12}^2
+   Z_{3,7,12}^2
+   Z_{3,11,12}^2\right]
-\frac{49}{2880}    \left[Y_{2,1,6}-Y_{2,2,6}\right] 
\nonumber\\ && 
-\frac{49}{33280}   \left[Y_{3,1,6}+Y_{3,2,6}-2Y_{3,3,6}\right]
+\frac{2071}{74880} \left[Y_{4,2,6}-Y_{4,4,6}\right]
-\frac{17}{20}  Y_{4,3,3}
\nonumber\\ && 
+\frac{5711}{2096640}\left[Y_{6,1,6} - Y_{6,2,6} - 2Y_{6,3,6}-Y_{6,4,6}+Y_{6,5,6}+2Y_{6,6,6}\right]
-\frac{41}{8190} Y_{4,3,3} \left[Y_{12,1,3} \right.
\nonumber\\ && 
\left. - Y_{12,11,3}\right]
-\frac{1759}{728} Y_{12,1,3}
-\frac{9}{8} Y_{12,2,3}
+\frac{269}{74880} \left[Y_{12,2,6} - Y_{12,4,6} \right]
-\frac{41}{4095} Y_{4,3,3} Y_{12,3,3}
\nonumber\\ && 
+\frac{5651}{8190} Y_{12,1,3} Y_{12,3,3}
-\frac{1759}{364} Y_{12,3,3}
+\frac{9}{32} Y_{12,2,3} Y_{12,4,3}
-\frac{9}{8} Y_{12,4,3}
-\frac{41}{8190} Y_{4,3,3}Y_{12,5,3}
\nonumber\\ && 
+\frac{5651}{16380} \left[Y_{12,1,3}+2Y_{12,3,3}  \right]Y_{12,5,3}
-\frac{1759}{728} Y_{12,5,3}
-\frac{269}{37440} Y_{12,6,6}
+\frac{41}{8190} Y_{4,3,3} Y_{12,7,3}
\nonumber\\ && 
-\frac{5651}{16380} \left[ Y_{12,1,3} + 2 Y_{12,3,3} + Y_{12,5,3}\right] Y_{12,7,3}
+\frac{1759}{728} Y_{12,7,3}
-\frac{9}{32} \left[Y_{12,2,3} +Y_{12,4,3} - 4\right]
\nonumber\\ && 
\times Y_{12,8,3}
-\frac{269}{74880} Y_{12,8,6}
+\frac{41}{4095} Y_{4,3,3} Y_{12,9,3}
-\frac{5651}{8190}\left[Y_{12,1,3}  + 2 Y_{12,3,3} + Y_{12,5,3} 
\right. \nonumber\\ && \left. 
- Y_{12,7,3} - Y_{12,9,3}  \right] Y_{12,9,3}
+\frac{1759}{364} Y_{12,9,3}
-\frac{9}{32} \left[  Y_{12,2,3} + Y_{12,4,3} - Y_{12,8,3}\right] Y_{12,10,3}
\nonumber\\ && 
+\frac{9}{8} Y_{12,10,3}
+\frac{269}{74880} Y_{12,10,6}
-\frac{5651}{16380} \left[Y_{12,1,3} + 2 Y_{12,3,3} + Y_{12,5,3} 
\right. \nonumber\\ && \left. 
- Y_{12,7,3} - 2 Y_{12,9,3}  \right] Y_{12,11,3}
+\frac{1759}{728} Y_{12,11,3}
+\frac{269}{37440} Y_{12,12,6}
+6 \left[
 Y_{12,1,3} + 2 Y_{12,3,3} 
\right. \nonumber\\ && \left.  
+  Y_{12,5,3}
-Y_{12,7,3} - 2 Y_{12,9,3} -  Y_{12,11,3} \right] Z_{3,1,4} 
-42 Z_{3,1,4}
-\frac{1161}{182} \left[
 Y_{12,1,3} +2 Y_{12,3,3} 
\right. \nonumber\\ && \left. 
+ Y_{12,5,3}
-Y_{12,7,3} -2 Y_{12,9,3} - Y_{12,11,3} \right] Z_{3,1,12}
+\frac{1161}{26} Z_{3,1,12}
-6 \left[
 Y_{12,1,3} +2 Y_{12,3,3} 
\right. \nonumber\\ && \left. 
+ Y_{12,5,3}
-Y_{12,7,3} -2 Y_{12,9,3} - Y_{12,11,3} \right] Z_{3,3,4}
+42 Z_{3,3,4}
-\frac{1161}{182} \left[
 Y_{12,1,3} +2 Y_{12,3,3} 
\right. \nonumber\\ && \left. 
+ Y_{12,5,3}
-Y_{12,7,3} -2 Y_{12,9,3} - Y_{12,11,3} \right] Z_{3,5,12}
+\frac{324}{5} Z_{3,1,12} Z_{3,5,12}
+\frac{1161}{26} Z_{3,5,12}
\nonumber\\ && 
+\frac{1161}{182} \left[
  Y_{12,1,3} + 2 Y_{12,3,3} +  Y_{12,5,3}
 -Y_{12,7,3} - 2 Y_{12,9,3} -  Y_{12,11,3}\right] Z_{3,7,12}
\nonumber\\ && 
-\frac{324}{5} \left[Z_{3,1,12} + Z_{3,5,12} \right] Z_{3,7,12}
-\frac{1161}{26} Z_{3,7,12}
+\frac{1161}{182}\left[
  Y_{12,1,3} + 2 Y_{12,3,3} +  Y_{12,5,3}
\right. \nonumber\\ && \left. 
 -Y_{12,7,3} - 2 Y_{12,9,3} -  Y_{12,11,3}\right] Z_{3,11,12}
-\frac{324}{5} \left[Z_{3,1,12} + Z_{3,5,12} - Z_{3,7,12} \right] Z_{3,11,12}
\nonumber\\ && -\frac{1161}{26}Z_{3,11,12}
-\frac{343}{320} Z_{6,1,1}
\Biggr\}.
\end{eqnarray}
For both (\ref{eq:den1}) and  (\ref{eq:den2}) 38 Lambert--series of the kind (\ref{eq:Y}, \ref{eq:Z}) 
contribute in our present basis representation. If expanded in $\Li_0(q^n)$ and the elliptic polylogarithms,
many more functions would appear. The expressions (\ref{eq:sqrt}, \ref{eq:den1}) and (\ref{eq:den2}) 
are rather large. Due to a large number of relations between modular forms we can currently not exclude that 
these expressions can be simplified. We leave this for a later study. Here our first goal has been to find  
valid representations algorithmically in all cases. 

Let us now turn to the harmonic polylogarithms appearing in the inhomogeneities.
We first change the measure for integral (\ref{eq:INHOM34}) to
\begin{eqnarray}
\frac{dx}{x} = \frac{dq}{q} J(q),~~~~~~\text{with}~~~J = \frac{d \ln(x)}{d \ln(q)}.
\end{eqnarray}
The Jacobian $J(q)$ is given by
\begin{eqnarray}
\label{eq:J0}
\frac{d \ln(x)}{d \ln(q)} =  -1 + \overline{E}_{0;-1}(\rho_3;i;q) + \overline{E}_{0;-1}(\rho_3;-i;q).
\end{eqnarray}
This is easy to see, since the relation
\begin{eqnarray}
\ln\left[\eta^b\left(a\tau\right)\right] = b \left[ \frac{a}{12} \ln(q) - \frac{1}{2a} 
\sum_{m=1}^{2a} \ELI_{0;-1;2}(1;\rho_{2a}^m;q)\right],~~~a,b \in \mathbb{N} \backslash \{0\}
\label{eq:LOGeta}
\end{eqnarray}
holds, which can be generalized to any $\eta$-ratio.

Integrating (\ref{eq:J0}) one obtains
\begin{eqnarray}
H_0(x) = - \ln\left(3q\right) + \overline{E}_{0;-1;2}(\rho_3;i;q) +   
           \overline{E}_{0;-1;2}(\rho_3;-i;q).
\end{eqnarray}
Since also other harmonic polylogarithms may occur in the inhomogeneities, let us briefly discuss the
next possible cases.

Similar to (\ref{eq:J0}), one has
\begin{eqnarray}
\label{eq:dlnomxdq}
\frac{d\ln(1+x)}{d\ln(q)} &=& 4\left[\overline{E}_{0; -1}(-1; 1; q^2) 
- \overline{E}_{0; -1}(\rho_6; 1; q^2)\right] - 
\left[\overline{E}_{0; -1}(-1; 1; q) - \overline{E}_{0;-1}(\rho_6;1; q)\right]
\nonumber\\ &&
- 1 + 4 \overline{E}_{0; -1}(\rho_3;-1;q^2)
\\
&=& -1 + \overline{E}_{0;-1}(-1;-1,q) 
- \overline{E}_{0;-1}(\rho_6;-1;q)
+ \overline{E}_{0;-1}(\rho_3;-i;q)
+ \overline{E}_{0;-1}(\rho_3;i;q)
\nonumber\\
\end{eqnarray}
and
\begin{eqnarray}
\label{eq:mir5}
\frac{d\ln(1-x)}{d\ln(q)} &=& \left.
\frac{d\ln(1+x)}{d\ln(q)}\right|_{q \rightarrow -q}. 
\end{eqnarray}
$H_{-1}(x)$ and $H_1(x)$ are obtained by integrating (\ref{eq:dlnomxdq}) and the relation (\ref{eq:mir5}) 
\begin{eqnarray}
H_{-1}(x) =  \ln(1+x) &=& -\ln(3q) -
  \overline{E}_{0;-1;2}(-1; -1; q) 
+ \overline{E}_{0;-1;2}(\rho_6; -1; q) 
\nonumber\\
&&
- \overline{E}_{0;-1;2}(\rho_3; -i; q)
- \overline{E}_{0;-1;2}(\rho_3; i; q)
\\ 
H_{1}(x) &= &  -\left.H_{-1}(x)\right|_{q \rightarrow -q} + 2\pi i, 
\end{eqnarray}
with
\begin{align}
&H_0(\xi) =  -H_0(x),~~~~~&H_1(\xi) = H_1(x) + H_0(x),~~~~~~~&H_{-1}(\xi) = H_{-1}(x) + H_0(x).
\end{align}
There are similar symmetry relations at higher weight. One also applies the shuffle algebra 
\cite{SHUF,Blumlein:2003gb} and it is therefore sufficient to calculate the $q$-representations for
$H_{0,1}, H_{1,-1}, H_{0,0,1}, H_{0,1,1}, H_{0,1,-1}$ and $H_{1,1,-1}$ up to weight {\sf w = 3}. 

In (\ref{eq:INHOM34}) we first transform to $\xi$ as the integration variable through which 
the HPLs $H_{\vec{a}}(x)$ are replaced by
\begin{eqnarray}
H_{\vec{a}}(x) = \sum_n a_n H_{\vec{b}_n}(\xi) + c_{\vec{a}},~~~a_n, c_{\vec{a}} \in \mathbb{C}.
\end{eqnarray}
By iteration, the different harmonic polylogarithms (\ref{eq:HPL}) are obtained as follows: 
\begin{eqnarray}
H_{0,\vec{a}}(\xi) &=& \int_0^\xi \frac{d\bar{\xi}}{\bar{\xi}} H_{\vec{a}}(\xi) 
                    =  \int_0^q \frac{d\bar{q}}
{\bar{q}} \frac{d\ln(\bar{\xi})}{d\ln(q)} H_{\vec{a}}(\bar{\xi}(\bar{q})).
\\
H_{1,\vec{a}}(\xi) &=&  -\int_0^q \frac{d\bar{q}}{\bar{q}} 
\frac{d\ln(1-\bar{\xi})}{d\ln(\bar{q})}
\overline{H}_{\vec{a}}(\bar{q}) 
\\
H_{-1,\vec{a}}(\xi) &=& \int_0^q \frac{d\bar{q}}{\bar{q}} 
\frac{d\ln(1+\bar{\xi})}{d\ln(\bar{q})}
\overline{H}_{\vec{a}}(\bar{q}),
\end{eqnarray}
with $\overline{H}_{\vec{a}}(\bar{q}) = H_{\vec{a}}(\bar{\xi}(\bar{q})$ and
\begin{eqnarray}
\frac{d\ln(\bar{\xi})}{d\ln(q)} &=& - \frac{d\ln(x)}{d\ln(q)}
\\
\frac{d\ln(1-\bar{\xi})}{d\ln(q)} &=& 
\frac{d\ln(1-x)}{d\ln(q)} - \frac{d\ln(x)}{d\ln(q)}
\\
\frac{d\ln(1+\bar{\xi})}{d\ln(q)} &=& 
\frac{d\ln(1+x)}{d\ln(q)} - \frac{d\ln(x)}{d\ln(q)}.
\end{eqnarray}

To express the solution $f_{9b}(x)$ one needs to differentiate $f_{8b}(x)$,
\begin{eqnarray}
\xi \frac{d}{d\xi} f(\xi) = \xi \frac{d}{d\xi} \int_0^q \frac{d\bar{q}}{\bar{q}} 
\frac{d\ln(\xi)}{d\ln(\bar{q})} \bar{f}(\bar{q}) = \bar{f}(q). 
\end{eqnarray}
For the solution of $f_{10b}(x)$, integrals of the type 
\begin{eqnarray}
\int_0^\xi \frac{d\bar{\xi}}{\bar{\xi}} [P(\bar{\xi}) f(\bar{\xi})] 
= \int_0^q \frac{d\bar{q}}{\bar{q}} 
\frac{d \ln(\xi)}
{d \ln(\bar{q})} 
[\overline{P} 
\cdot f](\bar{q})
\label{eq:INT10b}
\end{eqnarray}
are performed. Here, the integrand of (\ref{eq:INT10b}) has to be expressed in terms of
$q$. 

In the case of (\ref{eq:INHOM}), integrals of the kind
\begin{eqnarray}
\int dq \frac{q^m \ln^n(q)}{\eta^k(\tau(q))} 
\end{eqnarray}
contribute.  For $k = 0$ these integrals are given by polynomials of $q$ and $\ln(q)$ and 
the integration relations of the type  Eq.~(\ref{eq:ELP5}) can be used.

Because of
\begin{eqnarray}
\frac{1}{\eta(\tau)} = 
\frac{1}{q^{\frac{1}{12}}} 
\prod_{k=1}^\infty \frac{1}{1-q^{2k}} 
=  
\frac{1}{q^{\frac{1}{12}}} 
\Biggl(
1+ \sum_{k=1}^\infty \frac{q^{2k}}{\prod_{l=1}^k (1-q^{2l})} \Biggr)
= 
\frac{1}{q^{\frac{1}{12}}} 
\sum_{k=0}^\infty p(k) q^{2k}, 
\end{eqnarray} 
cf.~\cite{ERDELYI1,MART1}, $q$-Pochhammer symbols are appearing, which requires a corresponding generalization 
of the integration relation w.r.t. $q$. Here $p(k)$ denotes the partition function. There is no (known) finite 
rational closed form expression  for $p(n)$ \cite{MART1}, cf. also~\cite{PART}. 

In Ref.~\cite{Adams:2016xah}, Eqs.~(50, 69), only harmonic polylogarithms over the alphabet 
$\{0,1\}$ occurred, which all could be expressed in terms of elliptic polylogarithms. However, 
the kinematic variable in \cite{Adams:2016xah} is different from that in the present case. This
implies different representations for the harmonic polylogarithms in terms of $q$-series.

We finally remark that there is a multitude of equivalent representations of the $q$-series of a modular 
form, which obey many relations.\footnote{M.~Eichler stated~\cite{FERMAT} that 
there are five basic mathematical operations: addition, subtraction, multiplication, division and modular forms.} 
It would be worthwhile to find minimal representations. One criterion could be to 
minimize the number of elementary elliptic polylogarithms (\ref{eq:ELP1}) contributing. Still one would have to decide 
whether in this representation different arguments are synchronized or not, bearing in mind that the latter step is 
straightforward and only needed if the corresponding expression shall be integrated over $q$.
\section{Possible Extensions}
\label{sec:6}

\vspace*{1mm}
\noindent
In Section~\ref{sec:4} we have obtained a representation of new iterative integrals containing 
also letters which are impossible to be rewritten as integrals such that the next 
integration variable does only appear in one boundary of this integral. In the present study 
only the complete elliptic integrals were forming the new letters of this kind. Due to this, it 
is possible to express the corresponding integrands in terms of $\eta$-weighted Lambert-Eisenstein series,
given the type of inhomogeneities are of the class as in the present examples. For other irreducible 
differential equations of order {\sf o = 2} it may happen that we end up with $_2F_1$ solutions which 
cannot be reduced to complete elliptic integrals modulo some (ir)rational pre-factor.

In more general cases the $_2F_1$ solutions will not appear but other higher transcendental solutions
might be found, obeying higher order differential equations, which are the result of the corresponding
integration-by-parts reductions \cite{IBP}. They will usually have also definite integral representations
and appear as new letters other than the ones which we mentioned before. Whether or not a mathematical way 
exists to come up with an analogue to the case of the elliptic polylogarithm  will depend on the class 
of functions. In various cases the representation of Section~\ref{sec:4} will be the final one.

Still the case of the elliptic polylogarithm 
\begin{eqnarray}
\ELI_{n;m}(x;y;q) &=& \sum_{j=1}^\infty \sum_{k=1}^\infty \frac{x^j}{j^n} \frac{y^k}{k^m} q^{jk}
\nonumber
\end{eqnarray}
may get some generalizations in the case of Feynman integral calculations, as has been the case
before for the polylogarithms. The two summand terms 
\begin{eqnarray}
\frac{x^j}{j^n} 
\end{eqnarray}
appearing are those of the generalized harmonic sums, i.e. the Mellin transforms of Kummer-iterative integrals 
\cite{KUMMER1}, connected with the nome term $q^{jk}$. One may think of a cyclotomic extension 
\cite{Ablinger:2011te,Ablinger:2013cf} in the sense of real valued representations, where the two infinite
sums allow for periodic gaps choosing summands of the kind
\begin{eqnarray}
\label{eq:ext1}
\frac{x^{l_3}}{(l_1 j +l_2)^{l_3}},~~~l_i \in \mathbb{N}, l_1,l_3 > 0.
\end{eqnarray}
Further extensions, which occurred in the non elliptic case, may be binomially weighted  sums, 
cf.~\cite{Ablinger:2014bra}. Here, additional factors of the kind 
\begin{eqnarray}
\label{eq:ext2}
\frac{1}{(j+r)^l \displaystyle \binom{2j}{j}},~~~~~~~~~\frac{1}{(j+r)^l} \binom{2j}{j},~~~~2r,l \in \mathbb{N}
\end{eqnarray}
may occur in (\ref{eq:ext1}).

A reliable guide to find new structures consists in analyzing  the appearing integrals by applying dispersion 
relations \cite{DISPERSE,Laporta:2004rb}. The cuts immediately relate to a series of relevant Landau variables 
\cite{LANDAU} of the problem, which are usually only revealed at a much later stage using differential or 
difference equations directly to solve the same problem\footnote{While the dispersive technique can be applied 
to usual Feynman integral calculations directly, this is not the case for diagrams containing local operator 
insertions \cite{OP,OP1,Ablinger:2010ty,Ablinger:2014nga}. The latter short-distance representation would need to 
be re-derived after having performed the cut of the corresponding usual Feynman diagram.}. In higher order 
graphs one cannot exclude that hyperelliptic and Abel integrals \cite{ABEL1} are going to appear at some level, 
which are known to be multi-periodic compared to the double periodicity in the elliptic case; 
see Ref.~\cite{NEUMANN} for the corresponding theory. The corresponding integrals will require new classes 
of functions for the analytic representations.

We finally mention that in $\Phi^4$ theory at eight loops more complicated structures are occurring related to 
$K3$-surfaces\footnote{$K3$ stands for `Kummer, K\"ahler and Kodaira'. The term has been introduced by 
A.~Weil \cite{WEIL}.} \cite{BS}, compared to those implied by elliptic curves.
\section{Conclusions}
\label{sec:7}

\vspace*{1mm}
\noindent
A central problem in calculating higher loop Feynman integrals in renormalizable quantum field theories consists 
in solving the differential equations obtained from the IBPs, which rule the master integrals. In the present paper 
we have solved master integrals which correspond to irreducible differential equations of second order
with more than three singularities fully analytically. They appear in the calculation of the QCD corrections to 
the $\rho$-parameter at 3-loop order in \cite{Grigo:2012ji}. They form typical examples for structures which appear 
in solving IBP-relations for Feynman diagrams beyond the well understood case of singly factorizing integrals given 
as iterative integrals over a general alphabet. The latter case has been already algorithmized completely in 
Ref.~\cite{Ablinger:2015tua}, even not needing any special choice of basis. The second order structures can be 
mapped to $_2F_1$ solutions under conditions presented in this paper. We have outlined the algorithmic 
analytic solution in this case in terms of iterative integrals over partly non-iterative letters. Indeed this 
holds even for much more general solutions than those of the $_2F_1$ type. One is usually interested 
in representing the analytic solution for a certain interval of a (dimensionless) kinematic variable $x \in 
\mathbb{R}$, e.g. for $x \in [0,1]$. The solutions may have different singularities in this range, including
branch points. Yet piecewise analytic series expansions of the type
\begin{eqnarray}
\label{RES1}
\sum_{k = 0}^m \ln^k(x)\sum_{l=0}^\infty a_{k,l} (x-b)^l,~~~b \in [0,1], m \in \mathbb{N}
\end{eqnarray}
are possible, which overlap in finite regions allowing to obtain a very high accuracy by expanding to a sufficient finite 
degree. The simple form of (\ref{RES1}) is very appealing for many physics applications, despite the potentially 
involved structure described by the corresponding differential equations.

The question arises whether one may find a fully analytic diagonalization of the integral describing the solution
in the inhomogeneous case. If the $_2F_1$ solutions can be mapped to complete elliptic integrals using triangle 
group relations for the homogeneous solution and the inhomogeneity normalized by the Wronskian can be 
represented in terms of elliptic polylogarithms, the inhomogeneous solution is given in terms of elliptic 
polylogarithms of the nome $q$, solving the integral over the inhomogeneity. Also here, all necessary steps are 
known. The building blocks appearing in the present case are not all of this type, due to which modifications 
are necessary.

In the present case, the kinematic variable $x$ is determined from the rational function $k^2 = z(x)$  
appearing in the complete elliptic integral ${\bf K}(k^2)$. A related, but different approach has been followed 
in \cite{Adams:2014vja}. Our choice has the advantage to obey a mirror symmetry  for $x \leftrightarrow q$ by 
sign change in deriving the $q$-forms of the harmonic polylogarithms. The kinematic variable $x$ is obtained 
applying a cubic elliptic transformation. Next, one has to derive the elliptic integral representation of all 
factors appearing in the integrand of the inhomogeneous solution, and in some cases further integrals and 
derivations of the inhomogeneous solution in the $q$-representation. We map all building blocks to modular 
forms separating off a factor $1/\eta^k(\tau)$ if necessary, and obtain analytic solutions in terms of 
$\eta$-weighted  Lambert-Eisenstein series. As we have shown, in the present case the emergence of the elliptic
integral of the second kind, {\bf E}$(k^2)$, cannot be avoided in the solutions. This is one source of the 
$\eta$-factor mentioned. While the multiplication relation (\ref{eq:ELP4a}) 
allows to form the final elliptic polylogarithms in case of $\forall k = 0$, in general one obtains
$\eta$-weighted  elliptic polylogarithms. Because of the appearance of the  $q$-Pochhammer symbol in the denominator,
the occurring $q$-integrals are not of the class of the elliptic polylogarithms in general.

The main work went into the determination of elliptic polylogarithm representations of the $q$-series for 
the different building blocks. In the present case we had also to represent a square root term, which was 
possible using the structure of the rational function $z(x)$. In this way (functions of) Dedekind $\eta$-ratios 
are expanded into $q$-series trying to match them into linear combinations of elliptic polylogarithms. This is 
done for the most elementary factors, building the more complex ones using the relations (\ref{eq:ELP4a}). Here 
an essential issue is to prove the equality of two $q$-series, which can be done mapping to modular forms and
comparing a number of non-vanishing coefficients up to the predicted bound.

We have referred to a special choice for a basis in representing the occurring modular forms in ${\cal 
M}(k,N,x)$. In this way 
we were able to find the representation of every $\eta$-ratio for any modular form completely algorithmically. 
This has been our main goal here. As it is well-known, there is a very large number of relations between modular forms, 
which may be used to derive potentially shorter representations. One possible demand would be to find a minimal 
representation in terms of elliptic polylogarithms of e.g. the kind 
\begin{eqnarray}
\label{eq:ELLPm}
\ELI_{m;n}(x;y;q^j),
\ELI_{m;n}(x;q^k;q^j),~~~\text{and}~~~
\ELI_{m;n}(q^l;q^k;q^j),~~~m,n,k,j,l \in \mathbb{Z},~~~x,y \in \mathbb{C},
\end{eqnarray}
referring to the class of elliptic polylogarithms which appeared in the present paper. To synchronize the 
$q$-argument of the occurring elliptic polylogarithms is easily possible, but will usually lead to a 
proliferation of terms. 

We remark that the Mellin moments, in case of also elliptic contribution to the solutions contribute, cf.~\cite{OP},
map for fixed values of the Mellin variable $N$ to rational numbers and multiple zeta values. Large amounts of 
moments can be calculated using the algorithm of Ref.~\cite{Blumlein:2017dxp}, also providing a suitable method
to quantify the corresponding physical problem, cf. also \cite{Ablinger:2017tan}.

For higher topologies we envisage extensions to more general structures, as has been briefly discussed 
in Section~\ref{sec:6}. Structures of this kind are expected in solving differential equations of higher than 
second order, which may arise from Feynman diagrams, in the ongoing adventure to map out the mathematical beauty of the 
renormalizable quantum field theories of the microcosmos as initiated by Stueckelberg \cite{STUECKELBERG} and 
Feynman \cite{FEYNMAN}.

\vspace{5mm}
\noindent
{\bf Acknowledgment.}~We 
would like to thank A.~Behring, D.~Broadhurst, P.~Marquard, M.~Steinhauser, J.-A.~Weil and S.~Weinzierl for 
discussions, and G.~K\"ohler and P. Paule especially for mutual discussions on  $\eta$-functions and 
modular forms. This work was supported in part by the  European Commission through contract PITN-GA-2012-316704 
({HIGGSTOOLS}), by the Austrian Science Fund (FWF) grant SFB F50 (F5006-N15, F5009-N15),
by the strategic program ``Innovatives O{\"O} 2010 plus'' by the Upper Austrian Government in the frame of 
project W1214-N15-DK6 of the Austrian Science Fund (FWF), by the Austrian Science Fund FWF project no. P 27229,
and by NSF grant 1618657.

\newpage
\appendix
\section{\boldmath $_2F_1$ Solutions of Second Order Differential Equations with more than three
Singularities}
\label{sec:8}

\vspace*{1mm}
\noindent
In the following we describe an algorithm which allows to map an ordinary second order differential equation
into $_2F_1$ solutions. We are going to explain it referring to an extended example. For this reason we
consider the following homogeneous linear differential equation with rational function coefficients
\begin{align}
\label{diff_op}
	0 = &\, 256\,x \left( 3\,x+10 \right)  \left( 15\,x-4 \right)  \left( x+4\right) \frac{d^2}{dx^2}S(x) 
        \notag \\
	& + \left( 30240\,{x}^{3}+164160\,{x}^{2}+182784\,x-25600 \right)\frac{d}{dx}S(x) \notag \\
	& +4725\,{x}^{2}+17910\,x+6000 \equiv L[S(x)].
\end{align}
A hypergeometric solution of (\ref{diff_op}) is a closed form solution
\begin{align}\label{hyper_sol}
	& S(x)  = 
\exp{\left( \int r(x)dx \right)}
	\left( 
	r_0(x)\cdot \, _2F_1 \left[ \begin{matrix} a_1, a_2 \\ b_1 \end{matrix}; f(x) \right]
	+ r_1(x)\cdot \, {_2F_1}' \left[ \begin{matrix} a_1, a_2 \\ b_1 \end{matrix}; f(x) \right]
	\right),
\end{align}
where $r(x), r_0(x), r_1(x), f(x) \in \overline{\mathbb{Q}(x)}$ and $a_1,a_2,b_1 \in \mathbb{Q}$.
The algorithm in \cite{IVH} 
first tries to find solutions of (\ref{diff_op}) of the form:
\begin{equation}\label{special_hyper_sol}
	S(x)  = \exp{\left( \int r(x)dx \right)} 
	\cdot \, 
	_2F_1 \left[ \begin{matrix} a_1, a_2 \\ b_1 \end{matrix}; f(x) \right].
\end{equation}
If it finds no solutions of the form (\ref{special_hyper_sol}), 
then it tries to transform (\ref{diff_op}) to a simpler differential operator $\tilde{L}$ 
and tries to find solutions of $\tilde{L}$ of the type (\ref{special_hyper_sol}),
which then lead to solutions of (\ref{diff_op}) of type (\ref{hyper_sol}).

If (\ref{diff_op}) has solutions of type (\ref{special_hyper_sol}), 
then there exists a Gau{\ss} hypergeometric differential operator $L_B$ 
such that solutions of (\ref{diff_op}) can be obtained from solutions of $L_B$ 
via a \emph{change of variables} and an \emph{exp-product} transformations. 
This means that 
\begin{equation*}
	_2F_1 \left[ \begin{matrix} a_1, a_2 \\ b_1 \end{matrix}; x \right]
\end{equation*}
is a solution of $L_B$ and the change of variables $x \mapsto f(x)$ sends 
$L_B$ to an equation $L^f_B$ with 
a solution
\begin{equation*}
	_2F_1 \left[ \begin{matrix} a_1, a_2 \\ b_1 \end{matrix}; f(x) \right],
\end{equation*}
and the exp-product transformation sends $L^f_B$ to an 
equation with solutions (\ref{special_hyper_sol}).

The operator (\ref{diff_op}) has four non-removable singularities at $x = -4, -10/3, 0, 4/15$
and no removable singularities. 
The exponent-differences are $1/2, 1/4, 3/8$, and $1/4$ respectively. 
For example at $x=0$ there are formal solutions
(power series solutions)                                    
$x^0 \cdot  (1 + {\frac {15}{64}}x + \dots)$ 
and 
$x^{3/8} \cdot (1 + {\frac {249}{320}}x + \dots )$,
so the exponents are $0$ and $3/8$ and the exponent-difference is $3/8$.

Section 3.3 in \cite{IVH} gives relations between $\deg(f)$,
exponent-differences of $L_B$,
and exponent-differences of (\ref{diff_op}).
If $f(x)$ is a rational function, then sub-algorithm 3.2 in
Section 3.4 in \cite{IVH} produces candidates $L_B$'s compatible with 
those relations. 
The algorithm in \cite{IVH} finds the following candidates:
\begin{align} \label{candidates}
	(e_0, e_1, e_{\infty}) & = \left( \frac{3}{16},\frac{1}{4},\frac{1}{4} \right), \hspace{2mm} \deg(f(x)) = 2 \notag \\
	(e_0, e_1, e_{\infty}) & = \left( \frac{1}{8},\frac{1}{6},\frac{1}{2} \right), \hspace{2mm}  \deg(f(x)) = 3 \notag \\
	(e_0, e_1, e_{\infty}) & = \left(  \frac{3}{32}, \frac{1}{4}, \frac{1}{2} \right), \hspace{2mm}   \deg(f(x)) = 4 \notag \\
	(e_0, e_1, e_{\infty}) & = \left(  \frac{1}{8}, \frac{1}{4}, \frac{1}{2} \right), \hspace{2mm}    \deg(f(x)) = 5 \notag \\
	(e_0, e_1, e_{\infty}) & = \left(  \frac{1}{16}, \frac{1}{3}, \frac{1}{2} \right), \hspace{2mm}    \deg(f(x)) = 6 \notag \\
	(e_0, e_1, e_{\infty}) & = \left(  \frac{1}{8}, \frac{1}{3}, \frac{1}{2} \right), \hspace{2mm}   
\deg(f(x)) = 15 \notag.
\end{align} 
Here $e_0$, $e_1$, $e_{\infty}$ are 
the exponent-differences of a Gau{\ss} hypergeometric differential operator $L_B$ at 
$x=0$, $x=1$, and $x=\infty$. 
They determine $L_B$ 
upto an exp-product transformation.
The $\deg{(f(x))}$ of a rational function $f(x)$ 
is the maximum of the degree of its numerator 
and degree of its denominator. 
For (\ref{diff_op}), the algorithm finds six Gau{\ss} hypergeometric differential operators. 
Then the algorithm loops over each case and tries 
to recover $f(x)$ in (\ref{special_hyper_sol}). 

The fourth case $(e_0, e_1, e_{\infty})  = (1/8,1/4,1/2)$ and $\deg(f(x))=5$ 
gives a Gau{\ss} hypergeometric differential operator $L_B$ where
\begin{equation}\label{LB}
L_B = \frac{d^2}{dx^2}
	+{\frac { 3\left( 5\,x-3 \right)}{8\,x \left( x-1 \right) }}\frac{d}{dx}
	+{\frac {33}{256\,x \left( x-1 \right) }}	
\end{equation}
with an associated degree $5$ for $f(x)$.
One can always compute formal solutions of differential operators around a singular point. 
The algorithm in \cite{IVH} chooses a true singularity of (\ref{diff_op}), 
moves it to $x=0$, and then computes formal solutions of (\ref{LB}) and (\ref{diff_op}) at $x=0$. 
The point $x=0$ is a true singularity of (\ref{diff_op}). Formal solutions of (\ref{LB}) at $x=0$ are
\begin{align}
	y_1(x) 
	& = {_2F_1} \left[ \begin{matrix} \frac{3}{16}, \frac{11}{16} \\ \\ \frac{9}{8} \end{matrix}; x \right] 
	 = x^0 \left(  1+{\frac {11}{96}}x+{\frac {1881}{34816}}{x}^{2} + \dots \right) \label{solLB1} \\
	y_2(x) 
	& = {x}^{-\frac{1}{8}} \cdot {_2F_1} \left[ \begin{matrix} \frac{1}{16}, \frac{9}{16} \\ \\ \frac{7}{8} 
\end{matrix}; x \right] 
	 = { {x}^{-\frac{1}{8}} \left(1+{\frac {9}{224}}x
	+{\frac {255}{14336}}{x}^{2}+ \dots \right)}. \label{solLB2}
\end{align}
The exponents at $x=0$ are $0$ and $-1/8$, see (\ref{solLB1}) and (\ref{solLB2}).
Formal solutions of (\ref{diff_op}) at $x=0$ are
\begin{align*}
	Y_1(x) & = x^0 \left( 1+{\frac {15}{64}}x+{\frac {3825}{8192}}{x}^{2}+{\frac {3905875}{
	3670016}}{x}^{3}+ \dots \right) \\
	Y_2(x) & = {x}^{\frac{3}{8}} \left( 1+{\frac {249}{320}}x+{\frac {329697}{204800}}{x}^{2
	}+{\frac {774249529}{196608000}}{x}^{3}+ \dots  \right).
\end{align*}After a change of variables transformation $x \mapsto f(x)$ and 
an exp-product transformation one gets
\begin{align}
	Y_1(x) = \exp{\left( \int r(x)dx \right)} y_1(f(x)) \label{sol1} \\
	Y_2(x) = \exp{\left( \int r(x)dx \right)} y_2(f(x)). \label{sol2}
\end{align}\label{f_eq}If one takes the quotients of formal solutions of 
$Y_1(x)$, $Y_2(x)$ of (\ref{diff_op}) and $y_1(x)$, $y_2(x)$ of (\ref{LB}), 
then the effect of $\exp{\left( \int r(x)dx \right)}$ disappears: 
\begin{equation}\label{Q}
	Q(x) = \frac{Y_2{(x)}}{Y_1{(x)}}  
	\stackrel{?}{=} \frac{y_2{(f(x))}}{y_1{(f(x))}} = q(f(x)),
\end{equation}
where
\begin{equation}\label{q}
	q(x) = \frac{y_2{(x)}}{y_1{(x)}}.
\end{equation}
This suggests $f(x) = q^{-1}(Q(x))$, however,
the quotients of formal solutions, (\ref{Q}) and (\ref{q}), are unique up to a constant. 
So, the correct equation is:
\begin{equation}\label{series}
	f(x) = q^{-1}(c \cdot Q(x))
\end{equation}
where $c \in \mathbb{C}^*$ (here $c \in \mathbb{Q}^*$). 
If one knows the value of $c$, 
then (\ref{series}) gives
a power series expansion for
$f(x)$. 
That can be converted
to a rational function provided that
one has a degree (bound) for $f(x)$, which is $5$.
However, the value of $ c \in \mathbb{Q}$ is unknown and
there are infinitely many candidates for $c$. If one chooses a suitable 
prime number $p$ and works modulo $p$, then there are a finite number 
of candidates for the unknown constant $c$. 
The algorithm in \cite{IVH} chooses $p=13$ as the first 
suitable prime number and 
it loops over $c=1, \dots , p-1$ and for each $c$ tries to recover 
$f(x)$ modulo $p$ from its series expansion (\ref{series}) modulo $p$. 
If this succeeds for at least one $c$, then the algorithm uses 
Hensel lifting techniques \cite{HENSEL} to obtain $f(x)$ modulo higher powers of $p$. 
After that, it tries rational function and rational number reconstruction 
to find $f(x) \in \mathbb{Q}(x)$. 
After five Hensel lifting steps, the algorithm finds
\begin{equation}\label{pullback}
	f(x) = 
	{\frac {{x}^{3} \left( 3\,x+10 \right) ^{2}}{ \left( x+4 \right) 
	\left( 3\,{x}^{2}+4\,x-2 \right) ^{2}}}.
\end{equation}
Note that $\deg{(f(x))} = 5$. 

In order to find $r(x) \in \overline{\mathbb{Q}(x)}$, one can use Section 3.7 of \cite{IVH}.
The algorithm in \cite{IVH} finds
\begin{equation*}
	r(x) = -{\frac {45\,{x}^{4}+330\,{x}^{3}+690\,{x}^{2}+300\,x+480}{ \left( 16
	\,x+64 \right)  \left( 3\,x+10 \right)  \left( 3\,{x}^{2}+4\,x-2
	\right) x}}
\end{equation*}
and so
\begin{equation}\label{exp_part}
	\exp{\left( \int r(x)dx \right)} =
	{\frac {{x}^{3/8} {(3\,x+10)}^{1/4}}{ \left( x+4 \right) ^{3/16}
	\left( 3\,{x}^{2}+4\,x-2 \right) ^{3/8}}}.
\end{equation}

In the last step, the algorithm in \cite{IVH} forms solutions of (\ref{diff_op}) 
from solutions (\ref{solLB1}), (\ref{solLB2}) of (\ref{LB}). 
Then (\ref{sol1}) and (\ref{sol2}):
\begin{align}
\label{eq:h1}
	Y_1(x) & = \exp{\left( \int r(x)dx \right)} 
	\cdot {_2F_1} \left[ \begin{matrix} \frac{3}{16}, 
\frac{11}{16} \\ \\ \frac{9}{8} \end{matrix}; f(x) \right] \\
\label{eq:h2}
	Y_2(x) & = \exp{\left( \int r(x)dx \right)} 
	\cdot {f(x)}^{-\frac{1}{8}} \cdot {_2F_1} \left[ \begin{matrix} \frac{1}{16}, \frac{9}{16} \\ \\ \frac{7}{8} 
\end{matrix}; f(x) \right]
\end{align}
are solutions of (\ref{diff_op}) of type (\ref{special_hyper_sol})
with $\exp{\left( \int r(x)dx \right)}$ as in (\ref{exp_part}) and $f(x)$ as in (\ref{pullback}).

\begin{remark}
	The algorithm in {\rm \cite{IVH}} first simplifies the homogeneous parts of the differential equations 
studied,
        cf.~e.g. (\ref{eq:one}) and (\ref{eq:two}). Then it finds the hypergeometric solutions of the 
	simplified equations of type (\ref{special_hyper_sol}), and then uses this solutions 
	to form the solutions of their homogeneous parts of type (\ref{hyper_sol}).
\end{remark}
Since the differential equation (\ref{diff_op}) has more than three singularities, the argument
$f(x)$ of the $_2F_1$ solution has to have singularities. The expression in $_2F_1$ form has the advantage, that 
various properties of Gau\ss{}' hypergeometric functions can be used in subsequent calculations, 
would not be known otherwise.

The parameters $a,b,c$ of the solution 
are rational numbers and we will now investigate whether it is possible to map the homogeneous solutions 
(\ref{eq:h1}, \ref{eq:h2}) into 
complete elliptic integrals, which has been possible in all examples being discussed in Section~\ref{sec:3}.

\def\pFqnoargs#1#2{{}_#1F_#2}
\def\pFq#1#2#3#4#5{\pFqnoargs{#1}{#2}\biggl(\begin{matrix}{#3}\\{#4}\end{matrix}\,\bigg|\,#5\biggr)}

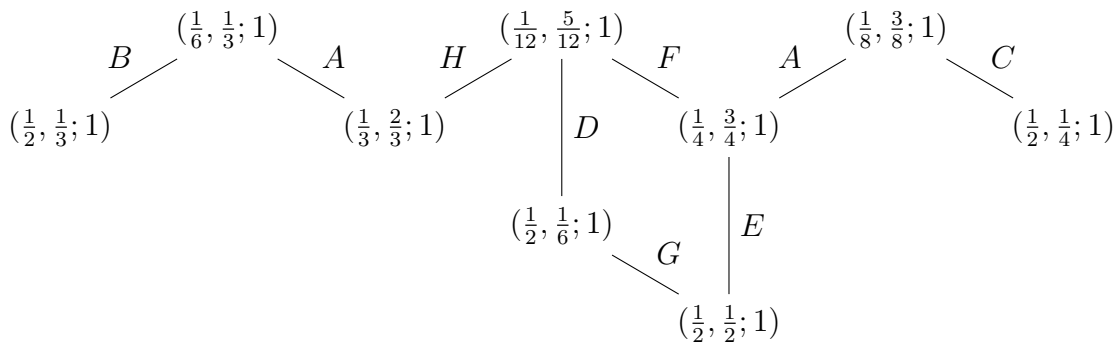
\begin{figure}
\centerline{%
\begin{tikzpicture}[baseline=(current bounding box.center),
  x={(22mm,0mm)},y={(0mm,13mm)}]
\node (F-12-13) {$(\frac12,\frac13;1)$};
\node (F-13-23) at (2,0) {{$(\frac13,\frac23;1)$}};
\node (F-14-34) at (4,0) {{$(\frac14,\frac34;1)$}};
\node (F-12-14) at (6,0) {$(\frac12,\frac14;1)$};
\node (F-16-13) at (1,1) {$(\frac16,\frac13;1)$};
\node (F-112-512) at (3,1) {$(\frac1{12},\frac5{12};1)$};
\node (F-18-38) at (5,1) {$(\frac18,\frac38;1)$};
\node (F-12-16) at (3,-1) {$(\frac12,\frac16;1)$};
\node (F-12-12) at (4,-2) {{$(\frac12,\frac12;1)$}};
\draw[-] (F-12-13)
  to node[above left] {${B}$} (F-16-13)
  to node[above right] {${A}$} (F-13-23)
  to node[above left] {$H$} (F-112-512)
  to node[above right] {${F}$} (F-14-34)
  to node[above left] {${A}$} (F-18-38)
  to node[above right] {${C}$} (F-12-14);
\draw[-] (F-112-512)
  to node[right] {${D}$} (F-12-16)
  to node[above right] {${G}$} (F-12-12)
  to node[right] {${E}$} (F-14-34);
\end{tikzpicture}
}
\caption[]{\label{FIG:DIA} The transformation of special $_2F_1$ functions under the triangle group.}
\label{fig:GR1}
\end{figure}

We would like to finally discuss a series of $_2F_1$ transformations in the case of the appearance of special rational 
parameters $a,b,c$, illustrated by the graph, Figure~\ref{fig:GR1}, cf. \cite{TAKEUCHI,VID2,BOSTAN2}.

\begin{center}
\begin{table}\centering
\renewcommand{\arraystretch}{1.6}
\begin{tabular}{|l|l|l|l|} \hline
$l$  & $d$  & $R$ & $f$ \\ \hline 

$A$ & $2$ & $1$ & $4x(1-x)$  \\ \hline
%


$B$ & $2$ & $(1-x)^{-1/6}$  & $\frac14 x^2 /(x-1)$ \\[1pt] \hline
%

$C$ & $2$ & $(1-x)^{-1/8}$  & $\frac14 x^2 /(x-1)$ \\[1pt] \hline
%

$D$ & $2$ & $(1-x)^{-1/12}$ & $\frac14 x^2 /(x-1)$ \\[1pt] \hline
%

$E$ & $2$ & $(1-x/2)^{-1/2}$ & $x^2 / (x-2)^{2}$ \\ \hline
%

$F$ & $3$ & $(1+3x)^{-1/4}$ &  $27x(1-x)^2 / (1+3x)^{3}$ \\ \hline
%

$G$ & $3$ & $(1 + \omega x)^{-1/2}$ & $1 - (x+\omega)^3/(x+\overline{\omega})^3$ \\ \hline
%

$H$ & $4$ & $(1-8x/9)^{-1/4}$ & $64x^3 (1-x) / (9-8x)^{3}$ \\ \hline
%

\end{tabular} \\[4pt]
\renewcommand{\arraystretch}{1}
\caption[]{\label{TAB1} The functions $R$ and $f$ for the different hypergeometric transformations of degree $d$ depicted in
Figure~\ref{FIG:DIA}.}
\end{table}
\end{center}

\noindent
If $(a,b;c)$ and $(a',b';c')$ are the endpoints of an edge labeled $l$ in the diagram,
with the latter endpoint above the former, then
\begin{equation}
{_2F_1} \left[ \begin{matrix} a, b \\ c \end{matrix}; x \right]
= R(x) \cdot {_2F_1} \left[ \begin{matrix} a', b' \\ c' \end{matrix}; f(x) \right]
\end{equation}
for $x$ sufficiently close to $0$, where $R,f$ are given in Table~\ref{TAB1}.

Here $\omega$ solves
\begin{eqnarray}
\omega^2+\omega+1=0 
\end{eqnarray}
and $d$ is the degree of $f$, the maximum of the degrees of the numerator and denominator.
The relations displayed in the above diagram can be used to map a wider class of $_2F_1$ solutions
to elliptic solutions. In various cases also the other relation obeyed by $_2F_1$ have to be applied
and one often ends up with complete elliptic integrals of the first and second kind, as in the cases
dealt with in the present paper.
\section{The Equal Mass Sunrise: from Kinematics to Elliptic Polylogarithms}
\label{sec:10}

\vspace*{1mm}
\noindent
In the following we summarize the necessary variable transformations in the case of the equal mass sunrise diagram,
dealt with in Refs.~\cite{BLOCH2,Adams:2014vja}. In the case of the kite diagram \cite{Adams:2016xah} the treatment is 
analogous. 
The intention is to represent the result in terms of the variable $q$, Eq.~(\ref{qeqa}). In different problems the module $k^2 = 
z(x)$ will refer to different expressions. Even dealing with the same case, different integration variables can be used,
with consequences for the form of $x(q)$. The inhomogeneity $N(x)$ will consequently have a different representation as a 
function 
of $q$, despite the final results are expressed in elliptic polylogarithms. In particular, all contributing functions, 
such as 
harmonic polylogarithms, may obtain a different representation in $q$.

We briefly discuss the results of Refs.~ \cite{BLOCH2,Adams:2014vja}, adding in some cases a few details.
\subsection{The treatment by Bloch and Vanhove, \cite{BLOCH2}.}
\label{sec:101}

\vspace*{1mm}
\noindent
Bloch and Vanhove \cite{BLOCH2} perform a treatment comparable to \cite{Adams:2014vja}, but with 
differences in the definition of the variable $t$ leading to a somewhat different expression for 
$I(q)$, also finally leading to elliptic polylogarithms. In obtaining their rational expressions
of $\eta$ functions they refer to the work of R.S.~Maier~\cite{MAIER} and obtain
\begin{eqnarray}
\label{eq:intB1}
I(q)= \frac{\eta(3\tau) \eta^5(\tau) 
\eta^4\left(\tfrac{3}{2}\tau\right)}{\eta^4\left(\tfrac{1}{2}\tau\right)}.
\end{eqnarray}
$I(q)$ is of {\sf w = 3} and belongs to $\Gamma_0(3)$. 
We first transform (\ref{eq:intB1}) using the relation by  M.D.~Rogers \cite{ROGERS}, 
Eq.~(4.21), and obtain
\begin{eqnarray}
\label{eq:intB1a}
I(q) = 
\frac{\eta^9\left(\tfrac{3}{2}\tau\right)}{\eta^3\left(\tfrac{1}{2}\tau\right)} +
   \frac{\eta^9\left(3\tau\right)}{\eta^3\left(\tau\right)}~.
\end{eqnarray}
A generating function representation in $q$, using the first terms, is given in \cite{OEIS} 
{\tt A106402} which finally yields
\begin{eqnarray}
I(q) = \sum_{k=1}^\infty k^2 \left(
  \frac{q^k}{1 + q^k + q^{2k}}
+ \frac{q^{2k}}{1 + q^{2k} + q^{4k}}\right).
\end{eqnarray}
This result is now transformed into a generalized Lambert series representation 
\cite{LAMBERT,BB,AGRAWAL,ARNDT} by using
\begin{eqnarray}
\label{eq:LAMB1}
L_0(x) = \frac{x}{1+x+x^2} = - \frac{i}{\sqrt{3}} \left[\Li_0(\rho_3 x) - \Li_0(\rho_3^2 x)\right]
\end{eqnarray}
with
\begin{eqnarray}
\label{eq:Iq2}
I(q) = \sum_{k=1}^\infty k^2 \left[L_0(q^k) + L_0(q^{2k})\right],
\end{eqnarray}
where one has
\begin{eqnarray}
\Li_0(\alpha q^k) = {\rm ELi}_{0;0}(\alpha;1;q).
\end{eqnarray}
Further logarithmic $q$-integrals, cf.~(\ref{eq:ELP5}), lead to higher weight elliptic polylogarithms.
Eq.~(\ref{eq:Iq2}) is closely related to corresponding expressions given in \cite{Adams:2014vja} to 
which we turn now.

\subsection{The treatment by Adams et al., \cite{Adams:2014vja}.}
\label{sec:102}

\vspace*{1mm}
\noindent
Choosing 
\begin{eqnarray}
\tau = i \frac{{\bf K}(k'^2)}{{\bf K}(k^2)} = \frac{1}{i\pi} \ln(q),
\end{eqnarray}
the integration variable $t = m^2 y$ is obtained from the product of the modules squared
\begin{eqnarray}
\label{eq:kk'}
k^2k'^2 = - \frac{16 y}{(1-y)^3 (9-y)} = 16 \left\{\frac{\eta\left(\tfrac{\tau}{2}\right) 
\eta(2\tau)}{\eta(\tau)^2}\right\}^{24} 
= \left\{\frac{\vartheta_2(\tau) \vartheta_4(\tau)}{\vartheta_3(\tau)}\right\}^4.
\end{eqnarray}
in Ref.~\cite{Adams:2014vja} while calculating the sunrise-integral.
Eq.~(\ref{eq:kk'}) is a modular function for $\Gamma_0(4)$ which 
is inverted for $y$
\begin{eqnarray}
\label{eq:skk'}
y = -9 \left\{\frac{\eta(\tau) \eta\left(\tfrac{3\tau}{2}\right) \eta(6\tau)} 
{\eta\left(\tfrac{\tau}{2}\right) \eta(2\tau) \eta(3\tau)}\right\}^4,
\end{eqnarray}
a modular function for $\Gamma_0(12)$. It is also the variable of the inhomogeneity, and in 
general of harmonic polylogarithms and related functions, depending on the complexity of the problem. The validity 
of (\ref{eq:skk'}) can be proven by applying a similar treatment as shown in Section~4.
Note that the cubic Legendre-Jacobi cubic transformation, cf.~\cite{Broadhurst:2008mx}, cannot be used directly,
unlike the case in (\ref{eq:cubic1}, \ref{eq:cubic2}).

The integrand of the special solution has been obtained by
\begin{eqnarray}
\label{eq:intW}
I(q)= 3 \sqrt{3}  \frac{\eta^{11}(\tau) \eta^7(3\tau)} 
{\eta^5\left(\tfrac{\tau}{2}\right) \eta\left(\tfrac{3\tau}{2}\right) \eta^5(2\tau)\eta(6\tau)}.
\end{eqnarray}

It is useful to consult Sloan's on-line encyclopedia of integer sequences \cite{OEIS} for this example. The 
corresponding solution has been given by D.~Zagier in 2009 \cite{ZAGIER2} by entry {\tt A214262} \cite{OEIS} 
for the series\footnote{It is interesting to note that this $q$-series 
is closely related to the series used by Beukers \cite{BEUKERS} in his series-proof of the irrationality of 
$\zeta_2$ and $\zeta_3$ related through an Eichler integral \cite{EICHLER}. Already his earlier proof based on 
integrals \cite{BEUKERS1} used functions playing a central role in the calculation of Feynman integrals.}
\begin{eqnarray}
\label{eq:intW1}
I(q)= -3 \sqrt{3} \sum_{n=1}^\infty \sum_{d|n} (-1)^{d-1} \legendre{-3}{n/d} d^2 (-q)^n,
\end{eqnarray}
where
$\legendre{a}{b}$ denotes the Legendre symbol 
\cite{HARDY}. The inner sum in (\ref{eq:intW1}) can be carried out resulting in 
\begin{eqnarray}
\label{eq:intW2}
I(q) =  3 \sqrt{3} \sum_{k=1}^\infty k^2 \frac{q^k}{1 + (-q)^k + q^{2k}}.
\end{eqnarray}
Note that somewhat different integrands $I(q)$ appear in the treatment in Ref. \cite{BLOCH2} and 
\cite{Adams:2014vja}, which are related, however. The modular form (\ref{eq:intW}) is of $\Gamma_0(12)$. 

Next the $q$-dependent part of (\ref{eq:intW2}) is again transformed into the Lambert form, cf.~(\ref{eq:LAMB1}),
and two integrals are performed to obtain special solution \cite{Adams:2014vja}:
\begin{eqnarray}
\label{eq:SspecW}
S_{\rm special} &=& \int_0^q \frac{dq_1}{q_1} \int_0^{q_1} \frac{dq_2}{q_2} I(q_2)
= \frac{3}{i} \sum_{k=1}^\infty (-1)^k \left[\Li_2\left(\rho_3(-q)^k\right) - \Li_2\left(\rho_3^{-1}(-q)^k\right)
\right]\nonumber\\
\\
&\equiv& \frac{3}{\pi} \overline{E}_{2;0}(\rho_3;-1;-q),
\end{eqnarray}
where we have dropped a common pre-factor.

To be able to incorporate the inhomogeneity into the solution it is necessary to express 
the harmonic polylogarithms depending on $y$ as a function of $q$. The lowest weight HPLs are in this 
case \cite{Adams:2014vja} 
\begin{eqnarray}
H_0(y) &=& \ln(-9q) -4 \overline{E}_{0;-1;2}(\rho_3;-1;-q)
\\
H_1(y) &=& 3 \left[\overline{E}_{1;0}(-1;1;-q) -\overline{E}_{1;0}(\rho_6;1;-q)\right]
\\
H_{0,1}(y) &=& 3 \left[\overline{E}_{2;1}(-1;1;-q) -\overline{E}_{2;1}(\rho_6;1;-q)\right]
\nonumber\\ &&
- 12  \left[
  \overline{E}_{0,1;-1,0;2}(\rho_3,-1;-1,1;-q) 
- \overline{E}_{0,1;-1,0;2}(\rho_3,\rho_6;-1,1;-q) \right],~~\text{etc.}
\end{eqnarray}
They are different to those obtained in the case presented in Section~\ref{sec:5}.
In \cite{Adams:2014vja} only HPLs over the alphabet $\{0,1\}$ occur. We note that the kinematic variable
(\ref{eq:skk'}) does not lead to a mirror symmetry like the one obtained in Section~\ref{sec-522}.
\section{A Series of Sums}
\label{sec:11}

\vspace*{1mm}
\noindent
In a recent paper \cite{Ananthanarayan:2016pos} on the sunrise graph, which belongs to the context
of the present paper, several sum-representations were presented, which could not yet be calculated.
In the following we give the solutions for all single infinite sums in terms of polylogarithmic expressions
with root arguments, limited to at most $\Li_2(z)$. They can be calculated with the techniques made available 
in the package {\tt HarmonicSums.m}, which were developed in the context of binomial 
sums~\cite{Ablinger:2014bra,Ablinger:2014yaa,Ablinger:2015tua}\footnote{For similar investigations in the case of 
infinite sums see \cite{Davydychev:2003mv,Weinzierl:2004bn}.}.

These sums may be represented referring to harmonic sums \cite{Blumlein:1998if,Vermaseren:1998uu} defined by
\begin{eqnarray}
S_{b,\vec{a}}(N) = \sum_{k=1}^N \frac{({\rm sign}(b))^k}{k^{|b|}} S_{\vec{a}}(k),~~~S_\emptyset(N) = 1;~~~
\left.S_{\vec{a}}(N)\right|_{N=0} = 0.
\end{eqnarray}
We use the replacements for the poly-gamma functions \cite{Blumlein:1998if}
\begin{eqnarray}
S_1(N)     &=& \psi(N+1) + \gamma_E \\
S_{k+1}(N) &=& \frac{(-1)^k}{k!} \psi^{(k)}(N+1) + \zeta_{k+1},~~k \in \mathbb{N}, k \geq 1~,
\end{eqnarray}
with $\gamma_E$ the Euler-Mascheroni number. Furthermore, single cyclotomic harmonic sums contribute \cite{Ablinger:2011te}.
They are defined by
\begin{eqnarray}
S_{\{a,b,c\}}(N)     &=& \sum_{k=1}^N \frac{({\rm sign}(c))^k}{(a k + b)^{|c|}},
\end{eqnarray}
with
\begin{eqnarray}
\psi\left(\frac{3}{2} +i\right) = 2 S_{\{2,1,1\}}(i) + \gamma_E
\end{eqnarray}
as one example.\\

\noindent
One obtains the following relations  
\begin{eqnarray}
&&s_1(x) = \sum_{i=0}^\infty \frac{i!(i+1)!}{(2i+3)!} x^i \left[-\frac{8 (1 + i)}{(1 + 2 i) (3 + 2 i)}  + 2 S_1(i)
- 2 S_1(2 i) +\ln(x)\right]  
\nonumber\\ 
&&=\frac{i}{2} \sqrt{\frac{4 - x}{x^3}}
\ln^2\left[\frac{1}{2}\left(2 + \sqrt{(-4 + x)x} - x\right)\right] 
- \frac{i}{(2 (-4 + x) x^2)} \ln\left[\frac{1}{2}\left(2 + \sqrt{(-4 + x)x} - x\right)\right] 
\nonumber\\ && \times
\Biggl\{3 \Biggl[-4 \sqrt{(4 - x) x} + \sqrt{(4 - x)^3 x} 
+  \sqrt{(4 - x) x^3}\Biggr] 
- 2 (4-x) \sqrt{(4 - x)}\ln(x) 
\nonumber\\ &&
- 4 (4 - x) \sqrt{\frac{4 - x}{x^3}} x^2 \ln\left[\frac{1}{2}\left(-\sqrt{(-4 + x)x} + x\right)\right]\Biggr\} 
+ \frac{1}{x^{3/2}}\Biggl\{
 2 i \sqrt{4 - x} \zeta_2 
\nonumber\\ &&
+ 2 \sqrt{x} \left(-2 + \ln(x)\right) 
- 2 i \sqrt{4 - x} \Li_2\left[\frac{1}{2}\left(2 + \sqrt{(-4 + x)x} - x\right)\right]\Biggr\},~~ 0 \leq x < 1,
\\  [5pt]
&&s_2(x) =\sum_{i=0}^\infty \frac{i! (2 + i)!}{3 +2i)!} x^i 
\Biggl\{
\frac{2 \left(-5 - 21 i - 39 i^2 - 32 i^3 + 12 i^5 + 4 i^6\right)}{(1 + i)^2 (2 + i)^2 (1 + 2 i)^2} 
+ \frac{\pi^2}{3} 
\nonumber\\ &&
+ \frac{4 (1 + 4 i + 2 i^2)}{(1+i)(2+i)} S_{\{2, 1, 1\}}(i) + 4 S^2_{\{2, 1, 1\}}(i)
- 2 S_{\{2, 1, 2\}}(i) 
+ \Biggl[-\frac{2 (1 + 4 i + 2 i^2)}{(1 + i) (2 + i)} 
\nonumber\\ &&
- 
    4 S_{\{2, 1, 1\}}(i) \Biggr] \ln(x) 
+ 
 \ln^2(x) + \Biggl[\frac{(2 (1 + 4 i + 2 i^2)}{(1 + i) (2 + i)} - 
    4 S_{\{2, 1, 1\}}(i) + 2 \ln(x) \Biggr] S_1(i) + S_1^2(i) 
\nonumber\\ &&
+ 
 \frac{3}{2} S_2(i) - 2 S_2(2i) \Biggr\}
\nonumber\\ && 
=\frac{\sqrt{\pi}}{24} x \Biggl\{-12 + \pi^2 \left(-2 + x - \sqrt{(-4 + x) x}\right) 
+ 3 \sqrt{(-4 + x) x} \ln^2(2) 
+ 3 \ln(x) [-4 
\nonumber\\ &&
+ (-2 + x) \ln(x)] 
- 3 \sqrt{(-4 + x) x} \left[
-2 \ln\left[1 - \sqrt{\frac{-4 + x}{x}}\right] 
+2 \ln\left[1 + \sqrt{\frac{-4 + x}{x}}\right]\right]
\nonumber\\ && \times
   \left[\ln(4 x) - 2 \ln\left[x - \sqrt{(-4 + x) x}\right]\right] + 
   \ln\left[-2 + x - \sqrt{(-4 + x) x}\right]
\nonumber\\ && \times 
   \ln\left[-2 + x + \sqrt{(-4 + x) x}\right] 
- 
   12 \sqrt{(-4 + x) x} \Li_2\left[ \frac{1}{2} \left(2 - x + \sqrt{(-4 + x) x}\right)\right]\Biggr\}
\\  [5pt]
&& s_3(x) = \sum_{i=0}^\infty \frac{i! (2+i)!}{(3+2i)!} x^i \Biggl[
-\frac{13+16 i+4 i^2}{(2+i) (1+2 i) (3+2 i)} + 2 S_1(i) -2 S_1(2i) + \ln(x)\Biggr] 
\nonumber\\ &&
=\frac{1}{x \sqrt{(4 - x) x}}
\Biggl\{
- 2 i (-2 + x) \zeta_2 
+ 2 i (-2 + x) \ln\left[1 - \frac{\sqrt{-4 + x} - \sqrt{x}}{\sqrt{-4 + x} + \sqrt{x}}\right] 
   \ln\left[\frac{\sqrt{-4 + x} - \sqrt{x}}{\sqrt{-4 + x} + \sqrt{x}}\right]
\nonumber\\ && 
+  \left(1- \frac{x}{2}\right)\ln^2\left[\frac{\sqrt{-4 + x} - \sqrt{x}}{\sqrt{-4 + x} + \sqrt{x}}\right]
+ \sqrt{(4 - x) x} \ln(x) 
- i (-2 + x) \ln\left[\frac{\sqrt{-4 + x} - \sqrt{x}}{
  \sqrt{-4 + x} + \sqrt{x}}\right] \ln(x) 
\nonumber\\ &&
- 2 \Biggl[\sqrt{(4 - x) x} 
 + i (2 - x) \Li_2\left[\frac{\sqrt{4 - x} + i \sqrt{x}}{\sqrt{4 - x} - i \sqrt{x}}\right]
\Biggr]
\Biggr\},~~0 \leq x < 1~.
\\ [5pt]
&& s_4(x) = \sum_{i=0}^\infty \frac{(2+2i)!}{i! (2+i)!} \frac{1}{x^i}
\Biggl\{
-\frac{6}{(2 + i)^2 (1 + 2 i)} + 2 \zeta_2 - \frac{6 \ln(x)}{(2 + i) (1 + 2 i)} + \ln^2(x) + 4 S_1^2(i)
\nonumber\\ &&
+ \left[
- \frac{12}{(2 + i) (1 + 2 i)} 
+ 4 \ln(x) - 8 S_1(2i)\right] S_1(i) + \left[\frac{12}{(2 + i) (1 + 2 i)} - 4 \ln(x)\right] S_1(2i)
\nonumber\\ &&
+ 4 S_1^2(2i) + 2 S_2(i) - 4 S_2(2i)
\Biggr\} 
\nonumber\\ &&
=\frac{x}{6 \sqrt{-4 + x}} \Biggl\{
12 \sqrt{-4 + x} + \pi^2 \sqrt{x} \left(-2 + x - \sqrt{(-4 + x) x}\right) 
-  3 (-2 + x) \sqrt{x} 
\nonumber\\ && \times
\Biggl[\ln\left(1 - i \sqrt{-1 + \frac{4}{x}}\right) - \ln\left(1 + i \sqrt{-1 + \frac{4}{x}}
\right)\Biggr]^2 
+  6 (-2 + x) \sqrt{x} \Biggl[
- \ln\left(1 - \sqrt{1 - \frac{4}{x}}\right) 
\nonumber\\ &&
+ \ln\left(1 + \sqrt{1 - \frac{4}{x}}\right)\Biggr] \ln(x) 
- 3 \sqrt{-4 + x} \ln(x) (-4 + x \ln(x)) - 
   12 (-2 + x) \sqrt{x} 
\nonumber\\ &&
\Biggl[
\ln\left(1 - i \sqrt{-1 + \frac{4}{x}}\right) 
   - \ln\left(1 + i \sqrt{-1 + \frac{4}{x}}\right) \Biggr] \ln\left[
     \frac{1}{2} \left(x - \sqrt{(-4 + x) x}\right) \right] 
\nonumber\\ &&
+ 12 (-2 + x) \sqrt{x} \Li_2\left[\frac{1}{2} \left(2 - x + \sqrt{(-4 + x) x}\right)\right]
\Biggr\},~~x > 9~.
\\ [5pt]
&& s_5(x) = \sum_{i=0}^\infty \frac{i! (1+i)!}{(2+2i)!} x^i
\Biggl[-\frac{2}{1 + 2 i} + 2 S_1(i) - 2 S_1(2i) + \ln(x) \Biggr]
\nonumber\\ && 
= \frac{1}{\sqrt{(4 - x) x}} \Biggl\{
2 i \ln\left[1 - \frac{\sqrt{-4 + x} - \sqrt{x}}{\sqrt{-4 + x} + \sqrt{x}}\right] 
   \ln\left[\frac{\sqrt{-4 + x} - \sqrt{x}}{\sqrt{-4 + x} + \sqrt{x}}\right] 
- \frac{i}{2} \ln^2\left[\frac{\sqrt{-4 + x} - \sqrt{x}}{\sqrt{-4 + x} + \sqrt{x}}\right]  
\nonumber\\ &&
- i \ln\left[\frac{\sqrt{-4 + x} - \sqrt{x}}{\sqrt{-4 + x} + \sqrt{x}}\right] \ln(x) 
- 2 i \Biggl\{ \zeta_2 - 
    \Li_2\left[\frac{\sqrt{4 - x} 
+ i \sqrt{x}}{\sqrt{4 - x} - i \sqrt{x}}\right]\Biggr\}\Biggr\},
0 \leq x < 1~.
\\ [5pt]
&& s_6(x) = \sum_{i=0}^\infty \frac{i!}{(2+i)!} \frac{1}{x^i} \left[\frac{1}{1 + i} + 2 S_1(i) - \ln(x)\right] 
\nonumber\\ &&
=(1 - x) x \ln^2\left(1 - \frac{1}{x}\right) 
- x \ln(x) 
- x(1-x) \ln\left(1 - \frac{1}{x}\right) (1 - (1 - x) x \ln(x)) 
\nonumber\\ &&
+ x \left[1 + \Li_2\left(\frac{1}{x}\right) \right],
x \geq 9~.
\\ [5pt]
&& s_7(x) = \sum_{i=0}^\infty \frac{(1 + 2 i)!}{i! (1 + i)!} \frac{1}{x^i}
\Biggl\{
-\frac{3 + 4 i}{(1 + i)^2 (1 + 2 i)^2} + \frac{\pi^2}{3} + \Biggl[
-\frac{1}{(1 + i) (1 + 2 i)} + 2 S_1(i)
\nonumber\\ &&
- 2 S_1(2 i) 
+ \ln(x)
\Biggr]^2 
+ 2 S_2(i) - 4 S_2(2i) \Biggr\} 
\nonumber\\ &&
= \frac{1}{6 \sqrt{-4+x}}
\Biggl\{
-\pi^2 x \sqrt{-4 + x}  - 
  3 x \sqrt{-4 + x}  \ln^2(x) + 
  x^{3/2} \Biggl\{ \pi^2 - 
     3 \Biggl[-\ln\left(1 - \sqrt{1 - \frac{4}{x}}\right) 
\nonumber\\ &&
+ \ln\left(1 + \sqrt{1 - \frac{4}{x}}\right) \Biggr]^2 + 
     6 \Biggl[-\ln\left(1 - \sqrt{1 - \frac{4}{x}}\right) + \ln\left(1+ \sqrt{1 - \frac{4}{x}}\right)\Biggr]
\Biggl[
2 \ln(2) + \ln(x) 
\nonumber\\ &&
- 2 \ln\left(x - \sqrt{(-4 + x) x}\right)
\Biggr] + 
   12 \Li_2\left[\frac{1}{2}\left(2 - x + \sqrt{(-4 + x) x}\right)\right]
\Biggr\} 
\Biggr\},~~~x \geq 9. 
\\ [5pt]
&&s_8(x) = - u_1 \sum_{i=0}^\infty \frac{i!((1+i)!)^2}{(3+i)!(3+2i)!} 
x^{i+2}
= - \frac{u_1 x^2}{36} 
{_3F_2} \left[ \begin{matrix} 1, 1, 2\\4,\tfrac{5}{2}  \end{matrix}; \frac{x}{4} \right]
\nonumber\\
&& = 
u_1 \Biggl\{-\frac{1}{4}(4+7x)
- i \frac{\sqrt{4-x}(2+x)}{2 \sqrt{x}} \ln\left[\frac{\sqrt{-4 + x} - \sqrt{x}}{\sqrt{-4 + x} + \sqrt{x}}\right] 
+ \frac{1-x}{x} \ln^2\left[\frac{\sqrt{-4 + x} - \sqrt{x}}{\sqrt{-4 + x} + \sqrt{x}}\right]
\Biggr\},
\nonumber\\ &&
\text{where}~~~~x \equiv u_2/u_1. 
\end{eqnarray}
The last sum does not form a genuine generalized hypergeometric function, but obeys a logarithmic representation.
All the yet uncalculated double sums in \cite{Ananthanarayan:2016pos} cannot be solved completely in terms of 
iterative integrals, as has been checked by the algorithms used in \cite{Ablinger:2015tua}, and will
therefore involve non-iterative integrals.

\vspace{5mm}
\noindent
{\sf Note added.} \\
After completion of the present paper the preprint \cite{Adams:2017ejb} appeared. In this paper
more special cases, compared to those in the present paper, are considered, which allow representations
in terms of modular forms and powers of $\ln(q)$ only. The latter terms, appearing also in the present case, are 
related to Eichler integrals \cite{EICHLER} in \cite{Adams:2017ejb}.

\newpage
 
\end{document}